\documentclass[fleqn,usenatbib]{mnras}

\usepackage{subfig}
\usepackage{graphicx}
\usepackage{amsfonts}
\usepackage{lastpage}
\usepackage{float}
\usepackage{newtxtext,newtxmath}
\usepackage{hyperref}
\graphicspath{{./}{./figures/}}
\usepackage{tcolorbox}

\usepackage[T1]{fontenc}
\usepackage{ae,aecompl}

\newcommand{\refresponse}[1]{{\color{black}{#1}}}

\title[PZ Impact for LSST $3\times2$pt Analysis]{Forecasting the Impact of Source Galaxy Photometric Redshift Uncertainties on the LSST $3\times2$pt Analysis}

\author[]{Tianqing Zhang$^{1,2}$\thanks{tq.zhang@pitt.edu}, Husni Almoubayyed$^2$, Rachel Mandelbaum$^2$, Markus Michael Rau$^{2,3,4}$, \newauthor Nikolina \v{S}ar\v{c}evi\'c$^{4,5}$, C. Danielle Leonard $^4$, Jeffrey A. Newman$^1$, Brett Andrews$^1$,  \newauthor Shuang Liang$^6$,    The LSST Dark Energy Science Collaboration \\
$^{1}$Department of Physics and Astronomy and PITT PACC, University of Pittsburgh, Pittsburgh, PA 15260\\
$^2$McWilliams Center for Cosmology and Astrophysics, Department of Physics, Carnegie Mellon University, 5000 Forbes Ave, Pittsburgh, PA 15213\\
$^{3}$High Energy Physics Division, Argonne National Laboratory, Lemont, IL 60439\\
$^{4}$School of Mathematics, Statistics and Physics, Newcastle University, Herschel Building, NE1 7RU Newcastle-upon-Tyne, UK \\
$^{5}$Department of Physics, Duke University, Durham NC 27708, USA
$^{6}$SLAC National Accelerator Laboratory, 2575 Sand Hill Road, Menlo Park, CA 94025\\
}

\date{\today}

\begin{document}

\label{firstpage}
\pagerange{\pageref{firstpage}--\pageref{LastPage}}
\maketitle

\begin{abstract}
Photometric redshifts \refresponse{of the source galaxies} are a key source of systematic uncertainty in the Rubin Observatory Legacy Survey of Space and Time (LSST)'s galaxy clustering and weak lensing analysis, i.e., the $3\times 2$pt analysis. This paper introduces a Fisher forecast code \texttt{FisherA2Z} for the LSST Year 10 (Y10) $3 \times 2$pt and cosmic shear analyses, utilizing a 15-parameter redshift distribution model, with one \refresponse{redshift} bias, variance, and outlier rate per tomographic bin.
\texttt{FisherA2Z} employs the Core Cosmology Library \texttt{CCL} to compute the large-scale structure power spectrum and incorporates a four-parameter nonlinear alignment model for intrinsic alignments. We evaluate the impact of marginalizing over redshift distribution parameters on weak lensing, forecast biases in cosmological parameters due to redshift errors, and assess cosmological parameter sensitivity to redshift systematic parameters using decision trees.
The sensitivity study reveals that for LSST $3 \times 2$pt analysis, $S_8$ is most sensitive to the mean redshift of the fourth out of the five source tomographic bins, while other cosmological parameters possess different sensitivities. Additionally, we provide cosmological analysis forecasts based on different scenarios of spectroscopic training datasets. We find that the figures-of-merit for the cosmological results increase with the number of spectroscopic training galaxies, and with the completeness of the training set above $z=1.6$, assuming the redshift information comes solely from the training set galaxies without other external constraints. 
\end{abstract}

\restylefloat{table}

\section{Introduction}

One major goal of the next generation of astronomical imaging surveys (e.g., the Vera C.~Rubin Observatory Legacy Survey of Space and Time, hereafter LSST; \citealt{2009arXiv0912.0201L,Overview}) is to increase our understanding of dark energy \citep{Weinberg}.
These surveys use a variety of methods to probe dark energy, such as weak gravitational lensing, galaxy clustering, Type~Ia supernovae, and baryon acoustic oscillations. 
Dark energy analyses combining multiple probes rely on the existence of accurate and well--characterized photometric redshifts \citep[e.g.,][]{Ma,2010MNRAS.401.1399B,2015APh....63...81N}.  

In this work, we focus on three observables, i.e., correlation functions involving the auto-- and cross--correlations of the positions and shears of galaxies.  The shear is the observed galaxy shape distortion due to the effect of gravitational lensing, the deflection of light from distant galaxies as it passes through the large--scale structure of the Universe \citep[for a review, see][]{weaklensingreview2}.  
These observables are (a) the galaxy two--point correlation function (`galaxy clustering'), (b) the shear--shear correlation function, which reveals the correlation between the shapes of galaxies due to weak gravitational lensing by the large--scale structure of the Universe, otherwise known as ``cosmic shear'' and (c) the cross--correlation between the source galaxy shapes and lens (foreground) galaxy positions, or `galaxy--galaxy lensing'. These auto-- and cross--correlation functions probe the large--scale structure of the Universe, and interpreting them requires an understanding of the redshift distributions of galaxies divided into redshift bins (hereafter referred to as tomographic bins). Joint analysis of the large--scale structure using these three two--point correlations is referred to as `$3 \times 2$pt' analysis \citep[e.g.,][]{2021A&A...646A.140H,2022PhRvD.105b3520A}. 

Separating the galaxies into tomographic bins allows us to extract more information on the growth of structure in the Universe from the measurements \citep[e.g.,][]{2002PhRvD..65b3003H,
  2002PhRvD..65f3001H, 2004PhRvD..70l3515B, 2004ApJ...600...17B,
  2004ApJ...601L...1T}, as well as provide a more precise treatment of systematics. Using these correlation functions measured within and across redshift bins, rather than the correlation functions of the galaxy sample overall, is particularly helpful in providing significantly more information about dark energy, through its impact on cosmic structure growth.

Future imaging surveys will enable measurements of cosmological parameters at unprecedented statistical precision, and therefore require tighter control of systematic biases and uncertainties than in precursor surveys \citep[e.g.,][]{Weinberg,rachelreview}. In this work, we focus on photometric redshift (or photo--$z$) uncertainties of the source galaxies, which have been explored in the context of weak lensing and/or $3 \times 2$pt analysis in the past \citep[e.g.,][]{Ma,2010MNRAS.401.1399B,2012JCAP...04..034H,2014MNRAS.444..129C,DESCSRD,schaan,Tessore2020,RuizZapatero2023, zuntz2021, moskowitz2023, moskowitz2024}.  Previous studies have often used somewhat realistic photo--$z$ error models in a simplified single probe analysis without accounting for some of the key systematic uncertainties for weak lensing (e.g., \citealt{Ma}), or used overly simplistic photo--$z$ error models with a full $3 \times 2$pt analysis including many sources of systematic uncertainty (such as a two--parameter photo--$z$ error model with constant redshift bias and variance error parameters, and no outlier error model, as in \citealt{DESCSRD}).  Since photometric redshift uncertainties can be degenerate with other sources of systematic uncertainty in weak lensing \citep[e.g.,][]{2021A&A...650A.148S, Leonard2024}, there is a strong motivation to consider requirements on a more complex photometric redshift error model in the context of a full $3 \times 2$pt analysis with other key sources of systematic uncertainty.

 In this work, we create a 15--parameter photo--$z$ model, with one bias, one variance and one outlier parameter in each of the five \refresponse{source} redshift bins.
 Moreover, the photometric redshift error distribution for the outliers is estimated using a photometric redshift catalog derived from a simulated extragalactic catalog. 
 We use a full $3 \times 2$pt analysis and include parameters to marginalize over other key sources of systematic uncertainty, such as intrinsic alignments \citep[for reviews, see][]{IA4,IA1,IA2,IA3,Lamman2024} and galaxy bias \citep[for a review, see][]{2018PhR...733....1D}. 
 We get approximate cosmological constraints for the $3 \times 2$pt analysis via Fisher forecasting, which is based on numerical differentiation of the data vector with regard to the parameters. 

 Fisher forecasting is a technique to approximate the parameter constraining power of an analysis with relatively cheap computing cost, and had been implemented in 3x2pt cosmology \citep[e.g.,][]{sarcevic2025,DESCSRD}.
 In this work, we present the Fisher forecast software \texttt{FisherA2Z}\footnote{\url{https://github.com/LSSTDESC/fisherA2Z}} as an open--source software. 
 Using the complex and realistic model for photo--$z$ errors and the Fisher forecasting, we study the effect of incorrect values for the photo--$z$ error parameters on cosmological parameters when inferring the ensemble redshift distributions for the source redshift bins. This allows us to (a) derive a greater understanding of which redshift bins and parameters are most important in determining the constraining power on cosmological parameters, and (b) set requirements on the minimum knowledge of photo--$z$ bias, variance, and outliers such that they are not a dominant factor in the error budget for the cosmological parameters. While we analyze an LSST-like survey setup, the methodology for this purpose is easily transferable to other survey setups.  Our results should help guide and prioritize efforts on obtaining more accurate photo--$z$ models by obtaining representative spectroscopic redshift samples \citep[e.g.,][]{2015APh....63...81N,2017ApJ...841..111M} and by cross--correlating with wide--area (non--representative) spectroscopic samples \citep[e.g.,][]{2008ApJ...684...88N,2013MNRAS.433.2857M,2020MNRAS.491.4768R}.

 In this work, we are assuming that the lens galaxies are a distinct population from the source galaxies and that the lens redshift distributions are precisely known. \refresponse{The rationale behind this choice is because (a) many $3 \times 2$pt analyses \citep[e.g.,][]{3x2pt_des} use lens galaxies with more secure photometric redshift because they occupy a smaller color space and are brighter; and (b) some $3 \times 2$pt analyses uses spectroscopic lens samples \citep[e.g.,][]{Sugiyama2023}, which have perfectly known lens redshift distributions. }
 We are assuming a specific tomographic binning scheme for the LSST Y10 $3 \times 2$pt analysis, which consists of ten lens bins and five source bins. Notice that although \texttt{FisherA2Z} has the ability to forecast LSST Y1 and Y10, this work focus on LSST Y10.

This work is structured as follows: In Section~\ref{sec:method}, we explain the methodology used to create the photometric redshift error model, carry out the cosmological parameter forecasts, and assess the importance of different photometric redshift error parameters. In Section~\ref{sec:results}, we present the results obtained from the forecast and feature importance method. Finally, in Section~\ref{sec:conclusion}, we summarize our findings and discuss future work.

\section{Background and Method}
\label{sec:method}

In this section, we describe the key concepts used for creating the source photometric redshift model, and computing the data vectors and covariance matrices, and carrying out the Fisher forecasts.  We also describe the methods used to identify which aspects of the photometric redshift error model have the largest impact on cosmological parameter inference.
 
\begin{figure*}
    \centering
    \includegraphics[width=0.8\textwidth]{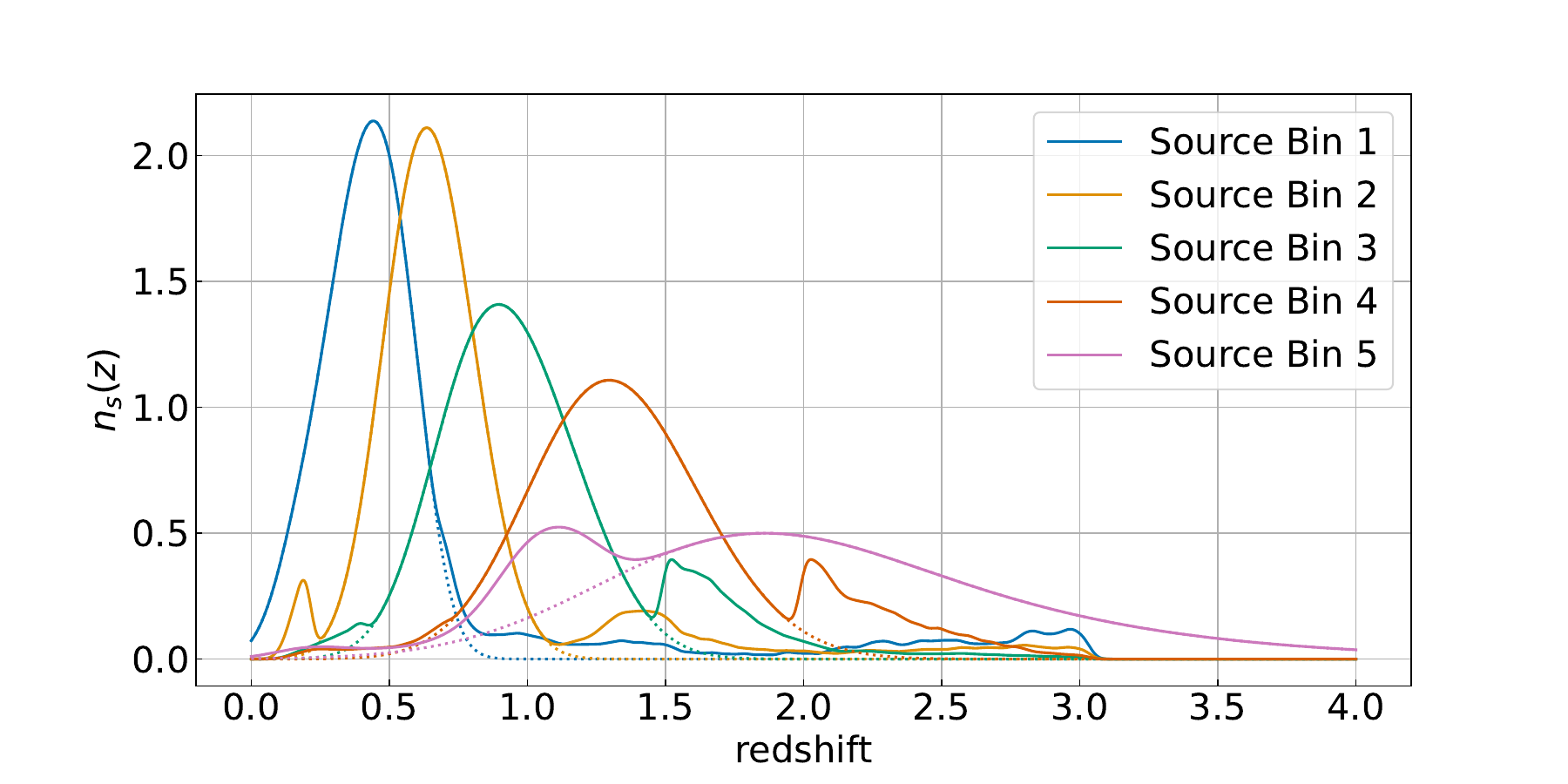}
    \caption{Redshift distribution of the source galaxy sample: The solid curves show the redshift distributions for the five tomographic source redshift bins, including both the core and outlier redshift distributions. The distributions are normalized such that the core and outlier rate combined integrate to unity, with the outliers comprising 15 per cent of the population. The core distributions follow the Gaussian model in Section~\ref{sec:core} and the outlier distributions follow the KDE model in Section~\ref{sec:outliers}. The dotted lines are the extensions to the core distributions if there were no outliers, shown to further illustrate the effects of the outliers. 
    }
    \label{fig:source}
\end{figure*}

\begin{table}
    \caption{\label{tab:pzparams}The fiducial values, priors, and the \citet{graham2020} values assumed in the Fisher information matrix analysis for the 15 photometric redshift error parameters: the biases $\delta z_i$, standard deviations $\sigma_i$, and outliers $f_{\text{out},i}$, subscripted by the bin number in order of increasing redshift. In Section~\ref{sec:results:cosmo_bias}, we assess the \refresponse{parameter}  biases in the $3 \times 2$pt Fisher Forecast induced by the values used in \citet{graham2020}.} 
    \centering
    \begin{tabular}{c|c|c|c}
        Parameter & $\!\!\!$Fiducial Value & $\!\!\!$Prior $\sigma$ & $\!\!\!$Graham 2020++ Value\\ \hline
        $\delta z_1 (1+z_{\rm center,1})$ & 0 & 0.1 & 0.0065 \\
        $\delta z_2 (1+z_{\rm center,2})$ & 0 & 0.1 & 0.001\\
        $\delta z_3 (1+z_{\rm center,3})$ & 0 & 0.1 & 0.0007\\
        $\delta z_4 (1+z_{\rm center,4})$ & 0 & 0.1 & 0.0016\\
        $\delta z_5 (1+z_{\rm center,5})$ & 0 & 0.1 & 0.0014\\
        $\sigma_1 (1+z_{\rm center,1})$ & 0.065 & 0.1& 0.0241 \\ 
        $\sigma_2 (1+z_{\rm center,2})$ & 0.0825 & 0.1& 0.0147 \\ 
        $\sigma_3 (1+z_{\rm center,3})$ & 0.0975 & 0.1& 0.0144 \\ 
        $\sigma_4 (1+z_{\rm center,4})$ & 0.1175 & 0.1& 0.022 \\ 
        $\sigma_5 (1+z_{\rm center,5})$ & 0.1875 & 0.1& 0.0391 \\
        $f_{\text{out},1}$ & 0.15 & 0.1& 0.1812 \\
        $f_{\text{out},2}$ & 0.15 & 0.1& 0.0701 \\
        $f_{\text{out},3}$ & 0.15 & 0.1& 0.0274 \\
        $f_{\text{out},4}$ & 0.15 & 0.1& 0.0424 \\
        $f_{\text{out},5}$ & 0.15 & 0.1& 0.379 \\
    \end{tabular}
\end{table}

\subsection{Source Photometric Redshift}\label{subsec:photozerror}

In imaging surveys such as LSST, the redshifts (a proxy for the line-of-sight distances) of most galaxies are determined by mapping the measured photometry in broad filters to the redshift of the galaxy. Such techniques are generally referred to as photometric redshifts, or photo--$z$. photo--$z$ can be estimated using data--driven techniques or template--fitting techniques \citep[for a review, see][]{2019NatAs...3..212S,2022arXiv220613633N}. Data--driven techniques, such as machine learning models, use a training, and/or calibration, dataset assembled from spatially overlapping spectroscopic (or multiband photometric) surveys that provide accurate redshift information. 
The mapping between photometry and redshift is then determined by fitting a flexible model and applied to the full photometric dataset. Template fitting uses models for the Spectral Energy Distributions (SEDs) of galaxies to find the redshift that yields the maximum likelihood between the predicted and measured photometry. 

Both classes of photometric redshift estimators produce uncertain redshift estimates since they are based on a limited number of broad--band fluxes or colors and, therefore, contain limited information about the galaxy SED \citep[e.g.,][]{pzdc1}.
In addition to the intrinsic uncertainty due to the reduced redshift information in the photometry compared with that in the full spectrum, photometric redshift estimators are also subject to sources of systematic uncertainty. These can arise, for example, because of model misspecification, degeneracies between SED type and the redshift of galaxies, and the impact of line--of--sight selection functions in spectroscopic training samples \citep[e.g.,][]{2020MNRAS.496.4769H}. These sources of systematic bias imprint characteristic error modes onto the estimated shape of the sample redshift distribution of the galaxy sample. As we will show in Section~\ref{sec:3x2pt}, the sample redshift distribution enters the modeling of two--point correlation functions through the lensing kernel. Biases in the estimated ensemble redshift distribution induced by inaccurate photometric redshift error estimates therefore propagate into biased correlation function predictions and, as a result, into biased cosmological parameter estimates.  

In the following subsections, we describe our models for the source redshift distribution. In short, we use a Gaussian `core' distribution defined by \refresponse{redshift} bias and variance in each tomographic bin (described in Section~\ref{sec:core}). The redshift distribution of the outlier galaxies (i.e., those outside of the Gaussian core) is defined empirically using a photo--$z$ catalog for a simulated galaxy sample (described in Section~\ref{sec:outliers}). We use the effective number density as a function of redshift
that was used in the DESC Science Requirements Document  \citep[DESC--SRD;][]{DESCSRD}, which was forecasted using the \texttt{WeakLensingDeblending} image simulation package \citep{wld, weaklensingdeblending}, and has an effective integrated number density for the entire sample of 26.94 arcmin$^{-2}$ for the ten--year LSST sample. 
Fig.~\ref{fig:source} shows our fiducial source $n(z)$ distribution for the five tomographic bins.

\subsubsection{Bias and Variance}
\label{sec:core}

Our model for the source sample redshift distributions is composed of a parametric model for a `core' distribution, parametrized by its first and second moments, 
and an empirical model for the outliers.
We assume that the source sample is split into five tomographic bins, with equal numbers of galaxies in each bin. The bin edges are defined by the total $n(z)$ distribution in the DESC--SRD. We also assume that the redshift distribution nuisance parameters for the tomographic bins are independent.  We use a Gaussian model for the core photometric redshift error distribution centered  $z_{\mathrm{center},i}$, with width $\sigma_i(1+z_{\rm center,i})$. 
We use one parameter per bin to parameterize the \refresponse{redshift} biases of these distributions, and another parameter per bin for the widths (standard deviation of the Gaussian kernels). Both the biases, $\delta z_i$, and standard deviations $\sigma_i$ are defined such that they are further multiplied by a $(1+z_{\rm center,i})$ factor using the redshift at the center redshift of the bin (not the average redshift of the galaxies in the bin). 
The bias is defined such that the center of the Gaussian core of the photometric redshift distribution will be $\mu = z_{\rm center,i} - \delta z_i (1+z_{\rm center,i}).$ The width of the core distribution will be $\sigma_i (1+z_{\rm center,i})$, where $\sigma_i$ is the width parameter.

\subsubsection{Outlier rate}
\label{sec:outliers}

\begin{figure}
    \centering
    \includegraphics[width=0.5\textwidth]{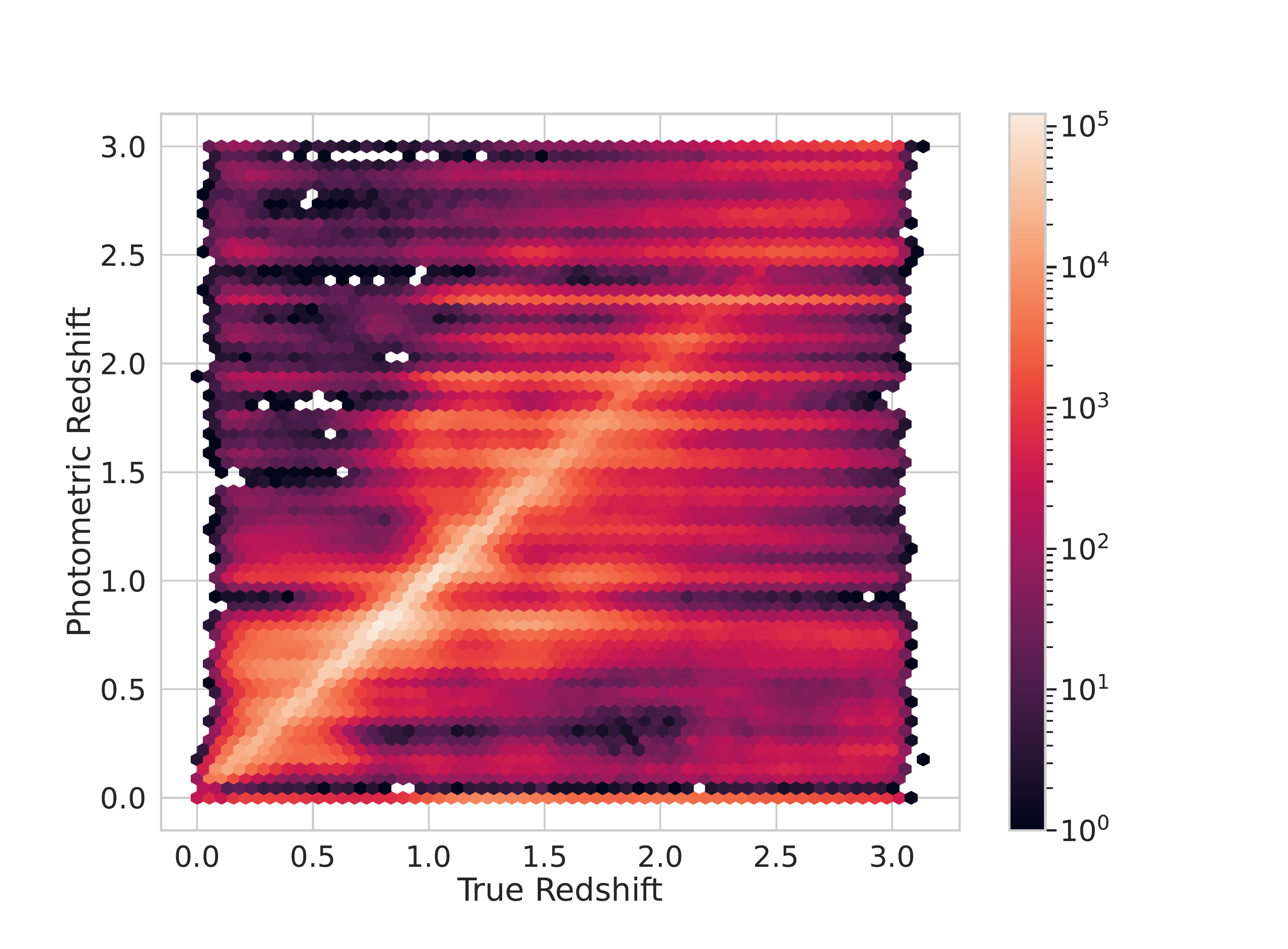}
    \includegraphics[width=0.5\textwidth]{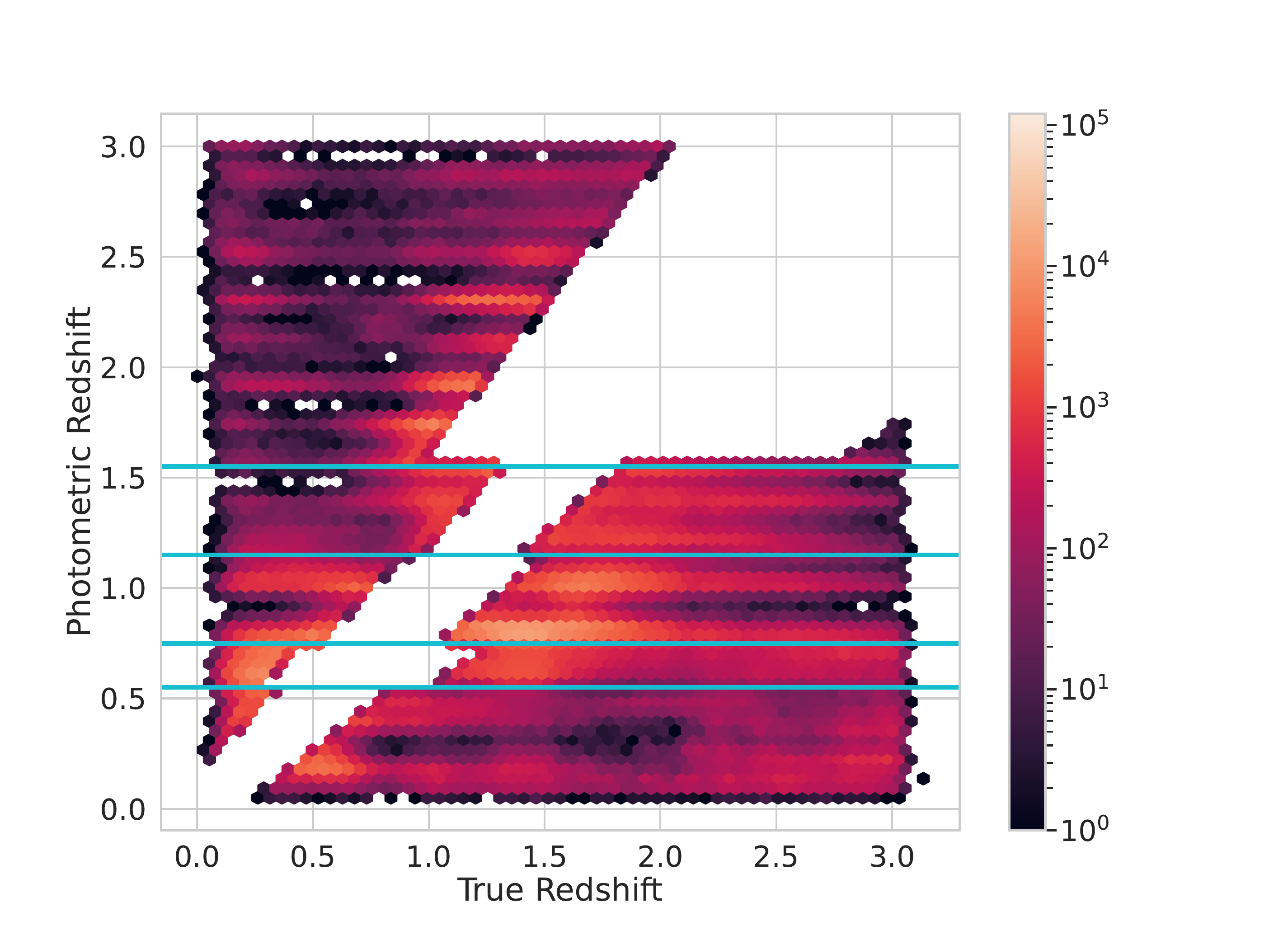}
    \caption{This figure compares the \texttt{FlexZBoost} photo--$z$ for the CosmoDC2 extragalactic catalog with $i$--mag$<26.5$, against the true redshifts of galaxies in that catalog; this catalog is used in the KDE to define the outlier populations in this work. The top panel shows a histogram of the true and photometric redshifts for the entire catalog, while the bottom panel shows a histogram of the outlier populations selected as described in Section~\ref{sec:outliers}. The teal horizontal lines in the bottom panel define the photometric redshift bin edges. There are some minor discontinuities at the bin edges, since the variances of the core distribution used to define the outlier population for the five bins are discrete values. The horizontal streak in the scatter is mostly from the faint galaxies with $i$--mag$>25$, the cause of which is suspected to be the discreteness in the SEDs for the training set galaxies, which map a certain region in the color space to a certain redshift mode. 
    }
    \label{fig:flexzboost}
\end{figure}

In this subsection, we define the photo--$z$ outlier distribution. We use a realistic photo--$z$ estimation on an LSST-like catalog to estimate the outlier population.
The shape of the outlier distribution $n_{\rm out}(z)$ is pre--determined before the analysis. In our case, we use the photo--$z$ from \texttt{FlexZBoost} \citep{10.1214/17-EJS1302}, as applied to the CosmoDC2 simulated extragalactic galaxy catalog \citep{2019ApJS..245...26K} with $i$--mag$<26.5$, with the same source binning. Fig.~\ref{fig:flexzboost} shows a 2--dimensional histogram of the \texttt{FlexZBoost} results, as well as a histogram of the outliers as defined here. 
The redshift distribution of outliers in each tomographic bin is modeled using a Kernel Density Estimation (KDE). 

For an individual galaxy, we define the statistical error quantity $\Delta_z$ as: 
\begin{equation}
    \Delta_z = \frac{z_\text{phot} - z}{1+z}
\end{equation}
If a galaxy in the source sample has a $|\Delta_z|$ that is larger than three times the sample's Interquantile Range (IQR) for that galaxy's tomographic bin, it is defined as an outlier. Following the outlier definition in \citealt{pzdc1}, the galaxy is an outlier if 
\begin{equation}
    \lvert \Delta_z \rvert > 3 \times \left[ \frac{\text{CDF}^{-1}(0.75) - \text{CDF}^{-1}(0.25)}{1.349} \right]
\end{equation}
where CDF is the cumulative distribution function of $\Delta_z$ inside one tomographic bin.

\begin{figure*}
    \centering
    \includegraphics[width=0.8\textwidth]{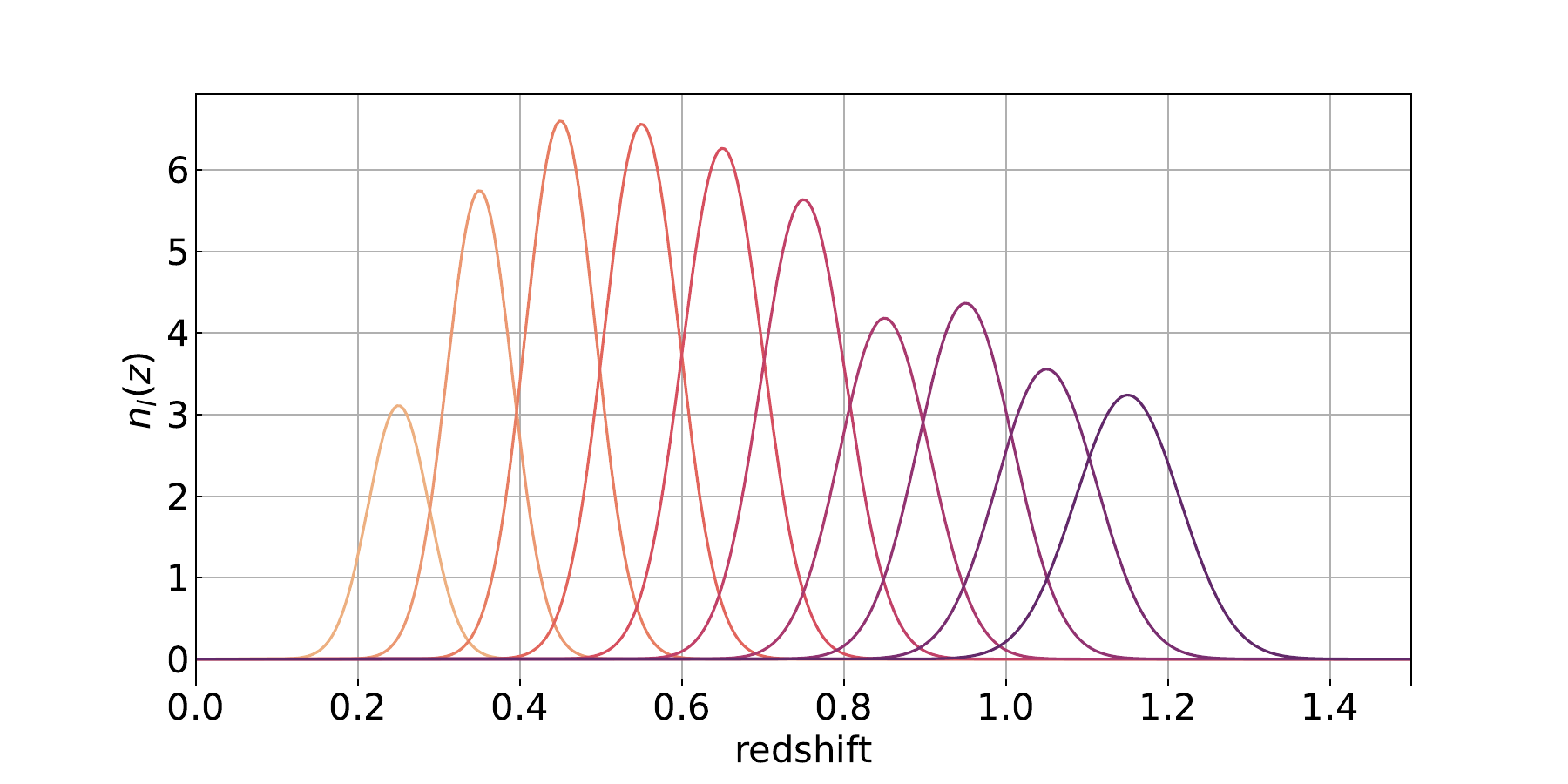}
    \caption{The redshift distributions of the lens sample in ten equally--spaced tomographic bins, described in Section~\ref{sec:3x2pt}. 
    }
    \label{fig:lens}
\end{figure*}

For each tomographic bin, we use the fraction of the outlier distribution in addition to its mean and variance to parameterize the uncertainty of the distribution. The fraction is chosen based on the assumption that we can have a fair understanding of the location of the typical outliers in each tomographic bin, but it is challenging to precisely decide the fraction of the outliers due to non--representativeness of the training galaxies. 
These five parameters control the outlier fraction in each of the five bins, $f_{\text{out},i}$. 
Overall, the redshift distribution $n_i(z)$ of the $i$--th tomographic is expressed as
\begin{align}
\label{eq:nz_parameters}
n_i(z) = (1-f_{\rm out, i})\mathcal{N}[\mu_i + \delta z_i (1+z_{\rm center,i}),& \sigma_i (1+z_{\rm center,i})]\\\nonumber &+ f_{\rm out, i} n_{\rm out}(z).
\end{align}
The definition of terms in Eq.~\ref{eq:nz_parameters} are as follows
\begin{itemize}
  \item $f_{\rm out, i}$: The outlier fraction for bin $i$ (fraction of galaxies not following the main distribution).
  \item $\mathcal{N}[\mu, \sigma]$: A normalized Gaussian distribution with mean $\mu$ and standard deviation $\sigma$.
  \item $\mu_i$: The mean redshift of the main distribution in bin $i$.
  \item $\delta z_i$: A shift parameter that adjusts the central redshift of bin $i$.
  \item $z_{\rm center,i}$: The reference or central redshift value for bin $i$ (typically the unperturbed mean).
  \item $\sigma_i$: The width (standard deviation) of the Gaussian distribution in bin $i$.
  \item $n_{\rm out}(z)$: The redshift distribution of the outlier population (assumed fixed across bins).
\end{itemize}

We note that $i$--mag$<26.5$ is a galaxy sample that includes many faint galaxies and is likely not what will be used for the LSST cosmological lensing analysis. However, we think the outlier distribution of this sample is an appropriately conservative choice for this study. 
We also note that the shapes of the outlier distributions can be significantly different from each other depending on the datasets and photo--$z$ algorithms. 
To confirm whether our conclusion is consistent with different outlier distributions, we conducted the Fisher forecast with an outlier distribution estimated from the CosmoDC2 galaxies with a different magnitude cut, $i$--mag$<25.3$, in Appendix~\ref{ap:outlier_distribution}. We find that our conclusions are overall unchanged given the different outlier distributions.

Table~\ref{tab:pzparams} shows the fiducial values and the standard deviations of the Gaussian priors for the photo--$z$ parameters. 
Fig.~\ref{fig:source} shows the sum of the core model (in dotted lines) and the KDE fits of the outliers used for the five redshift bins in the source sample. This plot was made with the fiducial parameters from Table~\ref{tab:pzparams}, including the default 15 per cent outlier rate in each bin. This 15 per cent value is consistent with the overall outlier rate in \texttt{FlexZBoost}, while the rate of outliers varies from one redshift bin to the next in \texttt{FlexZBoost}. We use the fiducial 15 per cent rate in each bin, as a conservatively large estimate, given that the outlier rate of the LSST golden sample is expected to be significantly lower than 15 per cent, according to Figure~\ref{fig:gold_redshift} according to the CosmoDC2 simulation. 
An outlier rate of 15 per cent is also comparable to the average outlier rate in, e.g., the deep sample of the Hyper Suprime--Cam (HSC) survey dataset in \citealt{hscpz}; different photo--$z$ estimation methods produce different patterns in outlier fraction as a function of redshift.

\begin{table}
    \caption{The fiducial value and the standard deviation of the Gaussian prior assumed in the Fisher information matrix for the seven cosmological parameters. 
    }
    \centering
    \begin{tabular}{c|c|c}
        Parameter & Fiducial Value & Prior $\sigma$ \\ \hline
        $\Omega_m$ & 0.3156 & 0.15\\
        $\sigma_8$ & 0.831 & 0.2\\
        $w_0$ & -1 & 0.8\\
        $w_a$ & 0 & 1.3\\
        $h$ & 0.6727 & 0.125\\
        $n_s$ & 0.9645 & 0.2\\
        $\Omega_b$ & 0.049 & 0.003 \\\hline
    \end{tabular}
    \label{tab:cosmoparams}
\end{table}

\subsection{$3 \times 2$pt Analysis}
\label{sec:3x2pt}

Three two--point correlation functions are included in a $3 \times 2$pt analysis. These are computed in different tomographic bin pairs, providing information about dark energy through its impact on the growth of the large--scale structure in the Universe. For recent $3 \times 2$pt methodology and analyses, see, e.g., \citealt{3x2pt_des,2021A&A...646A.140H,2022PhRvD.105b3520A,Sugiyama2023, Miyatake2023}, and for more details on the individual probes, see e.g., \citealt{weaklensingreview2}.

\emph{Galaxy--galaxy lensing} \citep[e.g.,][]{2022PhRvD.105h3528P} is the correlation function between the shape distortions of the galaxies in the source (background) sample, and the overdensities of the lens (foreground) galaxies that cause this distortion, traced by the position of galaxies in the lens sample. While galaxy--galaxy lensing is empirically computed using cross--correlation functions from observed galaxies, we can compute the galaxy--galaxy lensing theoretically by integrating 
the matter power spectrum with the product of two transfer functions that trace the matter density in the lens sample galaxies and that of the shear for galaxies in the source sample, along the line--of--sight.

The lens sample is split into ten tomographic redshift bins, equally separated between 0.2 and 1.2 in photometric redshift, shown in Fig.~\ref{fig:lens}. Each distribution follows a Gaussian model by convolving a Gaussian (centered in the middle of each bin with a standard deviation of 0.03) with the redshift distribution in each lens bin interval. The assumed lens redshift distribution is shown in Fig.~\ref{fig:lens}. 

In this work, the lens photometric redshift parameters are not varied and we do not marginalize over uncertainty in their values. This is a valid assumption for our purpose of studying the impact of source photo--$z$ uncertainties.\footnote{In some cases, for example in  \citet{2022PhRvD.105b3520A}, the lens sample is defined such that it has smaller and better--understood photometric redshift uncertainties than the source sample.} Another caveat about the redshift distribution of the lens galaxies is that our analysis assumes that there is some photometric noise in the redshift of the lens sample, resulting in overlap between adjacent tomographic bins. We directly include this uncertainty in the lens and source redshift distribution, although this is not perfectly refllective of a real analysis. 
However, as we previously mentioned, these photometric redshift uncertainties in the lens galaxies are precisely modeled and not marginalized over in the Fisher forecast.

\emph{Galaxy clustering} \citep[e.g.,][]{2022MNRAS.511.2665R} is the two--dimensional angular auto--correlation function between the positions of lens galaxies in each of the ten lens sample redshift bins.  Analysis of this measurement must take into account galaxy bias, the ratio of the overdensities between the observed positions of galaxies and the underlying dark matter overdensities, which is explained in more detail in Sec.~\ref{sec:gbias}.

\emph{Cosmic shear:} is the effect that distant galaxies gets deflected by the intervening large--scale structure of the Universe \citep[e.g.,][]{weaklensingreview2}. The shapes of the galaxies as observed are consequently sheared (among other changes to their size and flux).  Cosmic shear is empirically measured as the correlation function between the shapes of pairs of galaxies in the source sample \citep[e.g.,][]{2021A&A...645A.104A,2022PhRvD.105b3514A,2022PhRvD.105b3515S,2022arXiv220402396V,2023PhRvD.108l3519D,2023PhRvD.108l3518L}.

We briefly summarize the model for the $3 \times 2$pt data vector in this work as follows. We use the Core Cosmology Library \citep[\texttt{CCL}, version 2.1.0;][]{CCL}\footnote{\url{https://github.com/LSSTDESC/CCL}}  to compute the $3 \times 2$pt data vectors including contributions from intrinsic alignments and galaxy bias. \refresponse{The linear matter power spectrum is using computed using Eisenstein \& Hu transfer function \citep{eisenstein_hu}, the nonlinear power spectrum is computed by \texttt{Halofit} \cite{Smith2003,Takahashi2012}.}

The angular power spectra, $C_\ell$, for a Fourier mode $\ell$, between observable $A$ in redshift bin $i$, and observable $B$ in redshift bin $j$, across a comoving distance $\chi$ \citep[following, e.g.,][]{cosmolike, C_ells} is expressed as 
\begin{equation}
\label{eq:3x2pt}
C_{\ell, A B}^{i j}(\ell)=\int \mathrm{d} \chi \frac{q_{A}^{i}(\chi) q_{B}^{j}(\chi)}{\chi^{2}} P\left(\frac{\ell}{\chi}, z(\chi)\right)
\end{equation}
with $q$ being the kernel function for the different observables and $P$ is the three--dimensional nonlinear matter power spectrum. 
Equation~\eqref{eq:3x2pt} uses the Limber \citep{Limber1953} and flat--sky approximations to write a simple theoretical expression for the data vectors we use in this analysis.

The combinations ($A, B$) = ($\delta_g, \delta_g$), ($\delta_g, \kappa$), and ($\kappa, \kappa$) correspond to clustering, galaxy--galaxy lensing, and shear--shear power spectra respectively, where $\delta_g$ is the galaxy density contrast and $\kappa$ is the convergence field. The weight functions for $\delta_g$ and $\kappa$ for bin $i$ include both the redshift window and, for $\delta_g$, a linear galaxy bias parameter:
\begin{equation}\label{eq:clusteringq}
q_{\delta_{\mathrm{g}}}^{i}(\chi)=b_{i} \frac{n_{\mathrm{g}}^{i}\left(z(\chi)\right)}{\bar{n}_{\mathrm{g}}^{i}} \frac{\mathrm{d} z}{\mathrm{d} \chi}
\end{equation}
\begin{equation}\label{eq:lensingq}
q_{\kappa}^{i}(\chi)=\frac{3 H_{0}^{2} \Omega_{m}}{2 \mathrm{c}^{2}} \frac{\chi}{a(\chi)} \int_{\chi}^{\chi_{\mathrm{h}}} \mathrm{d} \chi^{\prime} \frac{n_{\kappa}^{i}\left(z\left(\chi^{\prime}\right)\right) \mathrm{d} z / \mathrm{d} \chi^{\prime}}{\bar{n}_{\kappa}^{i}} \frac{\chi^{\prime}-\chi}{\chi^{\prime}}
\end{equation}
where $b_i$ is the linear galaxy bias in the $i$th bin. $n^i(z)$ is the redshift distribution of the source (subscript $\kappa$) and lens (subscript g) sample, with $\bar{n}^i$ being the integrated number density in the bin. $H_0$ is the Hubble constant, $c$ is the speed of light, and $a$ is the scale factor. \refresponse{$\chi_{\mathrm{h}}$ is the comoving distance to the horizon. }

For LSST Y10, galaxy clustering power spectra are computed as auto--correlations in the same bin for each of the ten lens sample bins, resulting in ten spectra. Cosmic shear power spectra are computed using all possible bin pairs between the five source redshift bins, resulting in 15 spectra. Galaxy--galaxy lensing is computed in bin combinations where the source bin is at a higher redshift than the lens bin, resulting in 25 spectra. For LSST Y1, the number of clustering power spectra drops to five, and Galaxy--galaxy lensing drops to seven. 

The $3 \times 2$pt data vectors are computed in 20 $\ell$ bins spanning 20--15000. \refresponse{We follow the DESC--SRD \citep{DESCSRD} for the scale cuts.} Bins with $\ell>3000$ are masked for cosmic shear, whereas galaxy--galaxy lensing and clustering have bins with wavenumber $k>0.3 h/$Mpc 
masked (which effectively masks those two data vectors for values of $\ell$ larger than 200--900, depending on the tomographic bin). The unmasked $3 \times 2$pt data vector has 1000 elements (10 power spectra for clustering, 25 for galaxy--galaxy lensing, and 15 for cosmic shear, each with 20 angular bins) for LSST Y10. After masking, there are 96 data points for clustering, 228 for galaxy--galaxy lensing, and 225 for cosmic shear. The length of the masked $3 \times 2$pt data vector is 549.   

\subsection{Intrinsic Alignments}
\label{sec:intrinsicalignemtns}
The measurement of the coherent shear distortions using galaxy 
 shape correlation functions relies on the assumption of a random distribution for the intrinsic shapes of the galaxies. Hence, any other physical effect that causes coherent galaxy shape distortions can cause a systematic bias in these correlation functions and needs to be modeled. We use the term `intrinsic alignments' to refer to the non-isotropic galaxy shape alignment due to local physical effects in the environment of the galaxy \citep[for reviews, see][]{IA4,IA1,IA2,IA3, Lamman2024}.

\begin{table}   \caption{\label{tab:intrinsicalignments}The four intrinsic alignment parameters, their fiducial values and prior standard deviations used in the Fisher information matrix, adopted from the DESC--SRD.}
    \centering
    \begin{tabular}{c|c|c}
        Parameter &  Fiducial value & Prior $\sigma$\\ \hline
        Amplitude $A_0$ & 5.92 & 2.5\\
        Redshift dependence $\eta_l $ & -0.47 & 1.5\\
        High redshift dependence $\eta_h $ & 0.0  & 0.5\\
        Luminosity dependence $\beta$ &  1.1 & 1.0 \\
    \end{tabular}
\end{table}

We use \texttt{CCL}'s implementation \citep{CCL} of the nonlinear alignment model  \citep[NLA;][]{NLA} 
to calculate the intrinsic alignments contributions to the $3 \times 2$pt data vectors, given the alignment amplitude determined using the parameters in Table~\ref{tab:intrinsicalignments}. The fiducial values of the IA parameters are taken from the DESC--SRD's fiducial cosmology, following \cite{KEB}'s formalism. \refresponse{The IA amplitude is described by four parameters, $A_0$, $\beta$, $\eta_l$, and $\eta_h$
\begin{equation}
A_{\rm IA}(z) = \langle A_{\rm IA}(L, z|A_0, \beta)\rangle_L \left(\Theta(\tilde{z} - z)\frac{1+z}{1+\tilde{z}_l}^{\eta_l} + \Theta(z-\tilde{z}) \frac{1+z}{1+\tilde{z}}^{\eta_h}\right),
\end{equation}
where $\Theta$ is the step function, which returns 1 when the input is non--negative. $A_{\rm IA}(L, z|A_0, \beta)$ is the luminosity function-dependent amplitude of the intrinsic alignments, which is parameterized by $\beta$ and $A_0$, 
\begin{equation}
    A_{\rm IA}(L, z|A_0, \beta) = A_0 \frac{\int_{L(m_\text{lim})}^\infty \mathrm{d}L \left(\frac{L}{L_0}\right)^\beta \phi(L,z)}{\int_{L(m_\text{lim})}^\infty \mathrm{d}L \phi(L,z)},
\end{equation}
where $m_\text{lim}$ is the magnitude limit of the sample, $\phi(L,z)$ is the luminosity function, and $L_0$ is the pivot luminosity, which is set to $L_{\odot}$. 
The fiducial values and priors for the IA parameters are given in Table~\ref{tab:intrinsicalignments}. We set $\tilde{z}_l = 0.3$, and $\tilde{z} = 0.75$.} 

We use a relatively uninformative Gaussian priors on the IA parameters in this work. The main intention behind the priors is to ensure the numerical stability of the Fisher information matrix inversions and operations, rather than to be informative.

\subsection{Galaxy Bias}
\label{sec:gbias}

\refresponse{While galaxy--galaxy lensing and galaxy clustering are sensitive to the matter power spectrum, we measure those using luminous galaxies that trace the matter power spectrum.
In this work, we assume the galaxy density field is a linear tracer of the matter density field, modulated by a galaxy bias term per lens bin \citep{galaxybias}.} We use the same fiducial values and prior standard deviations used in DESC--SRD -- with the values presented in Table~\ref{tab:galaxybias}.  
The use of a linear galaxy bias motivates the relatively conservative scale cuts adopted in this work.

\begin{table}
    \caption{\label{tab:galaxybias}The ten lens galaxy bias parameters, and their fiducial values and prior standard deviations used in the Fisher information matrix. The ten parameters correspond to the galaxy biases of the ten clustering correlation functions in order of increasing redshift.
    }
    \centering
    \begin{tabular}{c|c|c}
        Parameter Term & Fiducial value & Prior $\sigma$\\ \hline
        $b_1$ & 1.38 & 0.9 \\
        $b_2$ & 1.45 & 0.9\\
        $b_3$ & 1.53 & 0.9\\
        $b_4$ & 1.61 & 0.9\\
        $b_5$ & 1.69 & 0.9\\
        $b_6$ & 1.78 & 0.9\\
        $b_7$ & 1.86 & 0.9\\
        $b_8$ & 1.94 & 0.9\\
        $b_9$ & 2.03 & 0.9\\
        $b_{10}$ & 2.12 & 0.9
    \end{tabular}
\end{table}

\subsection{Covariance Matrix}
\label{sec:method:cov}

Given our similar analysis setup, we use the covariance matrix from the $3 \times 2$pt DESC--SRD analysis forecast in  \citet{DESCSRD}, which was estimated numerically using \texttt{CosmoLike} \citep{cosmolike}. 

The covariance matrix includes both Gaussian and non--Gaussian contributions and is ordered by tomographic bin pair (50 total: 15 for cosmic shear, followed by 25 for galaxy--galaxy lensing, followed by 20 for clustering). The covariances are computed for the data vector with $\ell$ bins defined in Section~\ref{sec:3x2pt}. We also use the covariance matrix to effectively apply scale cuts: elements that should be masked due to scale cuts have the corresponding elements in the inverse covariance matrix set to 0. The scale cuts are described in Sec.~\ref{sec:3x2pt}.

\subsection{Fisher Information Matrix}
\label{sec:fisher}

In this work, we use Fisher forecasting code \texttt{FisherA2Z} to estimate the constraining power for the cosmological parameters given a fiducial cosmology and a covariance matrix for the data vectors. Here we explain the formalism of Fisher forecasting. 

For $p_\theta (X) = p(X | \theta)$, the probability of the random known variable $X$ given an unknown parameter $\theta$, we define the score function \begin{equation}
    s_\theta(X) = \frac{\partial \text{log} p_\theta (X)}{\partial \theta}.
\end{equation} 
The Fisher Information  \citep[e.g.,][]{Wasserman,Coe,naren} is then defined as the second moment of the score function:
\begin{equation}
\label{eq:fisher_def}
    \mathbb{I}(\theta) = \mathbb{E} [s_\theta(X)^2] = -\mathbb{E} \left[ \frac{\partial^2}{\partial \theta^2} \text{log} p_\theta(X)  \right].
\end{equation}
For i.i.d. samples $X_1, \dots, X_n$, the Fisher Information in terms of the overall sample of size $n$ as $I_n(\theta) = nI(\theta)$, such that for $X_i$, is $I(\theta) = I_{X_{i}}(\theta) = I_{X_{1}} (\theta)$.
From Eq.~\eqref{eq:fisher_def},the Fisher Information measures the curvature of the log--likelihood.   Given that the curvature of the log--likelihood quantifies the precision of the estimator, the Fisher Information can be used to quantify how well the parameter $\theta$ can be estimated.  \refresponse{Furthermore, the Cramér–Rao bound states that the variance derived from the Fisher matrix is a lower limit, making it a reasonable choice in our goal of putting a requirement on our knowledge of photo--$z$ errors such that they do not dominate the Figure of Merit error budget.}

Assuming a Gaussian likelihood function, we can rewrite the Fisher Information matrix elements as
\begin{equation}\label{eq:fishergauss}
    \mathbb{I}_{i,j} = \frac{\partial C_{\ell}}{\partial \alpha_i} ^T  V  \frac{\partial C_{\ell}}{\partial \alpha_j} + \frac{1}{\sigma_{\alpha_i}^2} \delta(i,j),
\end{equation}
where $\alpha$ is the model parameter vector and $V$ is the inverse of the covariance matrix. $\sigma_{\alpha_i}$ is the standard deviation of the Gaussian prior on parameter $\alpha_i$, and $\delta(i,j)$ is the Kronecker delta function.

The Fisher information matrix that we construct is a 36--dimensional matrix. It contains seven cosmological parameters, four intrinsic alignment parameters, 15 source photo--$z$ parameters, and ten galaxy bias parameters.  The seven cosmological parameters are: the total matter density parameter $\Omega_m$, the power spectrum normalization parameter $\sigma_8$, the baryonic matter density parameter $\Omega_b$, the dimensionless Hubble parameter $h$, the spectral index parameter $n_s$, and the dark energy equation of state parameters $w_0$ and $w_a$ \citep{Chevallier2001}.  The assumed fiducial values and prior standard deviations of the seven cosmological parameters are presented in Table~\ref{tab:cosmoparams}.

To obtain 2--dimensional confidence sets on pairs of model parameters, we marginalize over the rest of the parameters (cosmological and systematic).  
\refresponse{The marginalization refers to a specific mathematical operation. Given a full Fisher information matrix $\mathbb{I}$, the marginalized Fisher information matrix $\mathbb{I}_{\mathbf{p}}$ of a parameter subset $\mathbf{p}$ is 
\begin{equation}
\mathbb{I}_{\mathbf{p}} = \left\{ (\mathbb{I}^{-1})[\mathbf{p},\mathbf{p}]  \right\} ^{-1}.
\end{equation}
Here ``$[\mathbf{p},\mathbf{p}] $'' means taking the $\mathbf{p}$ column and rows of the original matrix. } 
We then obtain the semi--major and semi--minor axes (respectively, $a$ and $b$) along with the orientation ($\phi$) of the confidence ellipse from the marginalized $\mathbb{I}_{\mathbf{p}}$ \citep[see, e.g.,][]{Coe}.

In addition to the use of quantifying minimum expected uncertainties on cosmological model parameters, the Fisher matrix formalism can also be used to quantify biases in cosmological parameters if we make incorrect assumptions about model parameters (for the purpose of this work, the photo--$z$ error model parameters).  For any specific case of non--fiducial model parameters, we can calculate a new $C_\ell^\text{biased}$, and use it to calculate the bias in cosmological parameters that the new set of non--fiducial model parameters induces, under the assumption of small, linear changes in the $C_\ell$ \citep{biasformula2, centroidshift}: 
\begin{equation}
\label{eq:fisher_bias}
    f_\alpha = \mathbb{I}^{-1} \cdot \left( \frac{d C_\ell }{ d \alpha }  \cdot  V \cdot (C_\ell^{\textrm{biased}} - C_\ell) \right),
\end{equation}
where $C_\ell$ is the fiducial data vector. Eq.~\ref{eq:fisher_bias} is the first order Taylor expansion of the parameter bias, therefore, we only expect this to be accurate when the bias is relatively small. 

\refresponse{
Derivatives in the Fisher information matrix were numerically computed using the \texttt{numdifftools} Python library \citep{numdifftools} using a step size of 0.01 in most cases, which was found to lead to stable derivatives both in our tests and in other studies such as \citet{naren}. To ensure the stability of the derivative with regard to the step size, we inspected the derivative of the data vector with step size from $0.001$ to $0.02$ for all photo--$z$ and cosmological parameters, and confirmed that the derivatives are smooth and vary less than two per cent as the step size changed. 
We also changed the step size to 0.005 and 0.02, and computed the contour size by the Figure--of--Merit, or FoM. 
The FoM of parameter $\alpha_i$ and $\alpha_j$ ($i\neq j$) is defined as
\begin{equation}
\label{eq:FoM}
{\rm FoM} = \frac{1}{\sqrt{{\rm det}({\rm Cov}_{i,j})}},
\end{equation}
where the marginalized covariance matrix ${\rm Cov}_{i,j}$ is 
\begin{equation}
\label{eq:cov_fisher}
{\rm Cov}_{i,j} = \begin{bmatrix}
    (\mathbb{I}^{-1})_{ii} &(\mathbb{I}^{-1})_{ij}\\
    (\mathbb{I}^{-1})_{ij} & (\mathbb{I}^{-1})_{jj}
\end{bmatrix}.
\end{equation}
The differences in the FoMs in the $\Omega_m-\sigma_8$ plane and $w_0-w_a$ plane when varying the step size are less than  two per cent for this test. 
Besides these stability and convergence tests, we compare our forecasts with those from the DESC--SRD in Sec.~\ref{sec:results} as a form of numerical validation.}

As shown in Eq.~\eqref{eq:fishergauss}, information from the priors is added to the Fisher information matrix by adding the inverse variance, $1/\sigma^2$, listed in Tables~  \ref{tab:pzparams}, \ref{tab:cosmoparams}, \ref{tab:intrinsicalignments}, and~\ref{tab:galaxybias} to the diagonal element of the Fisher information matrix that corresponds to the parameter. 
Priors on cosmological parameters are based on previous experiments, while priors on nuisance parameters are uninformative. The priors on the cosmological parameters, intrinsic alignment parameters, and galaxy bias parameters are consistent with the DESC--SRD. \refresponse{In Appendix~\ref{sec:res:forecast_validation}, we validate the Fisher forecast by comparing the results to the DESC--SRD forecast. }

\subsection{Parameter Importance and Interpretability}
\label{sec:method:interp}

In this section, we briefly describe the metrics we use to understand the importance of the photo--$z$ parameters. We use the feature importance given by the decision tree to gauge the importance of a photo--$z$ parameter in determining each cosmological parameter. We also use the mean absolute fractional error (MAFE) to estimate the total impact on the data vector for each redshift parameter. 

The decision tree \citep[e.g., ][]{mlbook1,mlbook2,mlbook3} is a machine learning method developed for use in regression or classification.  It is widely used to learn the relative importance of each of the features used in training. Decision trees predict the value or class of a target $Y$, given a set of features $X_i$ that can be used to predict $Y$ by splitting features, one at a time, into smaller branches.
The most important feature will be the one that minimizes the conditional entropy.

For a target $Y$, and features $X_{i}$, the decision tree make the next split based on maximizing:
\begin{equation}
    \arg \max _{i} G\left(Y, X_{i}\right)=\arg \max _{i}\left[H(Y)-H\left(Y \mid X_{i}\right)\right]
\end{equation}
where $G$ is the information gain, and $H$ is the entropy. The initial entropy of the target is defined as
\begin{equation}
    H(Y)=-\sum_{y} P(Y=y) \log _{2} P(Y=y),
\end{equation}
where $P$ is the probability\refresponse{, and $y$ are the allowed values of $Y$.}

We use this methodology to identify the most important photo--$z$ parameters based on their effect on shifts in cosmological parameters, assuming an LSST Y10 cosmic shear or $3 \times 2$pt analysis. The initial step in doing so is to create training sets with a set of randomly sampled redshift parameters from bootstrapping our fiducial distribution (see detailed description in the next two paragraphs).  
and compute their biased data vector. We then compute the bias on the target cosmological parameter (Eq.~\ref{eq:fisher_bias}) assuming the biased data vector is analyzed with the fiducial redshift distribution. 
\refresponse{We use the \texttt{ExtraTreeRegressor} implementation in \texttt{scikit-learn} \citep{scikit-learn} 
for all training and cross--validation. We take the Gini importance metric as the feature importance after the training of each tree regressor. Gini importance in a decision tree measures the contribution of each feature to the model's predictive performance, calculated as the total decrease in Gini impurity across all trees in the forest when the feature is used for splitting \citep{Breiman2001}.} 

We create two types of training sets: the first type of training set samples redshift parameters by type (bias, variance, outlier rate), so the feature vector has a dimension of 5. The Gaussian distributions that sample $\delta z_i$, $\sigma_i$, and $f_{\rm out, i}$ have the same variance across all tomographic bins (before accounting for the $1+z_{\rm center}$ factor). 
This design choice is taken to compare the relative importance of each tomographic bin. 

The second type of training set samples redshift parameters jointly, creating a feature vector of dimension 15. To find the relative ranges of \refresponse{redshift} biases, standard deviations, and outlier fractions in each tomographic bin, we randomly sample 1000 galaxies 1000 times in each tomographic bin from the fiducial Y10 redshift distribution in Figure~\ref{fig:source}, and compute the corresponding scatter of the parameters across these realizations.
For each sample of 1000 galaxies, the $\delta z_i$ and $\sigma_i$ are computed as their mean redshift and standard deviation. The outlier rate is computed as the sum of probability evaluated by the core distribution versus the outlier distribution at the 1000 sampled redshifts.

To avoid overfitting, we use tree--pruning and 2--fold cross--validation to verify the sufficiency of the size of the dataset. We also test for the convergence of the decision tree results by confirming that the feature importance does not change when we increase the standard deviation of the Gaussian distribution from which we sample by factors of two and four. This is important because Eq.~\eqref{eq:fisher_bias} is an approximation that is only valid for small changes. 

To interpret the results of the decision tree feature importance, we compare the results with an illustrative quantity: the change in $C_\ell$ due to changing a photo--$z$ parameter in that bin. 
The MAFE is computed using the element--wise difference in the $C_\ell$ after changing a photo--$z$ parameter by a given amount (e.g., adding a \refresponse{redshift} bias of $0.01(1+\bar{z})$ to the core photo--$z$ error distribution in one tomographic bin), normalized by the original $C_\ell$.  
\begin{equation}
     \text{MAFE} =  \frac{1}{N} \sum_{\ell=1}^{N}\left\lvert \frac{C_\ell^\text{biased} - C_\ell}{C_\ell} \right\rvert
\end{equation} 
where the summation is taken over the $\ell$ bins defined in Section~\ref{sec:3x2pt}. \refresponse{Alternatively, one can compute the error as a fraction of the uncertainty of $C_\ell$, which would downweight data points with higher uncertainty.}

\begin{figure*}
    \centering
    \includegraphics[width=\textwidth]{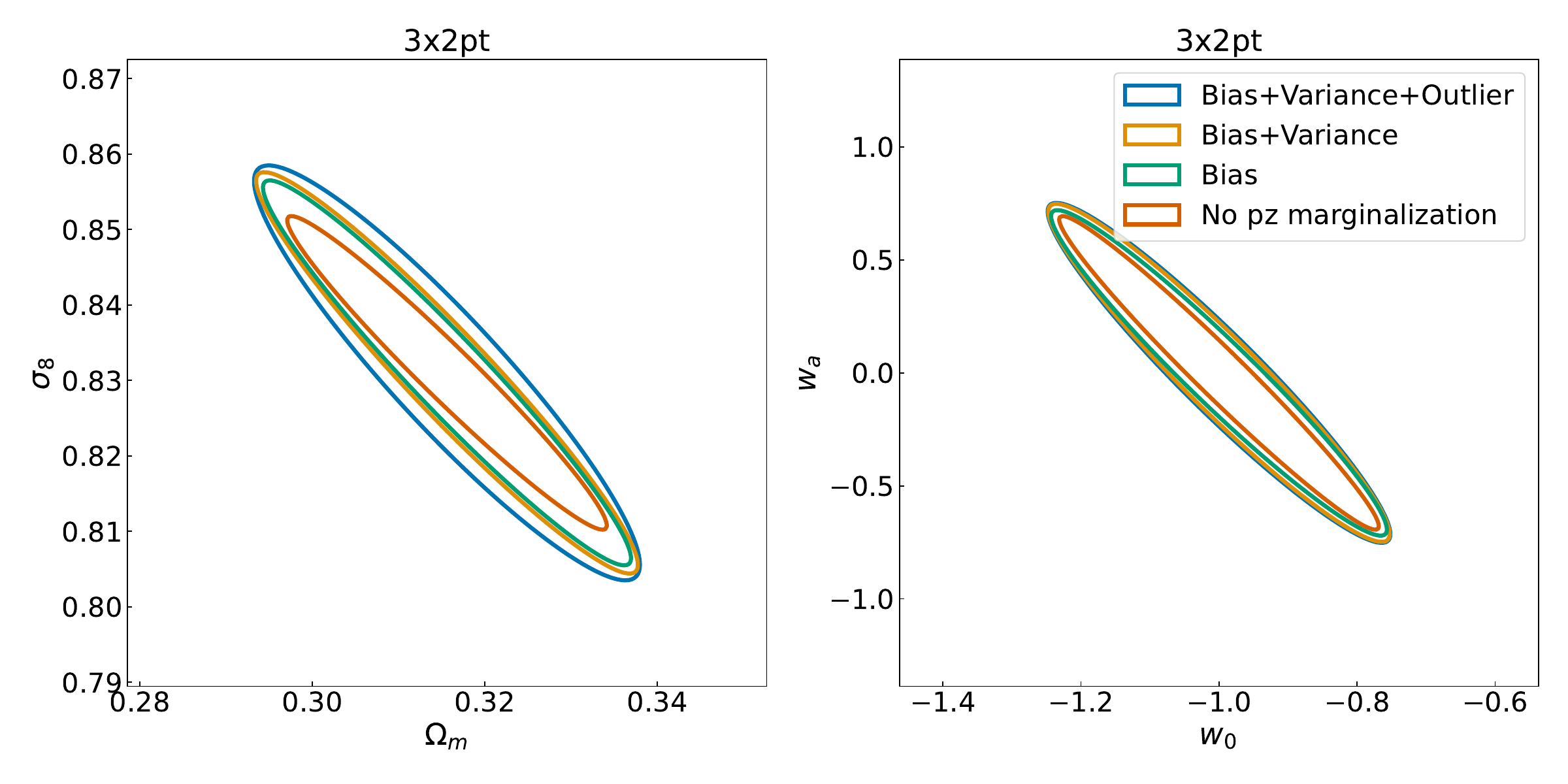}
    \includegraphics[width=\textwidth]{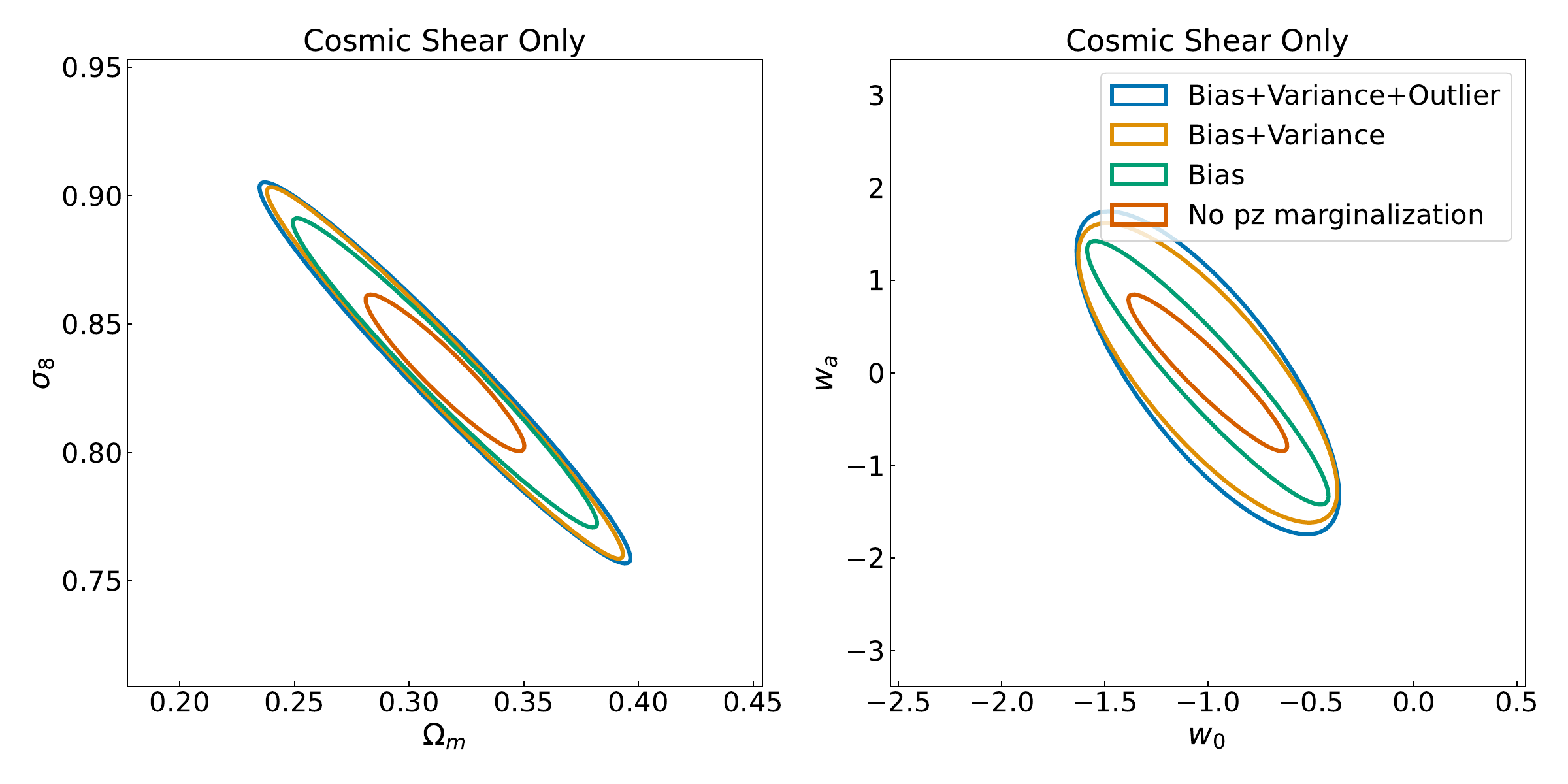}
    \caption{The 2$\sigma$ Fisher confidence sets in the ($\Omega_m, \sigma_8$) and ($w_0, w_a$) planes, before and after marginalizing over the 15 photo--$z$ parameters for the $3 \times 2$pt (upper two panels) and cosmic shear (lower two panels). The colors correspond to different models parameters that are marginalized when calculating the parameter contours.  Marginalizing over the photo--$z$ parameters increases the area 
    of the confidence sets. In all cases, the results include marginalization over the remaining seven cosmological parameters, as well as the ten galaxy bias parameters and four intrinsic alignments parameters.}
    \label{fig:marginalization}
\end{figure*}
  
\subsection{Priors Corresponding to Training Set Scenarios}
\label{sec:method:prior}

The Fisher forecast formalism takes the prior on its cosmological and nuisance parameters into account by adding the inverse variance of the Gaussian prior to the diagonal element of the Fisher matrix, expressed in Eq.~\eqref{eq:fishergauss}. As a result, we use the Fisher formalism to forecast the cosmological results for different priors on the parameters of redshift distributions. The priors developed in this section can be seen as additional variables to the Fisher forecast in addition to the fiducial prior shown in Table~\ref{tab:pzparams}. 

\refresponse{
For both representative and non--representative training sets, we generate priors of bias, variance, and outlier rates parameters corresponding to different scenarios. We compute Bayesian evidence \citep{Trotta2008} for models with different parameter sets. The $\log$--evidence of model with parameter set $\theta$, Gaussian likelihood and Gaussian prior can be estimated as 
\begin{align}
    \log \mathcal{Z} \approx \log \mathcal{L}(\theta_{\rm ML}) + \frac{n}{2}\log (2\pi)  - \frac{1}{2}\log |\mathbb{I}_{\rm likelihood} + \mathbb{I}_{\rm prior}| \\\nonumber + \frac{1}{2} \log |\mathbb{I}_{\rm prior}|, 
\end{align}
where $\mathbb{I}_{\rm tot} = \mathbb{I}_{\rm likelihood} + \mathbb{I}_{\rm prior}$ is the total Fisher matrix, the sum of the likelihood Fisher matrix and the prior Fisher matrix. $\log \mathcal{L}(\theta_{\rm ML})$ is the likelihood at maximum likelihood parameter $\theta_{\rm ML}$ is a constant in this equation, which leaves the relative difference in the evidence between models unchanged. The $\frac{n}{2}\log (2\pi)$ is the ``Occam's razor'' \citep{mackay2003information}, which penalize models with higher dimensionalities. Note we ignore the exponential term in Eq.~(25) in \citep{Trotta2008}, since the $\theta_{ML}$ is the same as the Maximum A Posteriori $\theta_{\rm MAP}$. 

When comparing model A and model B, we compute the $\log$ Bayes factor
\begin{equation}
    \log B_{\rm AB} = \log \mathcal{Z}_{\rm A} - \log \mathcal{Z}_{\rm B}.
\end{equation}
A positive $\log B_{\rm AB}$ means model $A$ is preferred by the Bayesian evidence, vice versa. We will use the Bayes factor to compare models with different complexity, under different scenario for the photo--$z$ priors. }

\subsubsection{Representative Training Set}
\label{sec:method:representative}

In this section, we are estimating the prior parameters of redshift distributions given a representative training set. Assuming we have a total of $N_{\rm samples}$ galaxies in our spectroscopic training set, split equally into five tomographic bins, we randomly draw $N_{\rm samples}/5$ samples in each tomographic bin from the fiducial number density as a probability distribution (shown in Fig.~\ref{fig:source}) $N_{\rm draw}$ times, denoted as $z_{ijk}$, where $i = 1 \dots 5$ (tomographic bins), $j = 1 \dots N_{\rm samples}/5$ (galaxies per bin), and $k = 1 \dots N_{\rm draw}$ (number of draws). We then compute the standard deviation of $\delta z_i$, $\sigma_i$, and $f_i$ between draws by
\begin{align}
    \label{eq:statistical_bias}\sigma_{\delta z_i} &= {\rm \sigma}(\mathbb{E}[z_{ijk}]_j)_k\\
    \sigma_{\sigma_i} &= {\rm \sigma}({\rm \sigma}[z_{ijk}]_j)_k\\
    \label{eq:statistical_outlier}\sigma_{f_i} &= {\rm \sigma}\left[\sum_{j=1}^{N_{\rm samples}/5}\frac{n_{\rm out}(z_{ijk})}{ n_{\rm total}(z_{ijk})}\right]_k.
\end{align}
Here the $n_{\rm out}(z)$ and $n_{\rm total}(z)$ are the outlier and total distribution as probability functions. 

We inspect the Fisher forecast for $N_{\rm samples} = 2000, 5000, 20000,$ and $100000$. For each scenario, we compute the Figure--of--Merit FoM of the $\Omega_m - \sigma_8$ plane and $w_0 - w_a$ plane. 

\subsubsection{Non--representative Training Set for $z>1.6$}
\label{sec:method:nonrepresentative}

In this section, we consider the non--representativeness of training galaxies over $z>1.6$. The optical spectroscopic success rate drops drastically at $z>1.6$, due to the lack of features in the rest--frame ultraviolet (UV) wavelength for galaxies while the features such as the O--II doublet at 372.7~nm are redshifted out of the visible wavelength \citep{2015APh....63...81N}. 

We draw galaxies similarly to Sec.~\ref{sec:method:representative}, with $N_{\rm samples} = 20000$. However, in each draw, the galaxies with $z>1.6$ has a $1-\mathcal{R}$ chance of being re--drawn once. We call this $\mathcal{R}$ the representativeness for galaxies over $z=1.6$. Essentially, the re--drawn mimics the scenario that the photo--$z$ model takes a random guess from the original distribution due to not having seen the high--$z$ galaxy.  

Because of the re--drawing for $z>1.6$ galaxies, we will not only have scattering in the \refresponse{redshift} bias, variance, and outlier rate for each draw, but also \refresponse{systematics shift} in those parameters. Therefore, in addition to Eqs.~\eqref{eq:statistical_bias}--\eqref{eq:statistical_outlier}, we compute another set of systematic uncertainty by,
\begin{align}
    \label{eq:systematic_bias} b[\delta z_i] &= {\mathbb{E}}(\mathbb{E}[z_{ijk}]_j)_k\\
    b[\sigma_i] &= {\mathbb{E}}({\rm \sigma}[z_{ijk}]_j)_k - \sigma_{i,\text{fid}}\\
    \label{eq:systematic_outlier}b[f_i] &= {\mathbb{E}}\left[\sum_{j=1}^{N_{\rm samples}/5}\frac{n_{\rm out}(z_{ijk})}{ n_{\rm total}(z_{ijk})}\right]_k - 0.15.
\end{align}
Here $\sigma_{i,\text{fid}}$ is the fiducial width of the distribution of $i$--th tomographic bin and $0.15$ is the fiducial outlier rate. The prior on parameters is computed as the $\sqrt{\sigma^2 + b^2}$, where we take the bias $b$ as the systematic uncertainty and the scatter $\sigma$ as the statistical uncertainty \refresponse{of the redshift parameters}.

\section{Results}
\label{sec:results}

In this section, we present the results of our investigations, beginning with understanding the impact of our photometric redshift error parametrization on the cosmological forecasts.  We then move to the investigation of which parameters most strongly connect to cosmological parameter biases if incorrect assumptions are made, using decision trees. Finally, we explore different training set scenarios and their impact on the forecasts.

\subsection{Impact of photo--$z$ Marginalization}
\label{sec:res:marginalization}

We use the Fisher information matrix to infer confidence intervals on sets of cosmological parameters following the steps in Sec.~\ref{sec:fisher}. Fig.~\ref{fig:marginalization} shows those 2$\sigma$ confidence intervals in the ($\Omega_m, \sigma_8$) and ($w_0, w_a$) planes, with fixed photometric redshift parameters (``no pz marginalization''), marginalizing over only the photo--$z$ bias parameters (``Bias''), marginalizing over photo--$z$ bias and variance (``Bias+Variance'') and after marginalizing over all photo--$z$ parameters at their fiducial values (``Bias+Variance+Outlier''). We show the results of $3 \times 2$pt  in the top panels, and cosmic shear only in the bottom panels. The ``no pz marginalization'' curve only marginalizes over the rest of the parameters in the Fisher matrix (i.e., four other additional cosmological parameters, four intrinsic alignment parameters, and ten galaxy bias parameters).

As expected, marginalizing over more parameters increases the size of the confidence intervals, or in other words, decreases the precision of our inference. In particular, for LSST Y10 $3 \times 2$pt, the area of the $2\sigma$ ($\Omega_m, \sigma_8$) confidence contours increase by a factor of 1.9, 2.6, 3.4 for the ``Bias'', ``Bias+Variance'', and ``Bias+Variance+Outlier'', and the $2\sigma$ ($w_0, w_a$) confidence contours increase by a factor of 2.0, 2.7, 3.2, correspondingly. For LSST Y10 cosmic shear, the $2\sigma$ ($\Omega_m, \sigma_8$) confidence contours increase by a factor of 2.5, 3.3, 3.8 for the ``Bias'', ``Bias+Variance'', and ``Bias+Variance+Outlier'', and the $2\sigma$ ($w_0, w_a$) confidence contours increase by a factor of 2.6, 5.4, 6.3,  correspondingly. Compared to the $3 \times 2$pt, the cosmic shear analysis is more heavily affected by the marginalization of the redshift distribution parameters. This is expected, because the $3 \times 2$pt analysis has more information to partially self--calibrate the redshift uncertainty parameters.  The results show that in the case where there is no informative redshift distribution prior, the LSST Y10 cosmic shear results will be redshift uncertainty-constrained.

Note that although the prior of the 15 photo--$z$ parameters is set to a Gaussian kernel with $\sigma = 0.1$, which is a really wide prior for photo--$z$, we are not marginalizing over such a wide uncertainty, due the self--constraining power of the probes. In Appendix~\ref{ap:self_constrain}, we show the posterior probability distribution for the photo--$z$ parameters for the cosmic shear and $3 \times 2$pt analysis. We see that because of the self--constraining power of the cosmological probes, the posteriors for the photo--$z$ parameters are significantly narrower than their priors.

In Robertson et al.\ \textit{in prep.}, the impact of marginalized uncertainty on cosmological and intrinsic alignment parameters in LSST Y1 and Y10 cosmic shear analyses is studied, with different redshift distribution parameterization including ones with or without outlier distributions. With different cosmological models, redshift distribution models and fiducial values and priors compared to this study.  They find that marginalizing over redshift calibration uncertainty will increase the uncertainty on $S_8$ by 45 per cent with a $\sigma_{\delta z} = 0.015$ prior, which is a much smaller prior compared to this study.

\begin{figure*} 
    \centering
    \includegraphics[width=\textwidth]{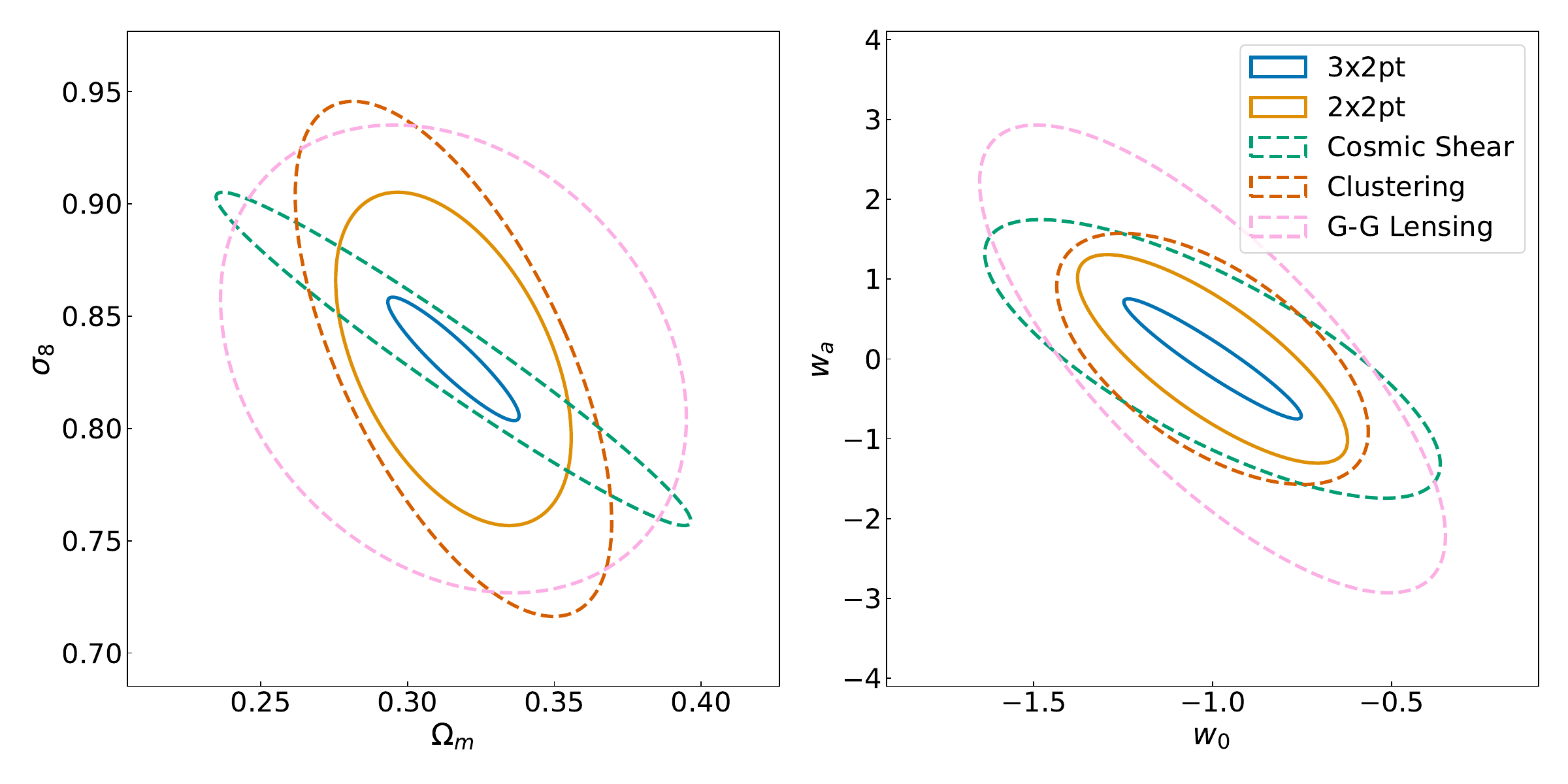}
    \caption{A comparison between the 2$\sigma$ Fisher confidence sets for the single--probe cases and for $3 \times 2$pt in the ($\Omega_m, \sigma_8$) and ($w_0, w_a$) planes. 
    The colors correspond to different probes according to the legend. The solid contours are from joint probes, while dashed lines contours are from individual probes. 
    Combining the three probes results in much tighter confidence contours on the cosmological parameters, with the actual increase, \refresponse{quantified by the FoM}, for each case listed in Table~\ref{tab:probe_comparison}. In all cases, the results include marginalization over the remaining 34 model parameters.}
    \label{fig:probe_comparison}
\end{figure*}

\subsection{Combining Constraints from Probes}

Fig.~\ref{fig:probe_comparison} compares the confidence intervals for the cases of single probe data vectors, i.e., the cases of cosmic shear, galaxy--galaxy lensing, and clustering separately; as well as for the $3 \times 2$pt and $2 \times 2$pt (combination of galaxy clustering and galaxy--galaxy lensing) combined probes in the ($\Omega_m, \sigma_8$) and ($w_0, w_a$) planes, after marginalizing over the rest of the parameters. 
Given that these three probes are covariant, the Fisher information matrix leverages the information in all of them and leads to a tighter bound on cosmological inference. Additionally, in some cases, the different orientations of the confidence contours break the degeneracies between different parameters, leading to better constraints on them, as seen more clearly in the ($\Omega_m, \sigma_8$) case. The ratios of the contour sizes \refresponse{in terms of Figure--of--Merit} for each of the single probes to the $3 \times 2$pt contour are shown in Table~\ref{tab:probe_comparison}. While the size of the confidence contours becomes 6--34 
times larger than the 3x2pt contours, \citealt{schaan} confirmed that marginalizing over photo--$z$ outliers in particular is important, as they found that a four per cent additive bias on a fiducial outlier fraction could lead to a $1\sigma$ bias in $w_0$ and $w_a$. We also show the marginalized $1\sigma$ uncertainty of $S_8 = \sigma_8 (\Omega_8/0.3)^{0.5}$, a useful way to quantify the constraining power on the large--scale structure amplitude, and its multiplying factor if only a subset of the data vector are used. 
\begin{table}
    \caption{\label{tab:probe_comparison}The increase in the area of the 1--$\sigma$ $\Omega_m-\sigma_8$ and $w_0-w_a$ contours (quantified by $1/$FoM), and $\sigma_{S_8}$ of probes compared to the LSST Y10 $3 \times 2$pt contours as shown in Fig.~\ref{fig:probe_comparison}, with the values of $3 \times 2$pt shown in the first row.}
    \centering
    \begin{tabular}{c|c|c|c}
        Probe & $1/$FoM $\Omega_m-\sigma_8$ & $1/$FoM $w_0-w_a$ & $\sigma_{S_8}$\\
        \hline
        $3 \times 2$pt & $1/$26070 & $1/$114 & 0.0017 \\ \hline
        Cosmic Shear & 5.9x & 9.2x & 1.8x \\
        Clustering & 20.3x & 10.4x & 7.5x\\
        G--G Lensing & 33.9x & 23.3x & 17.3x\\
        $2 \times 2$pt & 11.1x & 5.8x & 6.9x \\\hline
    \end{tabular}
\end{table}

\subsection{Impact of Incorrect photo--$z$ Error Models}
\label{sec:results:cosmo_bias}

One of the advantages of conducting Fisher forecasting is its ability to quickly estimate bias on cosmological parameters given a small change in the data vector.
Assuming values different than the fiducial values in our systematics model parameters will lead to biases in inferred cosmological parameters, as computed by Eq.~\eqref{eq:fisher_bias}. In addition, the size of the confidence intervals changes, because the derivatives that go into the Fisher information matrix are evaluated at different values. 

\begin{figure*}
    \centering
    \includegraphics[width=\textwidth]{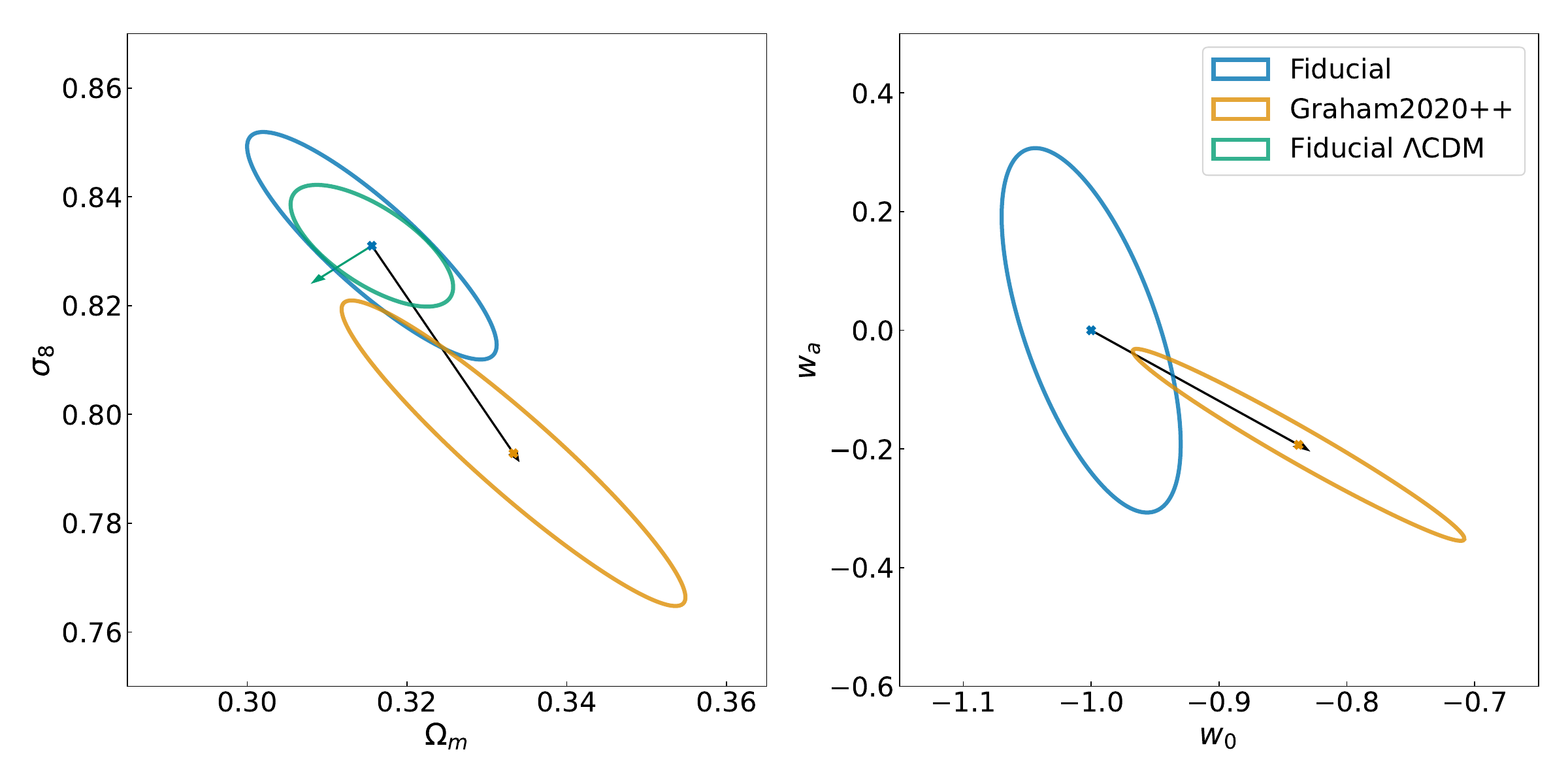}
    \caption{A comparison between the $3 \times 2$pt constraints for the 2--dimensional confidence sets in the ($\Omega_m, \sigma_8$) and ($w_0, w_a$) spaces for the case with fiducial values (blue contours) assumed in this paper, and the case with photo--$z$ model found in \citealt{graham2020} (orange contours). 
    The values used for the photo--$z$ error statistics in both cases are shown in Table~\ref{tab:pzparams}. As shown, there are changes in both the sizes of the contours, defined at the 2$\sigma$ level, and in the center of the inferred posterior (indicated by the arrow) due to different photo--$z$ error distribution values. Particularly, the bias induced in $\Omega_m, \sigma_8, w_0, w_a$ are $2.0\sigma_{\Omega_m}, -3.4\sigma_{\sigma_8}, 1.8\sigma_{w_0}, -0.73\sigma_{w_a}$  respectively, where $\sigma_\alpha$ is the standard deviation of parameter $\alpha$. In the left panel, the cyan arrow and countour shows the parameter bias and $2\sigma$ contour if the $\Lambda$CDM model is adopted. }
    \label{fig:graham}
\end{figure*}

To illustrate this effect, we choose a set of values for photo--$z$ error parameters from a realistic forecast for the LSST that includes calibration with an Euclid sample. In particular, we choose the `LSST Y10 + Euclid 5$\sigma$' case in Fig.~8 of \citealt{graham2020}, using the maximum error statistic in each source bin to input to the Fisher forecast model. The input values for the photo--$z$ error parameters are presented in the 4th column of Table~\ref{tab:pzparams}, while Fig.~\ref{fig:graham} shows the resulting Fisher confidence interval in the ($\Omega_m, \sigma_8$) and ($w_0, w_a$) planes. In both cases, we assume the prior of the photo--$z$ are incorrectly centered, and this biases the cosmological parameter inference and modifies the size of their uncertainties. 
In particular, the bias in the $\Omega_m, \sigma_8, w_0, w_a$ parameters are $2.0\sigma_{\Omega_m}, -3.4\sigma_{\sigma_8}, 1.8\sigma_{w_0}, -0.73\sigma_{w_a}$ respectively, where $\sigma_\alpha$ is the $1\sigma$ posterior of parameter $\alpha$ after marginalizing over all parameters.

We want to warn the readers that we expect the \refresponse{cosmological} parameter bias predicted by the Fisher forecast to be overestimated. As implicitly described in Eq.~\ref{eq:fisher_bias}, the Fisher forecast formalism predicts a cosmological bias using the posterior Gaussian likelihood centered at the center of the prior; therefore, the data vector generated by \cite{graham2020} can be $\sim 10\sigma$s away considering that the posteriors of the photo--$z$ parameters are much tighter than their priors (see Appendix~\ref{ap:self_constrain}). In a realistic MCMC analysis, when given a data vector produced by biased systematic parameters, the self--calibration will shift the posteriors \refresponse{of the nuisance parameters} toward the biased value, thereby reducing the bias in cosmological parameters.

The analysis in this work is carried out for the case of constraining $w_0-w_a$CDM models. However, it is also possible to carry out the $\Lambda$CDM analysis by fixing $w_0=-1$ and $w_a=0$. 
In the left panel of Figure~\ref{fig:graham}, we show the parameter bias and $2\sigma$ contour of the $\Omega_m, \sigma_8$ plane the $\Lambda$CDM model is applied. We find that the $2\sigma$ ($\Omega_m, \sigma_8$) contours are smaller by 48 per cent. The biases induced due to an incorrect photo--$z$ model are also smaller in both absolute and relative term. We notice that the direction of the bias in this 2--dimensional plane changes when a different cosmological model is assumed.

\subsection{Parameter Importance}
\label{sec:tests}

\begin{figure*} 
    \centering
    \includegraphics[width=0.8\textwidth]{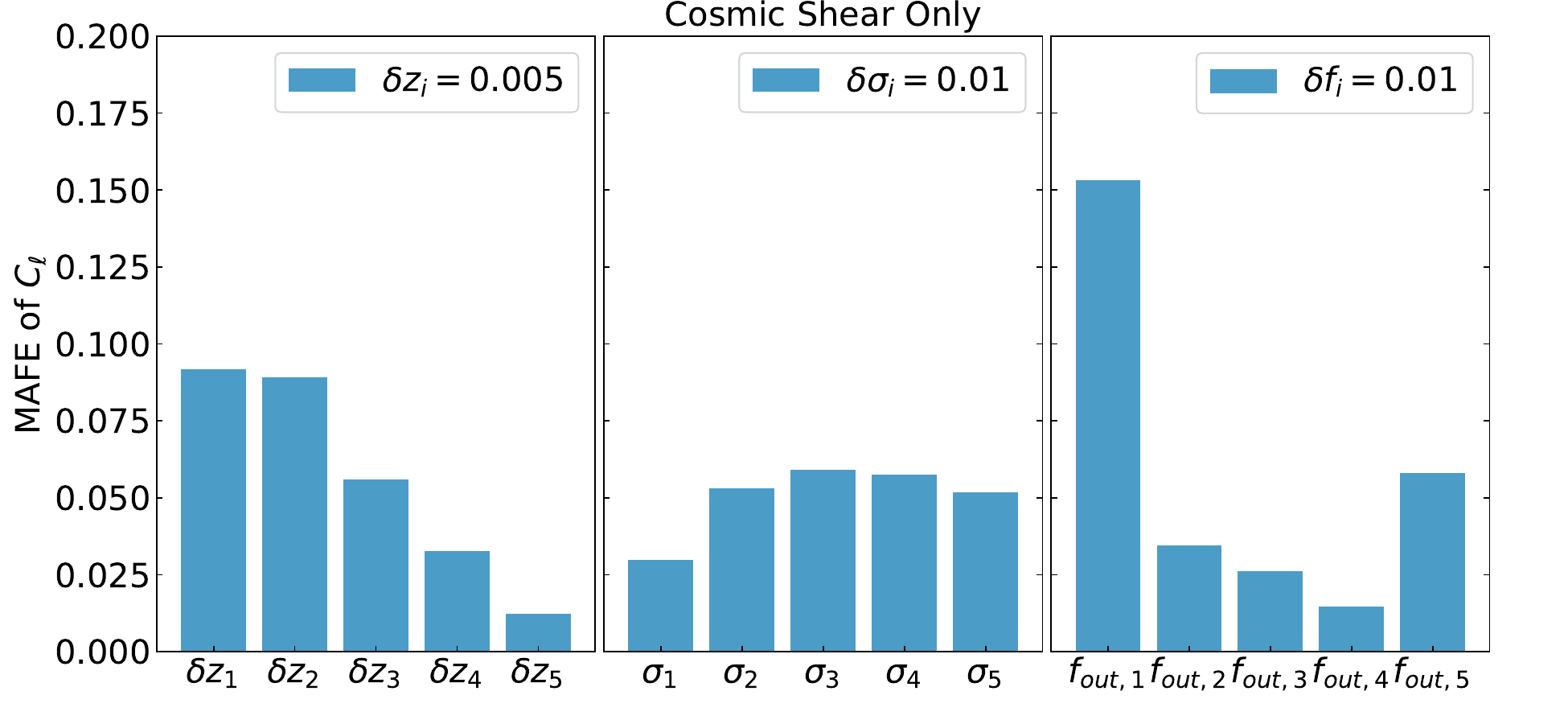}
    \caption{MAFE metrics for LSST Y10 cosmic shear analysis. We find the MAFE decrease with by redshift bias bins, stays approximately unchanged for the standard deviation, and is more sensitive to outlier rates of the first and last tomographic bin.  }
    \label{fig:mafe_cosmic_shear}
\end{figure*}

\begin{figure*} 
    \centering
    \includegraphics[width=0.8\textwidth]{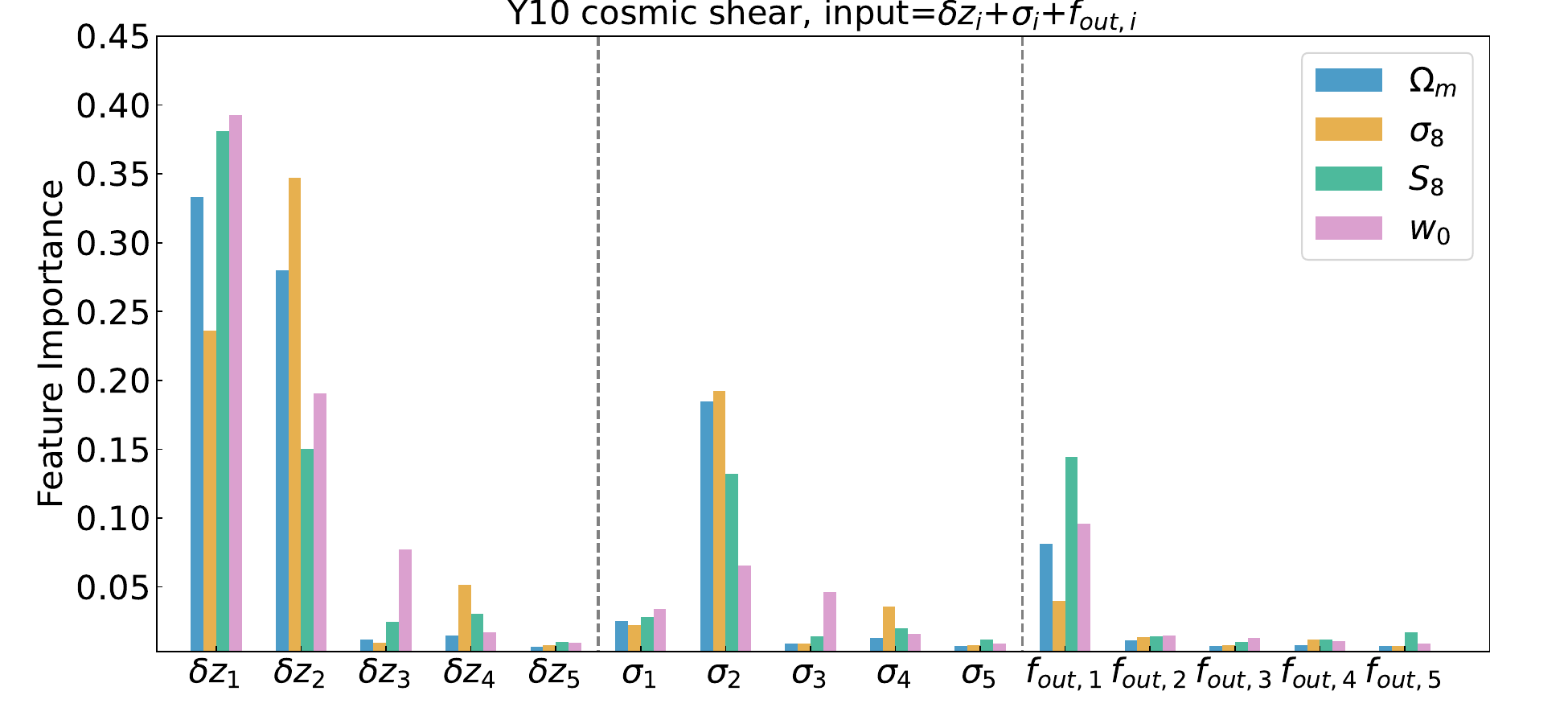}
    \includegraphics[width=0.8\textwidth]{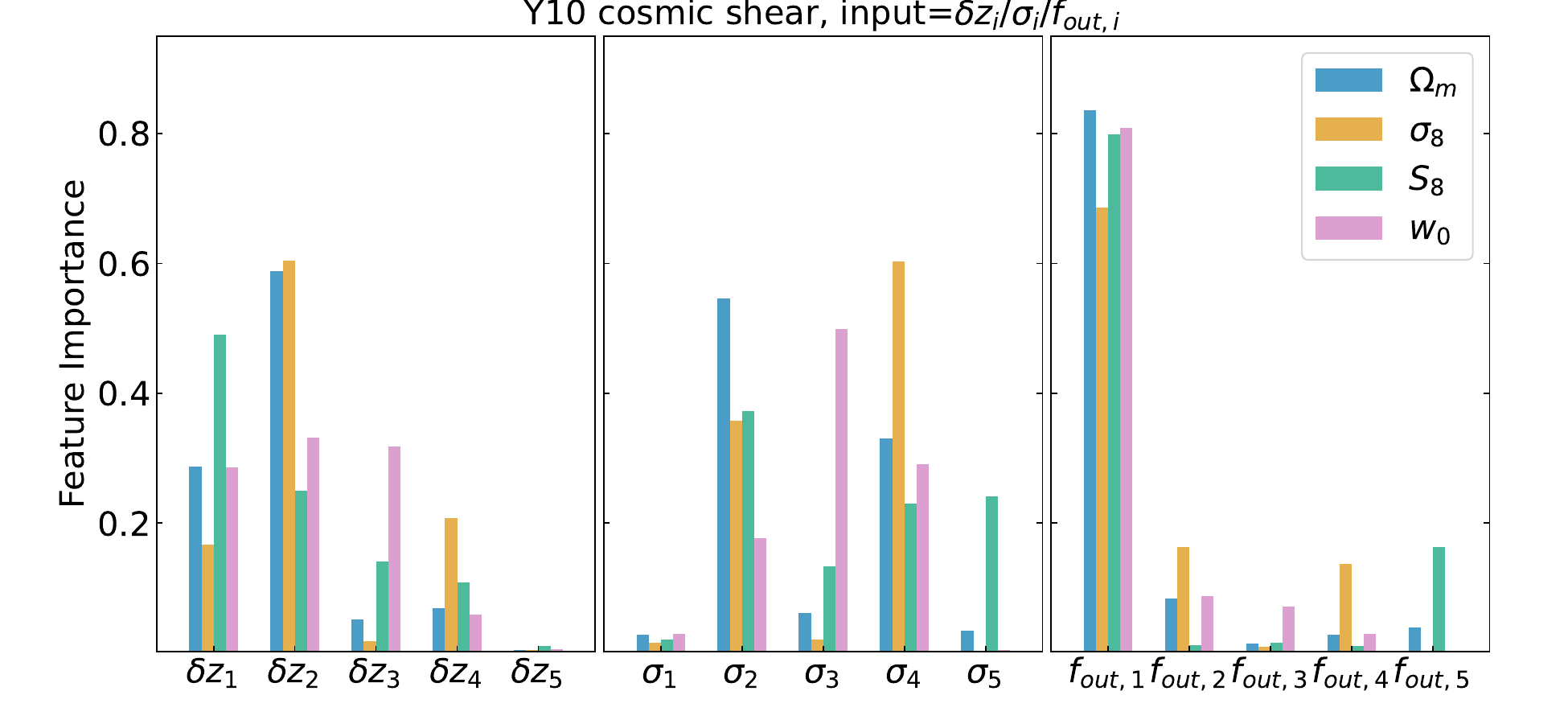}
    \caption{Feature importance for the cosmic shear analysis with the $w_0-w_a$ model. The top panel shows the importance metrics with all 15 redshift parameters as input, while the bottom panel shows the important metrics with redshift parameters grouped by types across redshift bins. 
    }
    \label{fig:feature_cosmic_shear}
\end{figure*}

\begin{figure*} 
    \centering
    \includegraphics[width=0.8\textwidth]{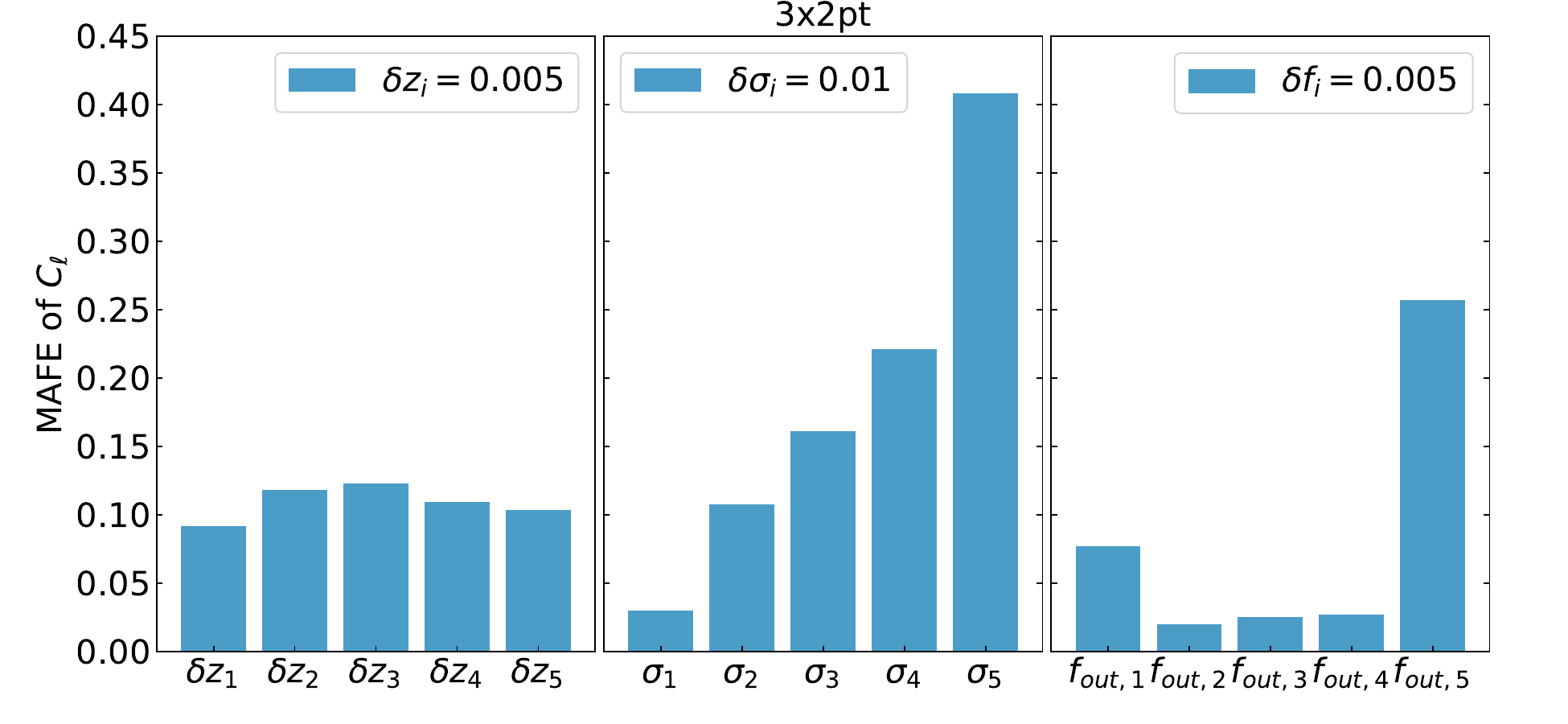}
    \caption{MAFE metrics for LSST Y10 $3 \times 2$pt analysis. We find that MAFE of $3 \times 2$pt analysis stays relatively unchanged with redshift bias $\delta z$, increase with redshift $\sigma$, and are more sensitive to the ourlier rates of the first and last tomographic bin. }
    \label{fig:mafe_3x2}
\end{figure*}

\begin{figure*} 
    \centering
    \includegraphics[width=0.8\textwidth]{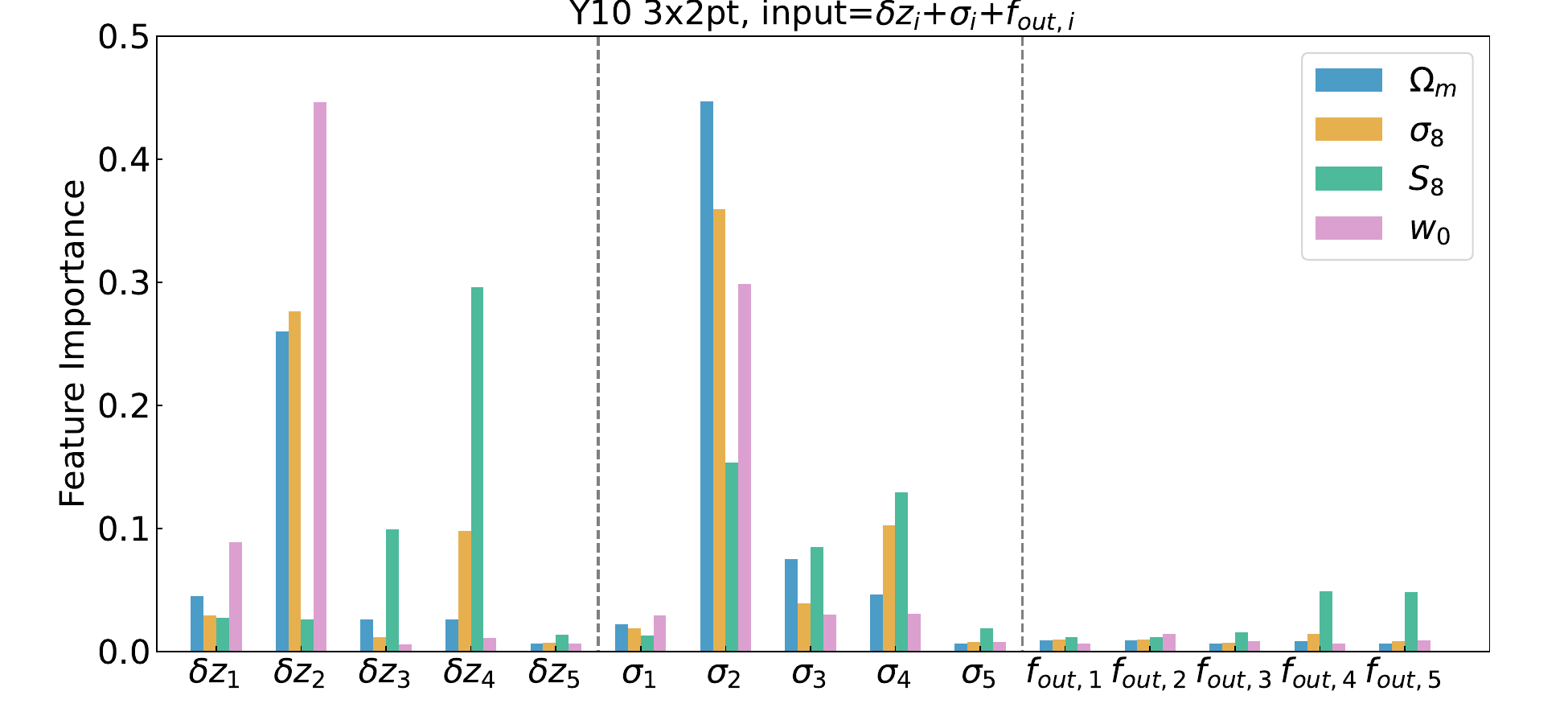}
    \includegraphics[width=0.8\textwidth]{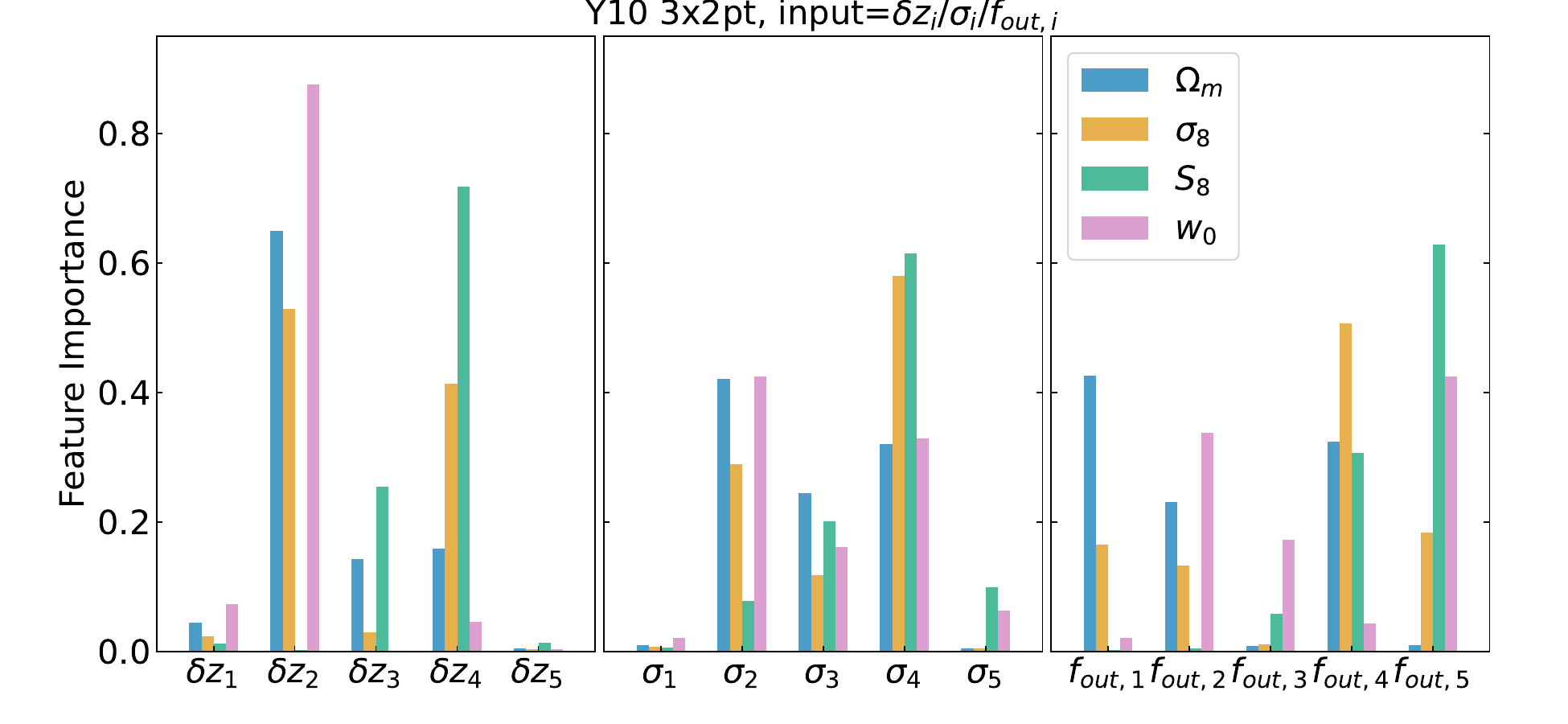}
    \caption{Feature importance for the $3 \times 2$pt analysis with the $w_0-w_a$ model. The top panel shows the importance metrics with all 15 redshift parameters as input, while the bottom panel shows the important metrics with redshift parameters grouped into three groups. }
    \label{fig:feature_3x2}
\end{figure*}

\begin{figure*} 
    \centering
    \includegraphics[width=1\textwidth]{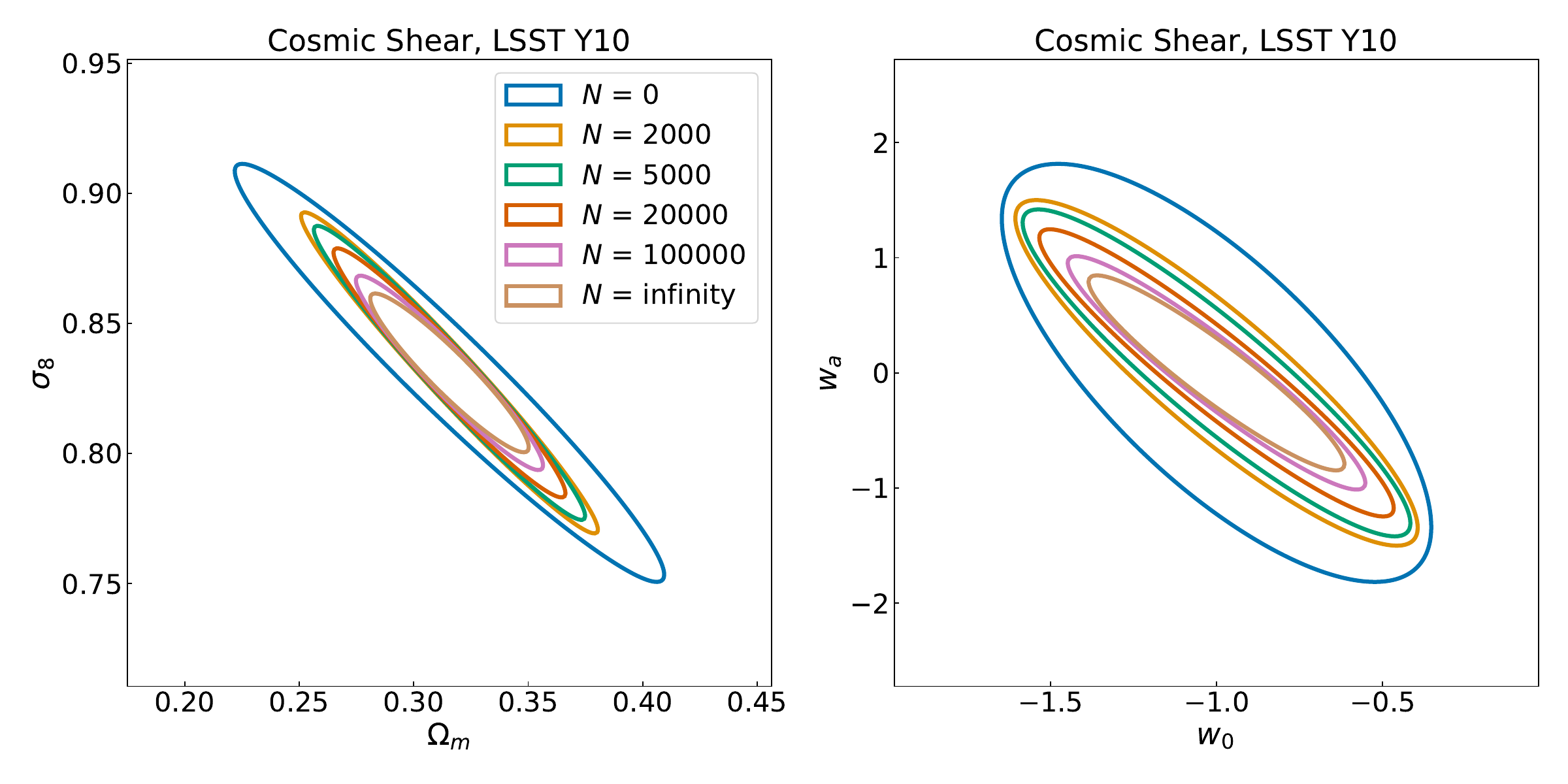}
    \caption{2--dimensional contour plot for the Fisher forecast of LSST Y10 cosmic shear only analysis with redshift priors corresponding to different numbers of galaxies in the spectroscopic training set for the photo--$z$ estimator. $N = 0$ corresponds to a totally uninformative prior, while $N = \infty$ corresponds to an infinitely small prior (or equivalent to not marginalizing over the redshift parameters).  }
    \label{fig:nsample_contour}
\end{figure*}

We use the decision tree feature importance method described in Sec.~\ref{sec:method:interp} to understand the importance of different redshift error model parameters with regard to each cosmological parameter. The `importance' of a feature is defined by the Gini importance, defined in Section~\ref{sec:method:interp}. We randomly sample a 5000 redshift parameters in each scenario and compute the biased data vector. We then use  Eq.~\eqref{eq:fisher_bias} to predict the bias in cosmological parameters assuming the redshift parameters are un-modeled.

In all cases, we carry out statistical tests to ensure the validity of our method. Eq.~\eqref{eq:fisher_bias} is a first--order approximation that works for a limited range of parameter deviations. To ensure its validity in the range of parameters we draw from, we use a convergence test where we lower the standard deviation of the Gaussian we draw from until we observe convergence in the resulting decision tree. More quantitatively, we consider the results to have converged when (a) the order of the bin importance stays the same after increasing the standard deviation by a further factor of two and 4, and (b) the relative bin importance does not change by more than ten percent.

We show the feature importance and the related interpretability figures for the cosmic shear and $3 \times 2$pt data vector from Fig.~\ref{fig:mafe_cosmic_shear}--\ref{fig:feature_3x2}. The feature importance is studied for the bias, variance, and outlier fraction both separately and jointly.

\begin{figure*} 
    \centering
    \includegraphics[width=0.49\textwidth]{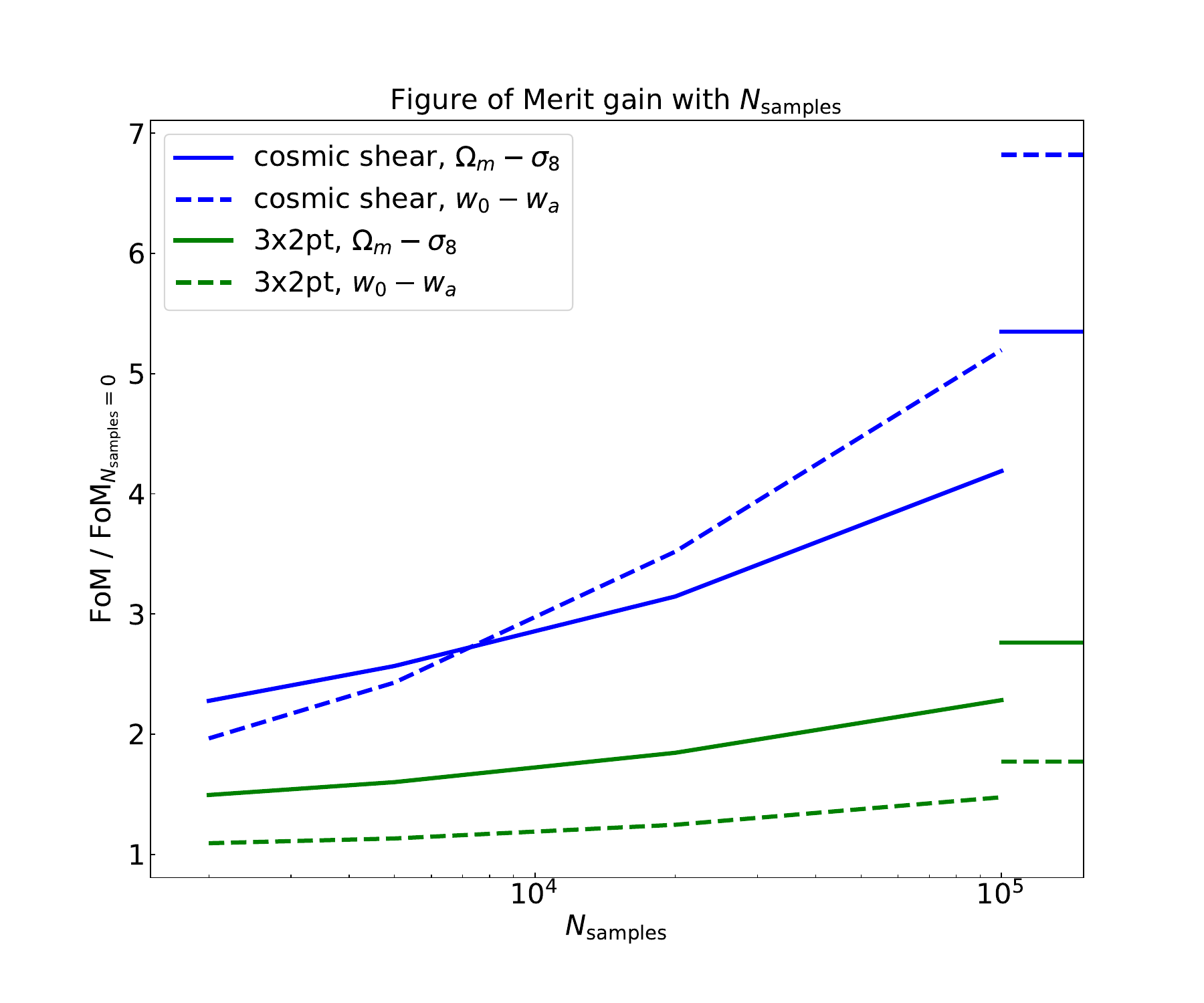}
    \includegraphics[width=0.49\textwidth]{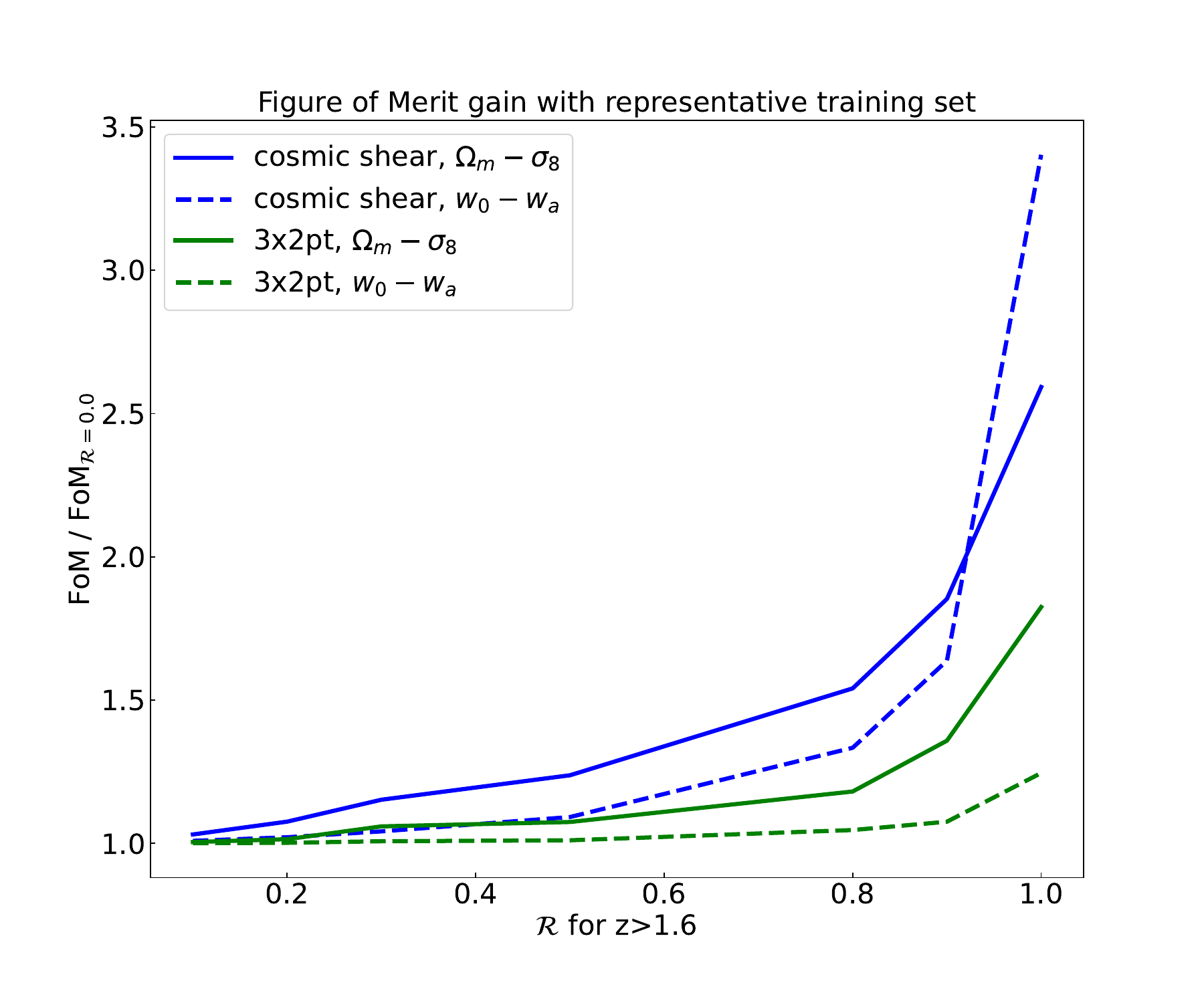}
    \caption{\textbf{Left}: The Figure--of--Merit (FoM) gain for $\Omega_m - \sigma_8$ and $w_0 - w_a$ plane compared to the scenario of $N_{\rm samples} = 0$, for LSST Y10 cosmic shear and $3 \times 2$pt experiments.  The dashed horizontal lines on the right side show the FoM improves with $N_{\rm samples} = \infty$. \textbf{Right}: The Figure--of--Merit (FoM) improve for the $\Omega_m - \sigma_8$ and $w_0 - w_a$ planes compared to the scenario of $\mathcal{R} = 0.0$. 
    }
    \label{fig:FoM_gain}
\end{figure*}

\begin{figure*} 
    \centering
    \includegraphics[width=0.49\textwidth]{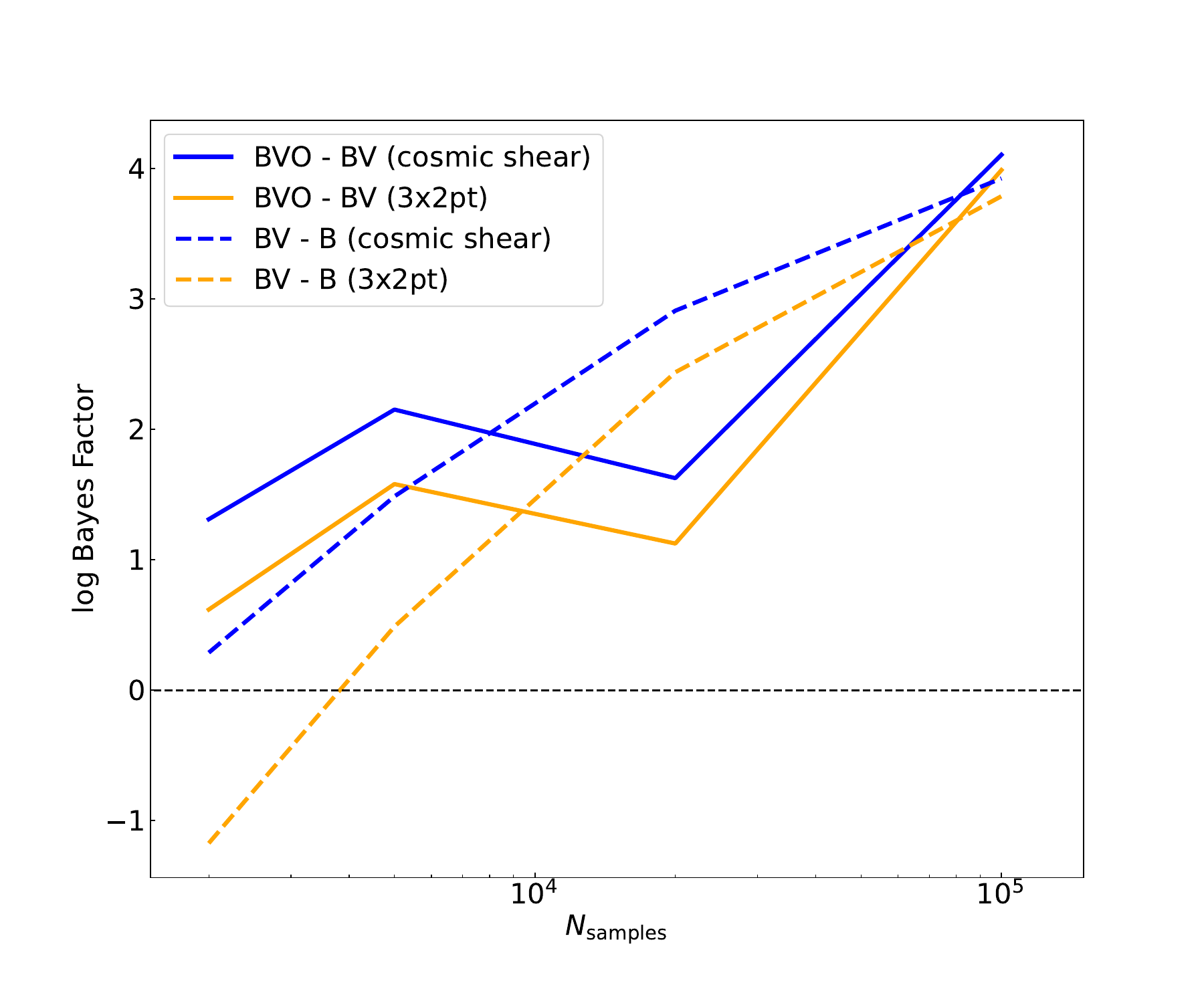}
    \includegraphics[width=0.49\textwidth]{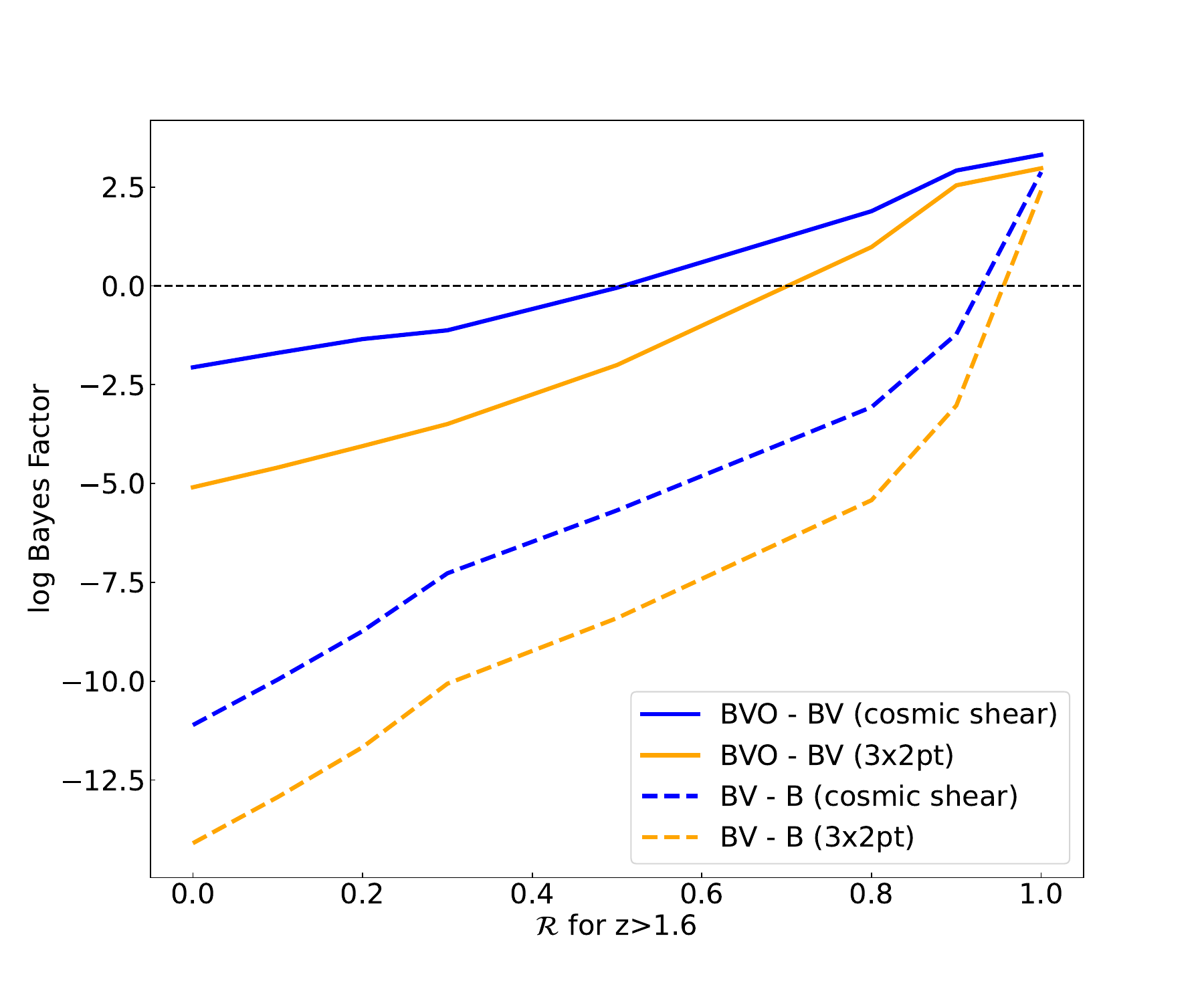}
    \caption{The log Bayes factor comparing different models given different photo--$z$ priors corresponding to $N_{\rm samples}$ and $\mathcal{R}$. ``BVO'', ``BV'' and ``B'' stands for bias+variance+outlier, bias+variance, and bias model correspondingly. A positive Bayes factor between model A and B means model A is preferred, vice versa. When given a representative training set, the Bayes factor almost always prefers the ``BVO'' model. When the high--redshift galaxies are not fairly represented, the Bayes factor may prefer a simpler model. 
    }
    \label{fig:bayes_factor}
\end{figure*}

\subsubsection{Cosmic Shear}

To learn the relative importance of photo--$z$ error parameters on cosmological parameter inference using cosmic shear, we present the results from the decision tree feature importance, as well as the results from the interpretability metric MAFE defined in Section~\ref{sec:method:interp}. 

We show two types of figures to help interpret the decision tree feature importance results. Figure~\ref{fig:mafe_cosmic_shear} shows the MAFE result described in Section~\ref{sec:method:interp}. Figure~\ref{fig:feature_cosmic_shear} shows the feature importance for the photo--$z$ parameters for the LSST Y10 cosmic shear analysis. The top panel shows the feature importance when all 15 redshift parameters are sampled jointly \refresponse{from their priors}. The bottom panels show the feature importance when all  bias, variance and outlier rate parameters \refresponse{across all tomographic bins are left to vary, while leaving other uncertainty parameters fixed}. The relative importance between different types of parameters (biases vs.\ variances vs.\ outliers) should not be compared in the MAFE results and the separated feature results. 

From the top panel of Fig~\ref{fig:feature_cosmic_shear}, we compare the importance of different types of parameters and their impact on cosmic shear analysis. Unsurprisingly, the $\delta z_i$ parameters pose the most cosmological bias to cosmic shear, across all four cosmological parameters, which is why many previous cosmic shear analyses use $\delta z_i$ as the redshift parameters \citep{2021A&A...645A.104A,2022PhRvD.105b3515S,2023PhRvD.108l3519D,2023PhRvD.108l3518L}. The results also show that the variance of the second bin and the outlier rate of the first bin have relatively high importance. Within the bias parameters, $\delta z_1$ and $\delta z_2$ are the most important parameters to consider. This matches the MAFE results, as the data vector is changed more drastically when the $\delta z_1$ and $\delta z_2$ are changed, compared to the others. 

When the redshift parameters are separately sampled \refresponse{from their priors}, the conclusion drawn from the feature importance might not be the same as the conclusion drawn from the joint samples. This is caused by the correlation between the change in data vector caused by different types of redshift parameters. We note that the feature importance shows that the first four shift parameters are all important, while for the outlier fraction, the first parameter is far more important than the other four. This may be caused by the fact that the outlier distribution of the first bin can change the mean and width of the distribution significantly.

\subsubsection{$3\times 2$pt analysis}
\label{sec:result:3x2pt}

We extend the feature importance analysis to the $3\times 2$pt analysis.  
Fig~\ref{fig:mafe_3x2} shows the MAFE results for the LSST Y10 $3 \times 2$pt analysis. Figure~\ref{fig:feature_3x2} shows the feature importance for the photo--$z$ parameters.

For the $3 \times 2$pt analysis, the joint feature importance analysis shows that the outlier rate parameters are generally less relevant in predicting the bias in cosmological parameters. The \refresponse{redshift} bias and variance parameters of bin two are extremely important for $\Omega_m$, $\sigma_8$ and $w_0$, and the bias parameter of bin four is the most important parameter when it comes to constraining $S_8$. 
In a $3 \times 2$pt analysis, the parameters of the higher redshift bins show a higher impact compared to those of cosmic shear analysis. The feature importance sampled separately also tends to match those sampled jointly better compared to the cosmic shear analysis.

\subsection{Training Set Scenarios}
\label{sec:result:scenario}

In this section, we show the results when the prior of the redshift parameters are changed due to different training set scenarios described in Section~\ref{sec:method:prior}. We consider two kinds of analysis: the LSST Y10 cosmic shear--only analysis and the LSST Y10 $3 \times 2$pt analysis. Due to the constraints of space, we only show one contour plot for the LSST Y10 cosmic shear--only analysis in the representative training set scenario, in Figure~\ref{fig:nsample_contour}. A larger $N_\text{samples}$ corresponds to a smaller contour.  
The Fisher forecasting still gives a reasonable contour for $N_\text{samples} = 0$ due to the self constraining power of the cosmic shear probe. 

In Fig.~\ref{fig:FoM_gain}, we compare the Figure--of--Merit (FoM) improvement from the base case for both the representative and non--representative training set scenarios, as functions of the number of training galaxies $N_{\rm samples}$ and the spectroscopic success rate for galaxies with $z \ge 1.6$, $\mathcal{R}$. 
The base case for the representative training set is when $N_{\rm samples} = 0.0$, 
while the base case for the non--representative training set is $\mathcal{R} = 0.0$. We present the FoM as a ratio with respect to the base case to show the relative improvement compared to the base case. 

In the case of a representative training set, a larger $N_{\rm samples}$ corresponds to higher FoM across all analyses and parameter sets. However, we see more benefit for the LSST Y10 cosmic shear by the increase of $N_{\rm samples}$, potentially due to its weaker self--calibrating power compared to the $3 \times 2$pt analysis. In the case of a non--representative training set, the increase of $\mathcal{R}$ slowly decreases the prior on the redshift parameters, until it is close to one while the prior quickly converges to statistical certainty. This argues that we need an almost representative training set for the photo--$z$ distribution to be able to drastically improve the cosmological constraining power of the LSST cosmic shear and $3 \times 2$pt analysis. 
This result emphasizes the importance of deep and complete medium band surveys or spectroscopic surveys at infrared wavelengths, which are suitable for determining the redshift of galaxies with $z>1.6$, in line with the arguments of \cite{2015APh....63...81N}. 
\refresponse{We also observe that the cosmic shear $w_0-w_a$ constraints improve more rapidly when $\mathcal{R}\rightarrow1$, which flags that the dark energy constraints rely more on redshift distribution calibration across the full redshift range.}

One caveat is that the baseline assumption that there is no high redshift calibration sample is quite extreme; in reality, infrared and median band surveys provide some high redshift galaxies that alleviate this problem \citep[e.g.,][]{cosmos}. Also, other samples over $z>1.6$ that can trace the large--scale structure, e.g., quasars, can potentially be used to calibrate the LSST redshift distribution via cross--correlation. 

\refresponse{
In Figure~\ref{fig:bayes_factor}, we compare the models under different training set scenarios using the logarithmic Bayes factor defined in Section~\ref{sec:method:prior}. To reiterate, a positive log Bayes factor between model A and model B $\log \mathcal{B}_{\rm AB} > 0$  means model A is preferred by the Bayesian evidence. We compare three models with decreasing complexity in the source photo--$z$ modeling: they are ``BVO'', ``BV'', and ``B'', which correspond to ``bias+variance+outlier rate'', ``bias+variance'', and ``bias only'' models for each tomographic bin. We find that the BVO model is nearly always preferred, except that the bias--only model is preferred when $N_{\rm samples} = 2000$ for the $3 \times 2$pt analysis. When the representativeness of $z>1.6$ galaxies is nonzero, the preferred model depends on the representativeness. For the cosmic shear, the bias--only model is preferred when $\mathcal{R} <0.5$, and the ``BVO'' is preferred otherwise. For the $3 \times 2$pt analysis, the bias--only model is preferred when $\mathcal{R}<0.7$, and the ``BVO'' model is preferred otherwise. This shows that a more complex model is preferred when the higher redshift representativeness improves.   }

\section{Conclusion}
\label{sec:conclusion}

Controlling systematic biases and uncertainties will be essential for the cosmological analyses for the upcoming surveys like the Rubin Observatory LSST and Roman. Photometric redshifts are one major source of systematic uncertainty in upcoming surveys. In this paper, we studied the impact of \refresponse{source} photometric redshift modeling errors on cosmological inference by creating a Fisher forecasting tool \texttt{FisherA2Z}. \texttt{FisherA2Z} can forecast the marginalized error and bias of LSST cosmic shear and $3 \times 2$pt results by modeling the redshift distribution $n(z)$ \refresponse{for the source galaxies} with a mean, variance and outlier parameter for each tomographic bin, with a total maximum of 15 parameters for the $n(z)$. 

For the $3 \times 2$pt analysis, we created a 36--parameter model incorporating four cosmological parameters, 15 $n(z)$ parameters, four intrinsic alignment parameters, and ten galaxy bias parameters, to forecast the uncertainty of cosmological parameters. We conducted a baseline Fisher forecast for the LSST Y10 clustering and weak lensing analysis and reach the following conclusions:
\begin{enumerate} 
    \item Marginalizing redshift distribution parameters increases the size of the cosmological parameters uncertainty contours in both the $3 \times 2$pt and cosmic shear analysis, more so in the cosmic shear analysis, since the $3 \times 2$pt analysis can partially self--calibrate these uncertainties. 
    \item We demonstrate the different constraining power of probes in the $3 \times 2$pt analysis, and show the sensitivity of different parameters to different probes. \refresponse{In particular, we find that the mean redshift of the fourth tomographic bin in a $3 \times 2$pt analysis is most important to $S_8$, while the mean redshift and variance of the second bin are important to $\Omega_m$, $\sigma_8$, and $w_0$. }
    \item \texttt{FisherA2Z} is able to make the Fisher forecast consistent with the DESC--SRD forecast within $20 $ per cent on key cosmological parameters such as $\Omega_m$, $\sigma_8$, $w_0$ and $w_a$. The difference may be caused by different implementations of the forward model.
    \item We showed that assuming an incorrect value for the photo--$z$ error parameters could lead to $\sim 2\sigma$ biases and changes in the uncertainty of inferred cosmological parameters. 
    \item \refresponse{We find that increasing the training set size and the representativeness of galaxies greater than $z=1.6$ can significantly help the constraining power of the LSST $3 \times 2$pt and cosmic shear analyses. Using the Bayes factor, we find that the bias+variance+outlier model is only preferred when the representativeness is greater than 0.5 for cosmic shear and greater than 0.7 for the $3 \times 2$pt analysis. The bias--only model is preferred otherwise.  }
\end{enumerate}

Mis--modeling in $n(z)$ parameters has different impacts on different cosmological parameters. We used conditional entropy minimization to learn the relative importance of photo--$z$ bias, scatter, and outlier fraction, in different bins, for $\Omega_m, \sigma_8$, $S_8$ and $w_0$, and defined two metrics to help qualitatively guide interpretability: one related to the change in $C_\ell$ due to the changes in $n(z)$ parameters, and one related to the sensitivity of the cosmological parameters to the information in each photo--$z$ bin, using the feature importance of the decision tree. These results show that the mean redshift of the first two bins and the scatter of the second bin are most important for the Y10 cosmic shear analysis, across most parameters. For the Y10 $3 \times 2$pt analysis, the parameters display different dependencies on redshift parameters, with $S_8$ depending mostly on the mean redshift of the fourth tomographic bin, and $w_0$ depending on the mean and scatter of the second bin, for instance.

There are multiple improvements and future work that can be made to \texttt{FisherA2Z} and this work. 
For the galaxy clustering, our model only considers auto--correlations in each lens sample bin. A future study could also include cross--correlations between different lens sample bins. Although these cross--correlations typically have negligible signals, it was shown in \citealt{schaan} that they can still provide useful information on photo--$z$ errors, as substantial photo--$z$ errors can lead to nonzero correlations between bins. Alternatives to the photo--$z$ error model used in our work also exist and can be compared to other models in future studies. While this paper uses the \texttt{FlexZBoost} photo--$z$ estimation method on the CosmoDC2 catalog with two different magnitude cuts to model the outliers, it would be valuable to test the robustness of the results against other redshift estimation methods by repeating this investigation with an outlier sample defined using the results of different photo--$z$ algorithms and datasets. There is an interesting opportunity to develop more advanced validation metrics and model comparison in the context of the apparent issues that arise in mis-specified cosmological and redshift distribution models.   Finally, it would be worth exploring the impact of lens photometric redshift uncertainties, in addition to those of the source sample. Another extension to this work could look at a combination of this bias and the size of the confidence interval from the Fisher information matrix, or possibly the Dark Energy Task Force (DETF) Figure--of--Merit \citep[FoM;][]{2009arXiv0901.0721A}, in order to construct a more comprehensive redshift calibration requirement. 

We will continue to make improvements to \texttt{FisherA2Z} by improving its capability and flexibility in the modeling choices. With \texttt{FisherA2Z} being completely open--source, we believe it is a great tool to forecast the cosmological impact of photo--$z$ systematics for future cosmological studies.

\section*{Contributors}

TZ: Lead and corresponding author, worked on running the analysis, research investigation, continuing software development, and writing the paper. 

\noindent HA: Led the work on the initial statistical and computational formal analysis, investigation, methodology development, and initial software development, visualization, and writing.

\noindent RM: Scientific oversight, paper editing. 

\noindent MMR: Scientific oversight, regular discussion with the project lead, paper editing.. 

\noindent NS: Initial testing for the code, contribute to development via bug--reporting and paper editing.

\noindent CDL: Developing the code of the intrinsic alignment modeling, paper editing.

 \noindent JAN: Scientific oversight around the prior scenarios, paper editing. 

\noindent BA: Scientific oversight around the prior scenarios, paper editing. 

\noindent SL: beta testing the \texttt{FisherA2Z}, paper editing. 

\subsection*{Acknowledgments}

TZ and RM are supported by Schmidt Sciences. 
TZ thanks SLAC National Accelerator Laboratory for providing hospitality and an excellent research environment during the course of this study. HA and RM are partially supported by the Department of Energy Cosmic Frontier program, grant DE--SC0010118.
NS is supported in part by the OpenUniverse effort, which is funded by NASA under JPL Contract Task 70-711320, ‘Maximizing Science Exploitation of Simulated Cosmological Survey Data Across Surveys.’

HA thanks the LSSTC Data Science Fellowship Program, which is funded by LSSTC, NSF Cybertraining Grant \#1829740, the Brinson Foundation, and the Moore Foundation; his participation in the program has benefited this work.

Figures in this paper were generated using the open--source Python libraries: matplotlib \citep{matplotlib}, seaborn \citep{seaborn}.

This paper has undergone internal review in the LSST Dark Energy Science Collaboration. 
The internal reviewers were Jaime Ruiz-Zapatero and Paul Rogozenski. We thank them for their valuable contribution.

The DESC acknowledges ongoing support from the Institut National de 
Physique Nucl\'eaire et de Physique des Particules in France; the 
Science \& Technology Facilities Council in the United Kingdom; and the
Department of Energy and the LSST Discovery Alliance
in the United States.  DESC uses resources of the IN2P3 
Computing Center (CC-IN2P3--Lyon/Villeurbanne - France) funded by the 
Centre National de la Recherche Scientifique; the National Energy 
Research Scientific Computing Center, a DOE Office of Science User 
Facility supported by the Office of Science of the U.S.\ Department of
Energy under Contract No.\ DE-AC02-05CH11231; STFC DiRAC HPC Facilities, 
funded by UK BEIS National E-infrastructure capital grants; and the UK 
particle physics grid, supported by the GridPP Collaboration.  This 
work was performed in part under DOE Contract DE-AC02-76SF00515.

\section*{Data Availability Statement}

The software and data to produce the results of this work are open source at \url{github.com/LSSTDESC/fisherA2Z}. 

\bibliographystyle{mnras}
\bibliography{main}

\begin{thebibliography}{}
\makeatletter
\relax
\def\mn@urlcharsother{\let\do\@makeother \do\$\do\&\do\#\do\^\do\_\do\%\do\~}
\def\mn@doi{\begingroup\mn@urlcharsother \@ifnextchar [ {\mn@doi@} {\mn@doi@[]}}
\def\mn@doi@[#1]#2{\def\@tempa{#1}\ifx\@tempa\@empty \href {http://dx.doi.org/#2} {doi:#2}\else \href {http://dx.doi.org/#2} {#1}\fi \endgroup}
\def\mn@eprint#1#2{\mn@eprint@#1:#2::\@nil}
\def\mn@eprint@arXiv#1{\href {http://arxiv.org/abs/#1} {{\tt arXiv:#1}}}
\def\mn@eprint@dblp#1{\href {http://dblp.uni-trier.de/rec/bibtex/#1.xml} {dblp:#1}}
\def\mn@eprint@#1:#2:#3:#4\@nil{\def\@tempa {#1}\def\@tempb {#2}\def\@tempc {#3}\ifx \@tempc \@empty \let \@tempc \@tempb \let \@tempb \@tempa \fi \ifx \@tempb \@empty \def\@tempb {arXiv}\fi \@ifundefined {mn@eprint@\@tempb}{\@tempb:\@tempc}{\expandafter \expandafter \csname mn@eprint@\@tempb\endcsname \expandafter{\@tempc}}}

\bibitem[\protect\citeauthoryear{{Abbott} et~al.,}{{Abbott} et~al.}{2022}]{2022PhRvD.105b3520A}
{Abbott} T.~M.~C.,  et~al., 2022, \mn@doi [\prd] {10.1103/PhysRevD.105.023520}, \href {https://ui.adsabs.harvard.edu/abs/2022PhRvD.105b3520A} {105, 023520}

\bibitem[\protect\citeauthoryear{{Albrecht} et~al.,}{{Albrecht} et~al.}{2009}]{2009arXiv0901.0721A}
{Albrecht} A.,  et~al., 2009, arXiv e-prints, \href {https://ui.adsabs.harvard.edu/abs/2009arXiv0901.0721A} {p. arXiv:0901.0721}

\bibitem[\protect\citeauthoryear{{Amon} et~al.,}{{Amon} et~al.}{2022}]{2022PhRvD.105b3514A}
{Amon} A.,  et~al., 2022, \mn@doi [\prd] {10.1103/PhysRevD.105.023514}, \href {https://ui.adsabs.harvard.edu/abs/2022PhRvD.105b3514A} {105, 023514}

\bibitem[\protect\citeauthoryear{{Asgari} et~al.,}{{Asgari} et~al.}{2021}]{2021A&A...645A.104A}
{Asgari} M.,  et~al., 2021, \mn@doi [\aap] {10.1051/0004-6361/202039070}, \href {https://ui.adsabs.harvard.edu/abs/2021A&A...645A.104A} {645, A104}

\bibitem[\protect\citeauthoryear{{Benabed} \& {van Waerbeke}}{{Benabed} \& {van Waerbeke}}{2004}]{2004PhRvD..70l3515B}
{Benabed} K.,  {van Waerbeke} L.,  2004, \mn@doi [\prd] {10.1103/PhysRevD.70.123515}, \href {http://adsabs.harvard.edu/abs/2004PhRvD..70l3515B} {70, 123515}

\bibitem[\protect\citeauthoryear{{Bernstein} \& {Huterer}}{{Bernstein} \& {Huterer}}{2010}]{2010MNRAS.401.1399B}
{Bernstein} G.,  {Huterer} D.,  2010, \mn@doi [\mnras] {10.1111/j.1365-2966.2009.15748.x}, \href {https://ui.adsabs.harvard.edu/abs/2010MNRAS.401.1399B} {401, 1399}

\bibitem[\protect\citeauthoryear{{Bernstein} \& {Jain}}{{Bernstein} \& {Jain}}{2004}]{2004ApJ...600...17B}
{Bernstein} G.,  {Jain} B.,  2004, \mn@doi [\apj] {10.1086/379768}, \href {https://ui.adsabs.harvard.edu/abs/2004ApJ...600...17B} {600, 17}

\bibitem[\protect\citeauthoryear{{Bhandari}, {Leonard}, {Rau}  \& {Mandelbaum}}{{Bhandari} et~al.}{2021}]{naren}
{Bhandari} N.,  {Leonard} C.~D.,  {Rau} M.~M.,   {Mandelbaum} R.,  2021, arXiv e-prints, \href {https://ui.adsabs.harvard.edu/abs/2021arXiv210100298B} {p. arXiv:2101.00298}

\bibitem[\protect\citeauthoryear{Bishop}{Bishop}{2006}]{mlbook1}
Bishop C.,  2006, Pattern Recognition and Machine Learning.
Springer, \url {https://www.microsoft.com/en-us/research/publication/pattern-recognition-machine-learning/}

\bibitem[\protect\citeauthoryear{{Breiman}}{{Breiman}}{2001}]{Breiman2001}
{Breiman} L.,  2001, \mn@doi [Machine Learning] {10.1023/A:1010933404324}, \href {https://ui.adsabs.harvard.edu/abs/2001MachL..45....5B} {45, 5}

\bibitem[\protect\citeauthoryear{{Bridle} \& {King}}{{Bridle} \& {King}}{2007}]{NLA}
{Bridle} S.,  {King} L.,  2007, \mn@doi [New Journal of Physics] {10.1088/1367-2630/9/12/444}, \href {https://ui.adsabs.harvard.edu/abs/2007NJPh....9..444B} {9, 444}

\bibitem[\protect\citeauthoryear{{Brodtkorb} \& {D'Errico}}{{Brodtkorb} \& {D'Errico}}{2019}]{numdifftools}
{Brodtkorb} P.~A.,  {D'Errico} J.,  2019, {numdifftools}, \url {https://numdifftools.readthedocs.io}

\bibitem[\protect\citeauthoryear{{Chevallier} \& {Polarski}}{{Chevallier} \& {Polarski}}{2001}]{Chevallier2001}
{Chevallier} M.,  {Polarski} D.,  2001, \mn@doi [International Journal of Modern Physics D] {10.1142/S0218271801000822}, \href {https://ui.adsabs.harvard.edu/abs/2001IJMPD..10..213C} {10, 213}

\bibitem[\protect\citeauthoryear{{Chisari} et~al.,}{{Chisari} et~al.}{2019}]{CCL}
{Chisari} N.~E.,  et~al., 2019, \mn@doi [\apjs] {10.3847/1538-4365/ab1658}, \href {https://ui.adsabs.harvard.edu/abs/2019ApJS..242....2C} {242, 2}

\bibitem[\protect\citeauthoryear{{Coe}}{{Coe}}{2009}]{Coe}
{Coe} D.,  2009, arXiv e-prints, \href {https://ui.adsabs.harvard.edu/abs/2009arXiv0906.4123C} {p. arXiv:0906.4123}

\bibitem[\protect\citeauthoryear{{Cunha}, {Huterer}, {Lin}, {Busha}  \& {Wechsler}}{{Cunha} et~al.}{2014}]{2014MNRAS.444..129C}
{Cunha} C.~E.,  {Huterer} D.,  {Lin} H.,  {Busha} M.~T.,   {Wechsler} R.~H.,  2014, \mn@doi [\mnras] {10.1093/mnras/stu1424}, \href {https://ui.adsabs.harvard.edu/abs/2014MNRAS.444..129C} {444, 129}

\bibitem[\protect\citeauthoryear{{Dalal} et~al.,}{{Dalal} et~al.}{2023}]{2023PhRvD.108l3519D}
{Dalal} R.,  et~al., 2023, \mn@doi [\prd] {10.1103/PhysRevD.108.123519}, \href {https://ui.adsabs.harvard.edu/abs/2023PhRvD.108l3519D} {108, 123519}

\bibitem[\protect\citeauthoryear{{Desjacques}, {Jeong}  \& {Schmidt}}{{Desjacques} et~al.}{2018}]{2018PhR...733....1D}
{Desjacques} V.,  {Jeong} D.,   {Schmidt} F.,  2018, \mn@doi [\physrep] {10.1016/j.physrep.2017.12.002}, \href {https://ui.adsabs.harvard.edu/abs/2018PhR...733....1D} {733, 1}

\bibitem[\protect\citeauthoryear{{Eisenstein} \& {Hu}}{{Eisenstein} \& {Hu}}{1998}]{eisenstein_hu}
{Eisenstein} D.~J.,  {Hu} W.,  1998, \mn@doi [\apj] {10.1086/305424}, \href {https://ui.adsabs.harvard.edu/abs/1998ApJ...496..605E} {496, 605}

\bibitem[\protect\citeauthoryear{{Graham} et~al.,}{{Graham} et~al.}{2020}]{graham2020}
{Graham} M.~L.,  et~al., 2020, \mn@doi [\aj] {10.3847/1538-3881/ab8a43}, \href {https://ui.adsabs.harvard.edu/abs/2020AJ....159..258G} {159, 258}

\bibitem[\protect\citeauthoryear{{Hartley} et~al.,}{{Hartley} et~al.}{2020}]{2020MNRAS.496.4769H}
{Hartley} W.~G.,  et~al., 2020, \mn@doi [\mnras] {10.1093/mnras/staa1812}, \href {https://ui.adsabs.harvard.edu/abs/2020MNRAS.496.4769H} {496, 4769}

\bibitem[\protect\citeauthoryear{Hastie, Tibshirani  \& Friedman}{Hastie et~al.}{2009}]{mlbook3}
Hastie T.,  Tibshirani R.,   Friedman J.,  2009, The Elements of Statistical Learning: Data Mining, Inference, and Prediction.
Springer series in statistics, Springer, \url {https://books.google.com/books?id=eBSgoAEACAAJ}

\bibitem[\protect\citeauthoryear{{Hearin}, {Zentner}  \& {Ma}}{{Hearin} et~al.}{2012}]{2012JCAP...04..034H}
{Hearin} A.~P.,  {Zentner} A.~R.,   {Ma} Z.,  2012, \mn@doi [\jcap] {10.1088/1475-7516/2012/04/034}, \href {https://ui.adsabs.harvard.edu/abs/2012JCAP...04..034H} {2012, 034}

\bibitem[\protect\citeauthoryear{{Heymans} et~al.,}{{Heymans} et~al.}{2021}]{2021A&A...646A.140H}
{Heymans} C.,  et~al., 2021, \mn@doi [\aap] {10.1051/0004-6361/202039063}, \href {https://ui.adsabs.harvard.edu/abs/2021A&A...646A.140H} {646, A140}

\bibitem[\protect\citeauthoryear{{Hu}}{{Hu}}{2002}]{2002PhRvD..65b3003H}
{Hu} W.,  2002, \prd, \href {http://adsabs.harvard.edu/cgi-bin/nph-bib_query?bibcode=2002PhRvD..65b3003H&db_key=AST} {65, 023003}

\bibitem[\protect\citeauthoryear{Hunter}{Hunter}{2007}]{matplotlib}
Hunter J.~D.,  2007, \mn@doi [Computing In Science \& Engineering] {10.1109/MCSE.2007.55}, 9, 90

\bibitem[\protect\citeauthoryear{{Huterer}}{{Huterer}}{2002}]{2002PhRvD..65f3001H}
{Huterer} D.,  2002, \mn@doi [\prd] {10.1103/PhysRevD.65.063001}, \href {https://ui.adsabs.harvard.edu/abs/2002PhRvD..65f3001H} {65, 063001}

\bibitem[\protect\citeauthoryear{{Huterer}, {Takada}, {Bernstein}  \& {Jain}}{{Huterer} et~al.}{2006}]{biasformula2}
{Huterer} D.,  {Takada} M.,  {Bernstein} G.,   {Jain} B.,  2006, \mn@doi [\mnras] {10.1111/j.1365-2966.2005.09782.x}, \href {https://ui.adsabs.harvard.edu/abs/2006MNRAS.366..101H} {366, 101}

\bibitem[\protect\citeauthoryear{{Ivezi{\'c}} et~al.,}{{Ivezi{\'c}} et~al.}{2019}]{Overview}
{Ivezi{\'c}} {\v{Z}}.,  et~al., 2019, \mn@doi [\apj] {10.3847/1538-4357/ab042c}, \href {https://ui.adsabs.harvard.edu/abs/2019ApJ...873..111I} {873, 111}

\bibitem[\protect\citeauthoryear{Izbicki \& Lee}{Izbicki \& Lee}{2017}]{10.1214/17-EJS1302}
Izbicki R.,  Lee A.~B.,  2017, \mn@doi [Electronic Journal of Statistics] {10.1214/17-EJS1302}, 11, 2800

\bibitem[\protect\citeauthoryear{{Joachimi} et~al.,}{{Joachimi} et~al.}{2015}]{IA4}
{Joachimi} B.,  et~al., 2015, \mn@doi [\ssr] {10.1007/s11214-015-0177-4}, \href {https://ui.adsabs.harvard.edu/abs/2015SSRv..193....1J} {193, 1}

\bibitem[\protect\citeauthoryear{{Kaiser}}{{Kaiser}}{1984}]{galaxybias}
{Kaiser} N.,  1984, \mn@doi [\apjl] {10.1086/184341}, \href {https://ui.adsabs.harvard.edu/abs/1984ApJ...284L...9K} {284, L9}

\bibitem[\protect\citeauthoryear{{Kiessling} et~al.,}{{Kiessling} et~al.}{2015}]{IA2}
{Kiessling} A.,  et~al., 2015, \mn@doi [\ssr] {10.1007/s11214-015-0203-6}, \href {https://ui.adsabs.harvard.edu/abs/2015SSRv..193...67K} {193, 67}

\bibitem[\protect\citeauthoryear{{Kilbinger}}{{Kilbinger}}{2015}]{weaklensingreview2}
{Kilbinger} M.,  2015, \mn@doi [Reports on Progress in Physics] {10.1088/0034-4885/78/8/086901}, \href {https://ui.adsabs.harvard.edu/abs/2015RPPh...78h6901K} {78, 086901}

\bibitem[\protect\citeauthoryear{{Kirk} et~al.,}{{Kirk} et~al.}{2015}]{IA1}
{Kirk} D.,  et~al., 2015, \mn@doi [\ssr] {10.1007/s11214-015-0213-4}, \href {https://ui.adsabs.harvard.edu/abs/2015SSRv..193..139K} {193, 139}

\bibitem[\protect\citeauthoryear{Kirkby, Mendoza  \& Sanchez}{Kirkby et~al.}{2020}]{weaklensingdeblending}
Kirkby D.,  Mendoza I.,   Sanchez J.,  2020, WeakLensingDeblending, \mn@doi{10.5281/zenodo.3975230}, \url {https://doi.org/10.5281/zenodo.3975230}

\bibitem[\protect\citeauthoryear{{Korytov} et~al.,}{{Korytov} et~al.}{2019}]{2019ApJS..245...26K}
{Korytov} D.,  et~al., 2019, \mn@doi [\apjs] {10.3847/1538-4365/ab510c}, \href {https://ui.adsabs.harvard.edu/abs/2019ApJS..245...26K} {245, 26}

\bibitem[\protect\citeauthoryear{{Krause} \& {Eifler}}{{Krause} \& {Eifler}}{2017}]{cosmolike}
{Krause} E.,  {Eifler} T.,  2017, \mn@doi [\mnras] {10.1093/mnras/stx1261}, \href {https://ui.adsabs.harvard.edu/abs/2017MNRAS.470.2100K} {470, 2100}

\bibitem[\protect\citeauthoryear{{Krause}, {Eifler}  \& {Blazek}}{{Krause} et~al.}{2016}]{KEB}
{Krause} E.,  {Eifler} T.,   {Blazek} J.,  2016, \mn@doi [\mnras] {10.1093/mnras/stv2615}, \href {https://ui.adsabs.harvard.edu/abs/2016MNRAS.456..207K} {456, 207}

\bibitem[\protect\citeauthoryear{{Krause} et~al.,}{{Krause} et~al.}{2017}]{C_ells}
{Krause} E.,  et~al., 2017, arXiv e-prints, \href {https://ui.adsabs.harvard.edu/abs/2017arXiv170609359K} {p. arXiv:1706.09359}

\bibitem[\protect\citeauthoryear{{Krause} et~al.,}{{Krause} et~al.}{2021}]{3x2pt_des}
{Krause} E.,  et~al., 2021, arXiv e-prints, \href {https://ui.adsabs.harvard.edu/abs/2021arXiv210513548K} {p. arXiv:2105.13548}

\bibitem[\protect\citeauthoryear{{LSST Dark Energy Science Collaboration} et~al.,}{{LSST Dark Energy Science Collaboration} et~al.}{2018}]{DESCSRD}
{LSST Dark Energy Science Collaboration} et~al., 2018, arXiv e-prints, \href {https://ui.adsabs.harvard.edu/abs/2018arXiv180901669T} {p. arXiv:1809.01669}

\bibitem[\protect\citeauthoryear{{LSST Science Collaboration} et~al.,}{{LSST Science Collaboration} et~al.}{2009}]{2009arXiv0912.0201L}
{LSST Science Collaboration} et~al., 2009, arXiv e-prints, \href {https://ui.adsabs.harvard.edu/abs/2009arXiv0912.0201L} {p. arXiv:0912.0201}

\bibitem[\protect\citeauthoryear{{Lamman}, {Tsaprazi}, {Shi}, {{\v{S}}ar{\v{c}}evi{\'c}}, {Pyne}, {Legnani}  \& {Ferreira}}{{Lamman} et~al.}{2024}]{Lamman2024}
{Lamman} C.,  {Tsaprazi} E.,  {Shi} J.,  {{\v{S}}ar{\v{c}}evi{\'c}} N.~N.,  {Pyne} S.,  {Legnani} E.,   {Ferreira} T.,  2024, \mn@doi [The Open Journal of Astrophysics] {10.21105/astro.2309.08605}, \href {https://ui.adsabs.harvard.edu/abs/2024OJAp....7E..14L} {7, 14}

\bibitem[\protect\citeauthoryear{{Leonard}, {Rau}  \& {Mandelbaum}}{{Leonard} et~al.}{2024}]{Leonard2024}
{Leonard} C.~D.,  {Rau} M.~M.,   {Mandelbaum} R.,  2024, \mn@doi [\prd] {10.1103/PhysRevD.109.083528}, \href {https://ui.adsabs.harvard.edu/abs/2024PhRvD.109h3528L} {109, 083528}

\bibitem[\protect\citeauthoryear{{Li} et~al.,}{{Li} et~al.}{2023}]{2023PhRvD.108l3518L}
{Li} X.,  et~al., 2023, \mn@doi [\prd] {10.1103/PhysRevD.108.123518}, \href {https://ui.adsabs.harvard.edu/abs/2023PhRvD.108l3518L} {108, 123518}

\bibitem[\protect\citeauthoryear{{Limber}}{{Limber}}{1953}]{Limber1953}
{Limber} D.~N.,  1953, \mn@doi [\apj] {10.1086/145672}, \href {https://ui.adsabs.harvard.edu/abs/1953ApJ...117..134L} {117, 134}

\bibitem[\protect\citeauthoryear{{Ma}, {Hu}  \& {Huterer}}{{Ma} et~al.}{2006}]{Ma}
{Ma} Z.,  {Hu} W.,   {Huterer} D.,  2006, \mn@doi [\apj] {10.1086/497068}, \href {https://ui.adsabs.harvard.edu/abs/2006ApJ...636...21M} {636, 21}

\bibitem[\protect\citeauthoryear{MacKay}{MacKay}{2003}]{mackay2003information}
MacKay D.~J.,  2003, Information Theory, Inference and Learning Algorithms.
Cambridge University Press

\bibitem[\protect\citeauthoryear{{Mandelbaum}}{{Mandelbaum}}{2018}]{rachelreview}
{Mandelbaum} R.,  2018, \mn@doi [Annual Review of Astronomy and Astrophysics] {10.1146/annurev-astro-081817-051928}, \href {http://adsabs.harvard.edu/abs/2017arXiv171003235M} {56, 393}

\bibitem[\protect\citeauthoryear{{Masters}, {Stern}, {Cohen}, {Capak}, {Rhodes}, {Castander}  \& {Paltani}}{{Masters} et~al.}{2017}]{2017ApJ...841..111M}
{Masters} D.~C.,  {Stern} D.~K.,  {Cohen} J.~G.,  {Capak} P.~L.,  {Rhodes} J.~D.,  {Castander} F.~J.,   {Paltani} S.,  2017, \mn@doi [\apj] {10.3847/1538-4357/aa6f08}, \href {https://ui.adsabs.harvard.edu/abs/2017ApJ...841..111M} {841, 111}

\bibitem[\protect\citeauthoryear{{McQuinn} \& {White}}{{McQuinn} \& {White}}{2013}]{2013MNRAS.433.2857M}
{McQuinn} M.,  {White} M.,  2013, \mn@doi [\mnras] {10.1093/mnras/stt914}, \href {https://ui.adsabs.harvard.edu/abs/2013MNRAS.433.2857M} {433, 2857}

\bibitem[\protect\citeauthoryear{Mitchell}{Mitchell}{1997}]{mlbook2}
Mitchell T.,  1997, Machine Learning.
McGraw-Hill International Editions, McGraw-Hill, \url {https://books.google.com/books?id=EoYBngEACAAJ}

\bibitem[\protect\citeauthoryear{{Miyatake} et~al.,}{{Miyatake} et~al.}{2023}]{Miyatake2023}
{Miyatake} H.,  et~al., 2023, arXiv e-prints, p. arXiv:2304.00704

\bibitem[\protect\citeauthoryear{{Moskowitz}, {Gawiser}, {Bault}, {Broussard}, {Newman}, {Zuntz}  \& {LSST Dark Energy Science Collaboration}}{{Moskowitz} et~al.}{2023}]{moskowitz2023}
{Moskowitz} I.,  {Gawiser} E.,  {Bault} A.,  {Broussard} A.,  {Newman} J.~A.,  {Zuntz} J.,   {LSST Dark Energy Science Collaboration} 2023, \mn@doi [\apj] {10.3847/1538-4357/accc88}, \href {https://ui.adsabs.harvard.edu/abs/2023ApJ...950...49M} {950, 49}

\bibitem[\protect\citeauthoryear{{Moskowitz}, {Gawiser}, {Crenshaw}, {Andrews}, {Malz}, {Schmidt}  \& {LSST Dark Energy Science Collaboration}}{{Moskowitz} et~al.}{2024}]{moskowitz2024}
{Moskowitz} I.,  {Gawiser} E.,  {Crenshaw} J.~F.,  {Andrews} B.~H.,  {Malz} A.~I.,  {Schmidt} S.,   {LSST Dark Energy Science Collaboration} 2024, \mn@doi [\apjl] {10.3847/2041-8213/ad4039}, \href {https://ui.adsabs.harvard.edu/abs/2024ApJ...967L...6M} {967, L6}

\bibitem[\protect\citeauthoryear{{Newman}}{{Newman}}{2008}]{2008ApJ...684...88N}
{Newman} J.~A.,  2008, \mn@doi [\apj] {10.1086/589982}, \href {https://ui.adsabs.harvard.edu/abs/2008ApJ...684...88N} {684, 88}

\bibitem[\protect\citeauthoryear{{Newman} \& {Gruen}}{{Newman} \& {Gruen}}{2022}]{2022arXiv220613633N}
{Newman} J.~A.,  {Gruen} D.,  2022, arXiv e-prints, \href {https://ui.adsabs.harvard.edu/abs/2022arXiv220613633N} {p. arXiv:2206.13633}

\bibitem[\protect\citeauthoryear{{Newman} et~al.,}{{Newman} et~al.}{2015}]{2015APh....63...81N}
{Newman} J.~A.,  et~al., 2015, \mn@doi [Astroparticle Physics] {10.1016/j.astropartphys.2014.06.007}, \href {https://ui.adsabs.harvard.edu/abs/2015APh....63...81N} {63, 81}

\bibitem[\protect\citeauthoryear{{Nishizawa}, {Hsieh}, {Tanaka}  \& {Takata}}{{Nishizawa} et~al.}{2020}]{hscpz}
{Nishizawa} A.~J.,  {Hsieh} B.-C.,  {Tanaka} M.,   {Takata} T.,  2020, arXiv e-prints, \href {https://ui.adsabs.harvard.edu/abs/2020arXiv200301511N} {p. arXiv:2003.01511}

\bibitem[\protect\citeauthoryear{Pedregosa et~al.,}{Pedregosa et~al.}{2011}]{scikit-learn}
Pedregosa F.,  et~al., 2011, Journal of Machine Learning Research, 12, 2825

\bibitem[\protect\citeauthoryear{{Prat} et~al.,}{{Prat} et~al.}{2022}]{2022PhRvD.105h3528P}
{Prat} J.,  et~al., 2022, \mn@doi [\prd] {10.1103/PhysRevD.105.083528}, \href {https://ui.adsabs.harvard.edu/abs/2022PhRvD.105h3528P} {105, 083528}

\bibitem[\protect\citeauthoryear{{Rau}, {Hoyle}, {Paech}  \& {Seitz}}{{Rau} et~al.}{2017}]{centroidshift}
{Rau} M.~M.,  {Hoyle} B.,  {Paech} K.,   {Seitz} S.,  2017, \mn@doi [\mnras] {10.1093/mnras/stw3338}, \href {https://ui.adsabs.harvard.edu/abs/2017MNRAS.466.2927R} {466, 2927}

\bibitem[\protect\citeauthoryear{{Rau}, {Wilson}  \& {Mandelbaum}}{{Rau} et~al.}{2020}]{2020MNRAS.491.4768R}
{Rau} M.~M.,  {Wilson} S.,   {Mandelbaum} R.,  2020, \mn@doi [\mnras] {10.1093/mnras/stz3295}, \href {https://ui.adsabs.harvard.edu/abs/2020MNRAS.491.4768R} {491, 4768}

\bibitem[\protect\citeauthoryear{{Rodr{\'\i}guez-Monroy} et~al.,}{{Rodr{\'\i}guez-Monroy} et~al.}{2022}]{2022MNRAS.511.2665R}
{Rodr{\'\i}guez-Monroy} M.,  et~al., 2022, \mn@doi [\mnras] {10.1093/mnras/stac104}, \href {https://ui.adsabs.harvard.edu/abs/2022MNRAS.511.2665R} {511, 2665}

\bibitem[\protect\citeauthoryear{{Ruiz-Zapatero}, {Hadzhiyska}, {Alonso}, {Ferreira}, {Garc{\'\i}a-Garc{\'\i}a}  \& {Mootoovaloo}}{{Ruiz-Zapatero} et~al.}{2023}]{RuizZapatero2023}
{Ruiz-Zapatero} J.,  {Hadzhiyska} B.,  {Alonso} D.,  {Ferreira} P.~G.,  {Garc{\'\i}a-Garc{\'\i}a} C.,   {Mootoovaloo} A.,  2023, \mn@doi [\mnras] {10.1093/mnras/stad1192}, \href {https://ui.adsabs.harvard.edu/abs/2023MNRAS.522.5037R} {522, 5037}

\bibitem[\protect\citeauthoryear{{Salvato}, {Ilbert}  \& {Hoyle}}{{Salvato} et~al.}{2019}]{2019NatAs...3..212S}
{Salvato} M.,  {Ilbert} O.,   {Hoyle} B.,  2019, \mn@doi [Nature Astronomy] {10.1038/s41550-018-0478-0}, \href {https://ui.adsabs.harvard.edu/abs/2019NatAs...3..212S} {3, 212}

\bibitem[\protect\citeauthoryear{{Sanchez}, {Mendoza}, {Kirkby}, {Burchat}  \& {LSST Dark Energy Science Collaboration}}{{Sanchez} et~al.}{2021}]{wld}
{Sanchez} J.,  {Mendoza} I.,  {Kirkby} D.~P.,  {Burchat} P.~R.,   {LSST Dark Energy Science Collaboration} 2021, \mn@doi [\jcap] {10.1088/1475-7516/2021/07/043}, \href {https://ui.adsabs.harvard.edu/abs/2021JCAP...07..043S} {2021, 043}

\bibitem[\protect\citeauthoryear{{Schaan}, {Ferraro}  \& {Seljak}}{{Schaan} et~al.}{2020}]{schaan}
{Schaan} E.,  {Ferraro} S.,   {Seljak} U.,  2020, \mn@doi [\jcap] {10.1088/1475-7516/2020/12/001}, \href {https://ui.adsabs.harvard.edu/abs/2020JCAP...12..001S} {2020, 001}

\bibitem[\protect\citeauthoryear{{Schmidt} et~al.,}{{Schmidt} et~al.}{2020}]{pzdc1}
{Schmidt} S.~J.,  et~al., 2020, \mn@doi [\mnras] {10.1093/mnras/staa2799}, \href {https://ui.adsabs.harvard.edu/abs/2020MNRAS.499.1587S} {499, 1587}

\bibitem[\protect\citeauthoryear{{Scoville} et~al.,}{{Scoville} et~al.}{2007}]{cosmos}
{Scoville} N.,  et~al., 2007, \mn@doi [\apjs] {10.1086/516585}, \href {https://ui.adsabs.harvard.edu/abs/2007ApJS..172....1S} {172, 1}

\bibitem[\protect\citeauthoryear{{Secco} et~al.,}{{Secco} et~al.}{2022}]{2022PhRvD.105b3515S}
{Secco} L.~F.,  et~al., 2022, \mn@doi [\prd] {10.1103/PhysRevD.105.023515}, \href {https://ui.adsabs.harvard.edu/abs/2022PhRvD.105b3515S} {105, 023515}

\bibitem[\protect\citeauthoryear{{Smith} et~al.,}{{Smith} et~al.}{2003}]{Smith2003}
{Smith} R.~E.,  et~al., 2003, \mn@doi [\mnras] {10.1046/j.1365-8711.2003.06503.x}, \href {https://ui.adsabs.harvard.edu/abs/2003MNRAS.341.1311S} {341, 1311}

\bibitem[\protect\citeauthoryear{{St{\"o}lzner}, {Joachimi}, {Korn}, {Hildebrandt}  \& {Wright}}{{St{\"o}lzner} et~al.}{2021}]{2021A&A...650A.148S}
{St{\"o}lzner} B.,  {Joachimi} B.,  {Korn} A.,  {Hildebrandt} H.,   {Wright} A.~H.,  2021, \mn@doi [\aap] {10.1051/0004-6361/202040130}, \href {https://ui.adsabs.harvard.edu/abs/2021A&A...650A.148S} {650, A148}

\bibitem[\protect\citeauthoryear{{Sugiyama} et~al.,}{{Sugiyama} et~al.}{2023}]{Sugiyama2023}
{Sugiyama} S.,  et~al., 2023, \mn@doi [\prd] {10.1103/PhysRevD.108.123521}, \href {https://ui.adsabs.harvard.edu/abs/2023PhRvD.108l3521S} {108, 123521}

\bibitem[\protect\citeauthoryear{{Takada} \& {White}}{{Takada} \& {White}}{2004}]{2004ApJ...601L...1T}
{Takada} M.,  {White} M.,  2004, \mn@doi [\apjl] {10.1086/381870}, \href {https://ui.adsabs.harvard.edu/abs/2004ApJ...601L...1T} {601, L1}

\bibitem[\protect\citeauthoryear{{Takahashi}, {Sato}, {Nishimichi}, {Taruya}  \& {Oguri}}{{Takahashi} et~al.}{2012}]{Takahashi2012}
{Takahashi} R.,  {Sato} M.,  {Nishimichi} T.,  {Taruya} A.,   {Oguri} M.,  2012, \mn@doi [\apj] {10.1088/0004-637X/761/2/152}, \href {https://ui.adsabs.harvard.edu/abs/2012ApJ...761..152T} {761, 152}

\bibitem[\protect\citeauthoryear{{Tessore} \& {Harrison}}{{Tessore} \& {Harrison}}{2020}]{Tessore2020}
{Tessore} N.,  {Harrison} I.,  2020, \mn@doi [The Open Journal of Astrophysics] {10.21105/astro.2003.11558}, \href {https://ui.adsabs.harvard.edu/abs/2020OJAp....3E...6T} {3, 6}

\bibitem[\protect\citeauthoryear{{Trotta}}{{Trotta}}{2008}]{Trotta2008}
{Trotta} R.,  2008, \mn@doi [Contemporary Physics] {10.1080/00107510802066753}, \href {https://ui.adsabs.harvard.edu/abs/2008ConPh..49...71T} {49, 71}

\bibitem[\protect\citeauthoryear{{Troxel} \& {Ishak}}{{Troxel} \& {Ishak}}{2015}]{IA3}
{Troxel} M.~A.,  {Ishak} M.,  2015, \mn@doi [\physrep] {10.1016/j.physrep.2014.11.001}, \href {https://ui.adsabs.harvard.edu/abs/2015PhR...558....1T} {558, 1}

\bibitem[\protect\citeauthoryear{Waskom et~al.,}{Waskom et~al.}{2017}]{seaborn}
Waskom M.,  et~al., 2017, mwaskom/seaborn: v0.8.1 (September 2017), \mn@doi{10.5281/zenodo.883859}

\bibitem[\protect\citeauthoryear{Wasserman}{Wasserman}{2004}]{Wasserman}
Wasserman L.,  2004, All of Statistics.
Springer, \mn@doi{https://doi.org/10.1007/978-0-387-21736-9_9}

\bibitem[\protect\citeauthoryear{{Weinberg}, {Mortonson}, {Eisenstein}, {Hirata}, {Riess}  \& {Rozo}}{{Weinberg} et~al.}{2013}]{Weinberg}
{Weinberg} D.~H.,  {Mortonson} M.~J.,  {Eisenstein} D.~J.,  {Hirata} C.,  {Riess} A.~G.,   {Rozo} E.,  2013, \mn@doi [\physrep] {10.1016/j.physrep.2013.05.001}, \href {http://adsabs.harvard.edu/abs/2013PhR...530...87W} {530, 87}

\bibitem[\protect\citeauthoryear{{Zuntz} et~al.,}{{Zuntz} et~al.}{2021}]{zuntz2021}
{Zuntz} J.,  et~al., 2021, \mn@doi [The Open Journal of Astrophysics] {10.21105/astro.2108.13418}, \href {https://ui.adsabs.harvard.edu/abs/2021OJAp....4E..13Z} {4, 13}

\bibitem[\protect\citeauthoryear{{{\v{S}}ar{\v{c}}evi{\'c}}, {Leonard}, {Rau}  \& {the LSST Dark Energy Science Collaboration}}{{{\v{S}}ar{\v{c}}evi{\'c}} et~al.}{2025}]{sarcevic2025}
{{\v{S}}ar{\v{c}}evi{\'c}} N.,  {Leonard} C.~D.,  {Rau} M.~M.,   {the LSST Dark Energy Science Collaboration} 2025, \mn@doi [\mnras] {10.1093/mnras/staf156}, \href {https://ui.adsabs.harvard.edu/abs/2025MNRAS.537.1924S} {537, 1924}

\bibitem[\protect\citeauthoryear{{van den Busch} et~al.,}{{van den Busch} et~al.}{2022}]{2022arXiv220402396V}
{van den Busch} J.~L.,  et~al., 2022, arXiv e-prints, \href {https://ui.adsabs.harvard.edu/abs/2022arXiv220402396V} {p. arXiv:2204.02396}

\makeatother
\end{thebibliography}

\appendix

\section{Forecast Validation}
\label{sec:res:forecast_validation}

\begin{figure*}
    \centering
    \includegraphics[width=\textwidth]{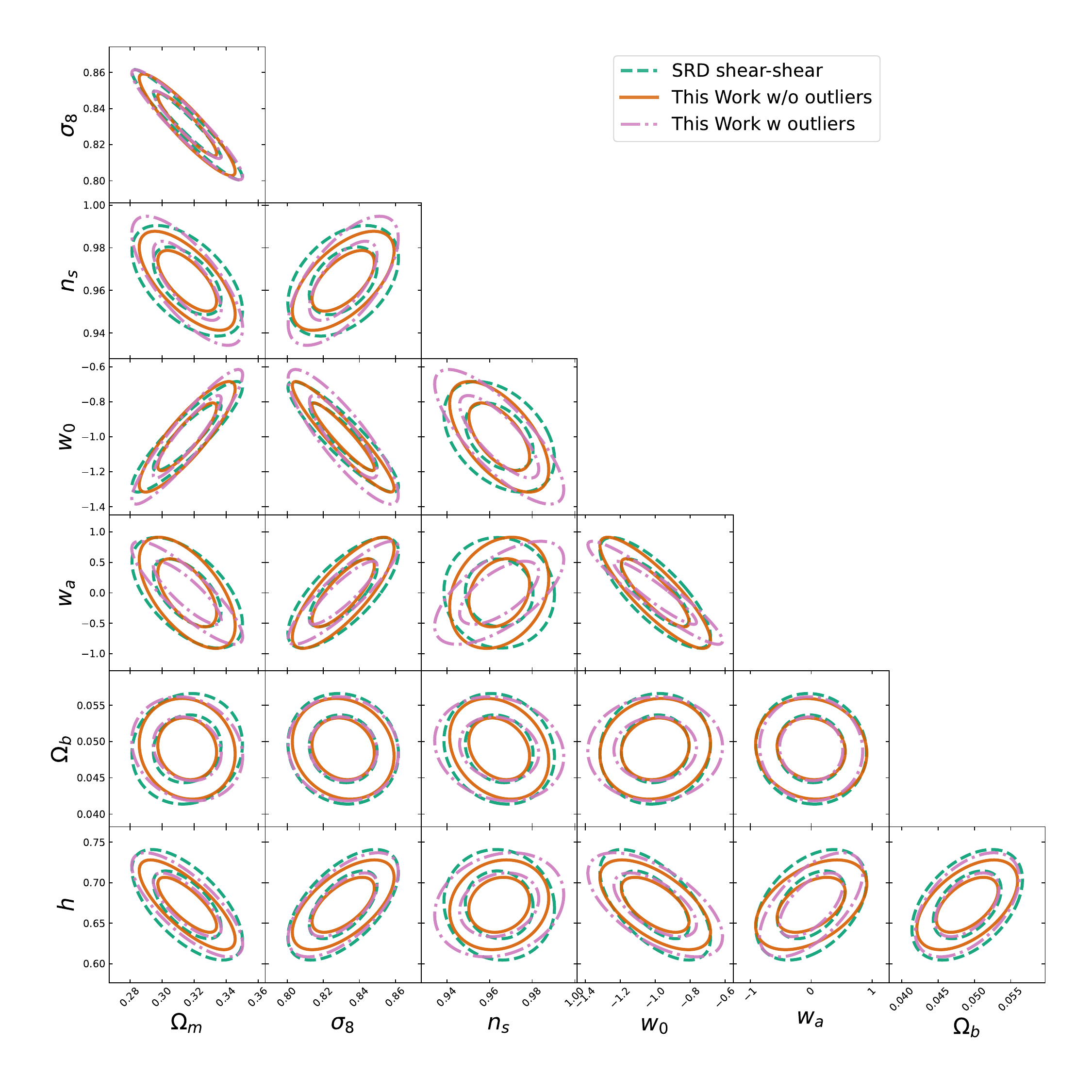}
    \caption{Comparison of the Fisher forecast on the seven cosmological parameters in this work to the results given in the DESC--SRD, for the cosmic shear analysis in LSST Y10. We see that the forecasting gives comparable results, importantly for the ($\Omega_m, \sigma_8$) and ($w_0, w_a$) planes. The minor difference is caused by the difference in implementation, especially due to the addition of the outlier distribution.  }
    \label{fig:SRD}
\end{figure*}

We validate our Fisher forecast by comparing the results with the previous DESC--SRD forecast. Due to our use of  a different parameter space for the redshift distribution (with the addition of an outlier fraction), we only compare forecasts that marginalize over the cosmological, intrinsic alignment parameters, and galaxy bias when comparing the $3 \times 2$pt results. 

In Fig.~\ref{fig:SRD}, we compare the cosmic shear forecast from the Fisher information matrix in this work without photo--$z$ marginalization to previous work in the DESC--SRD, and found them to be consistent with minor differences due to implementation differences, \refresponse{which may comes from differences between \texttt{CCL} and \texttt{CosmoLike}, and differences in redshift distribution models. }
For the cosmic shear, the 2$\sigma$ ($\Omega_m, \sigma_8$) contours have the same orientation and 93 per cent of the FoM of that in the DESC--SRD, and the $w_0 - w_a$ contour has 119 per cent of the FoM of that in the DESC--SRD. \refresponse{The direction of the $\Omega_m, \sigma_8$ degeneracy is compatible between the two analyses, while the directions of the $w_0 - w_a$ degeneracy are slightly different. }

For the $3 \times 2$pt analysis, the $\Omega_m- \sigma_8$ contour also has the same orientation and 106 per cent of the FoM of that in the DESC--SRD, and the $w_0 - w_a$ contour has 124 per cent of the FoM of that in the DESC--SRD. \refresponse{We expect these differences are due to slight differences in the implementation of the forward model. }

\section{Sensitivity to different outlier distribution}
\label{ap:outlier_distribution}

In Section~\ref{subsec:photozerror}, we use the photo--$z$ results estimated on the simulated catalog equivalent to LSST ten--year full depth catalog. We acknowledge that the actual outlier distribution of an analysis strongly depends on the combination of depth in all bands, and the redshift inference methods deployed. 

\begin{figure}
    \centering
    \includegraphics[width=0.5\textwidth]{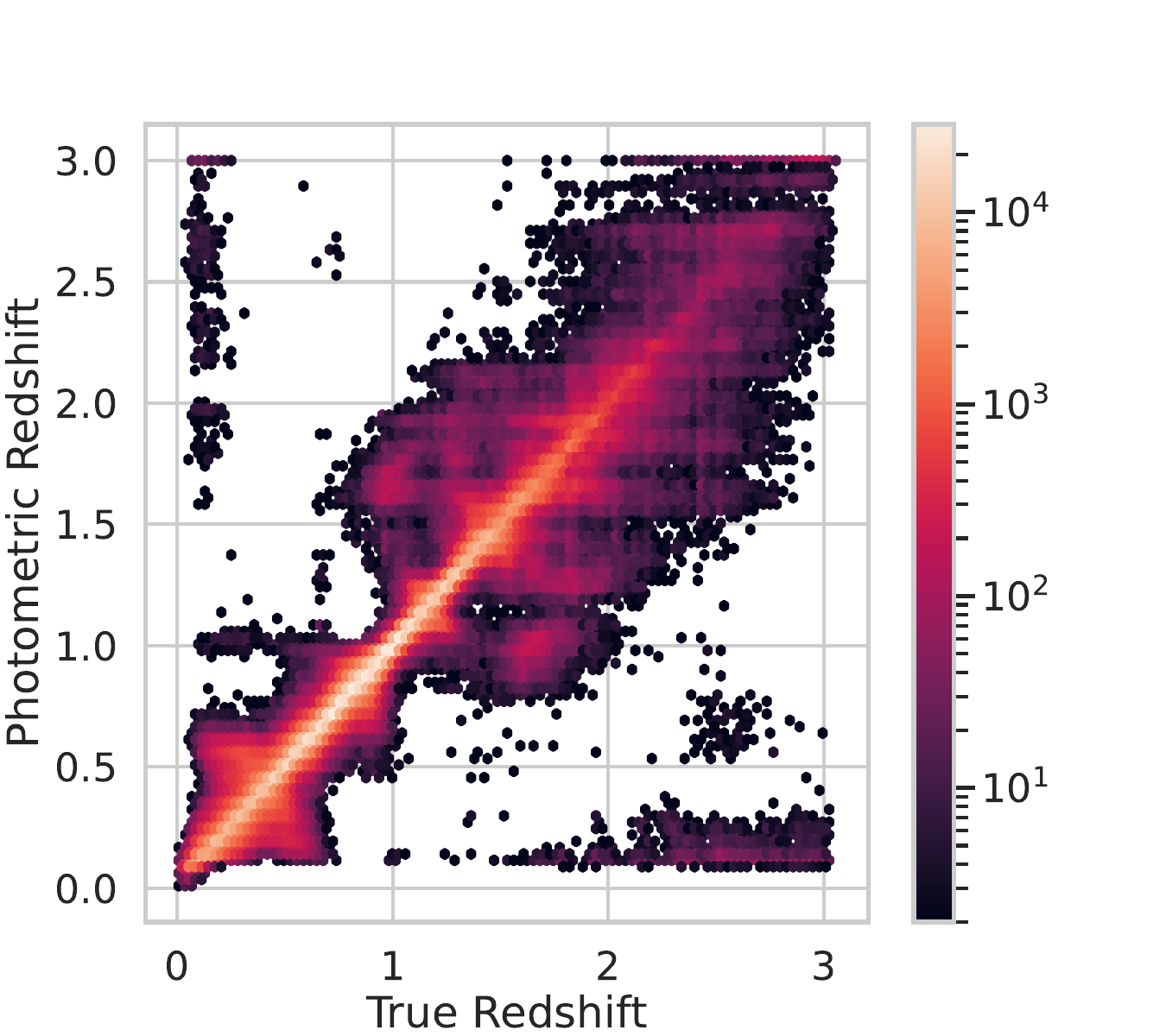}
    \includegraphics[width=0.5\textwidth]{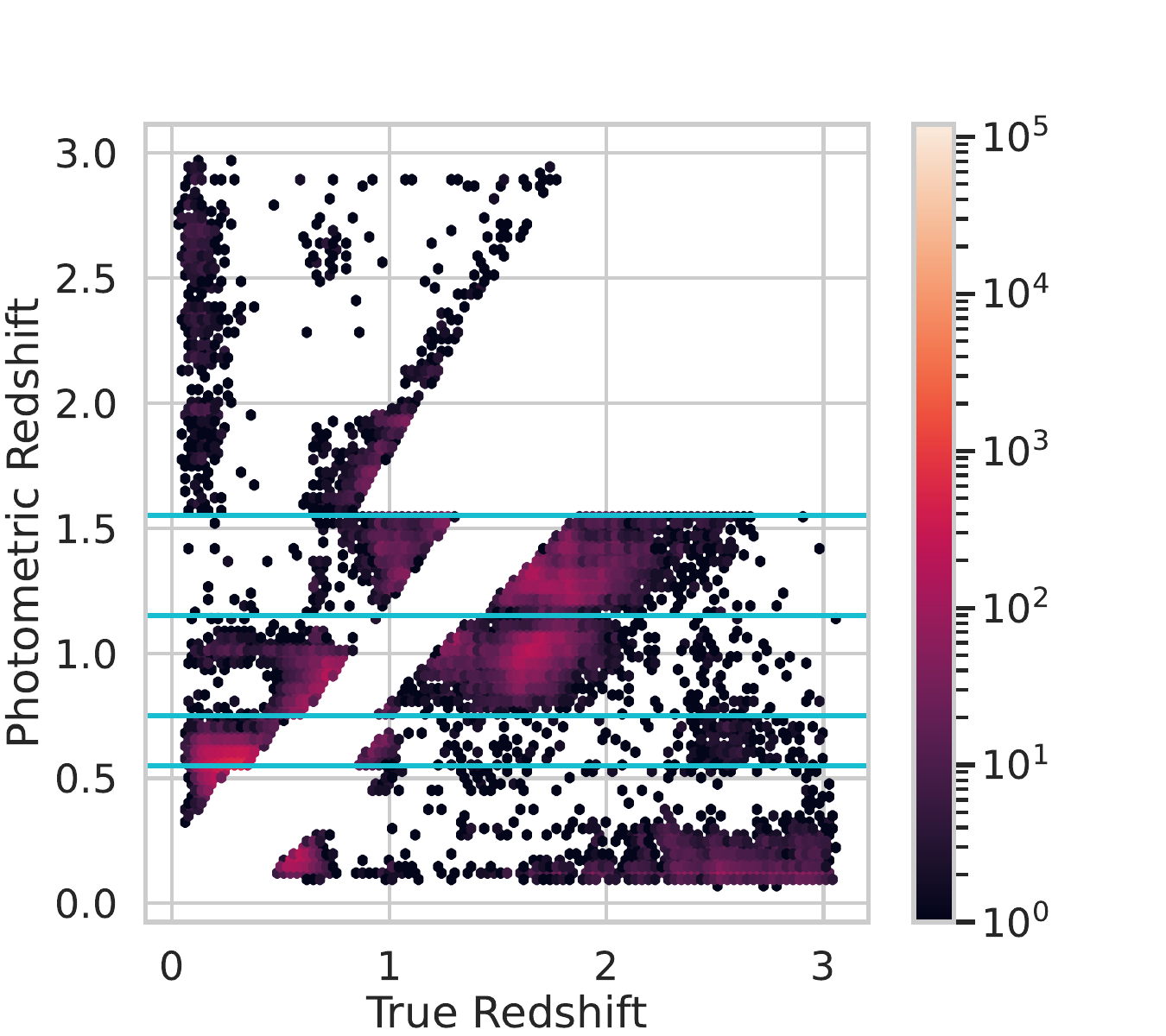}
    \caption{This figure shows the \texttt{FlexZBoost} results on the LSST Y10 gold sample, which is defined by $i$--mag $<25.3$. Compare to Figure~\ref{fig:flexzboost}, the number of outlier photo--$z$ estimates are much lower. However, there are now distinct degeneracy between galaxy at $z\sim0.3$ and $z \in [1.5 ,3.0]$, which is shown in the top left and bottom right of the second panel. 
    }
    \label{fig:gold_redshift}
\end{figure}

To ensure that our conclusion remain consistent with different experimental setup, we modified our sample selection from LSST ten--year full depth to the LSST gold sample, which corresponding to galaxies with $i$--mag $<25.3$, still estimated by \texttt{FlexZBoost}. In Figure~\ref{fig:gold_redshift}, we see that the photo--$z$ estimation are drastically better compared to the results of the full--depth sample. Therefore, for this setup we lower the fiducial outlier rate to $0.05$, a much lower value compared to $0.15$. 

We rerun our analysis from Section~\ref{sec:res:marginalization} to Section~\ref{sec:results:cosmo_bias} on this dataset, and find that our main conclusion are unchanged, even though the outlier distributions between the two experimental setup are drastically different. We suspect that different outlier distributions impact the cosmological analysis in a similar way since they mainly change the effective mean and width of the redshift distribution, thus, a similar prior will results in similar increase of contour. 

\section{Constraining power on the redshift parameters}
\label{ap:self_constrain}

\begin{table}
    \caption{\label{tab:self_constrain} The self--constraining posterior of the LSST Y10 cosmic shear and $3 \times 2$pt analysis, predicted by Fisher forecast.
    }
    \centering
    \begin{tabular}{c|c|c|c}
        Parameter & Prior & CS Y10 Posterior & $3 \times 2$pt Y10 Posterior \\\hline
        $\delta z_1 (1+z_{\rm center,1})$ &0.1 & 0.012 & 0.012  \\
        $\delta z_2 (1+z_{\rm center,2})$ &0.1& 0.021 & 0.009 \\
        $\delta z_3 (1+z_{\rm center,3})$ &0.1& 0.037 & 0.014 \\
        $\delta z_4 (1+z_{\rm center,4})$ &0.1& 0.055 & 0.021\\
        $\delta z_5 (1+z_{\rm center,5})$ &0.1& 0.15 & 0.041 \\
        $\sigma_1 (1+z_{\rm center,1})$ &0.1& 0.015 & 0.008\\ 
        $\sigma_2 (1+z_{\rm center,2})$ &0.1& 0.020 & 0.007 \\ 
        $\sigma_3 (1+z_{\rm center,3})$ &0.1& 0.027 & 0.009\\ 
        $\sigma_4 (1+z_{\rm center,4})$ &0.1& 0.024 & 0.010\\ 
        $\sigma_5 (1+z_{\rm center,5})$ &0.1& 0.048 & 0.021\\
        $f_{\text{out},1}$ &0.15& 0.004 & 0.005 \\
        $f_{\text{out},2}$ &0.15& 0.015 & 0.009\\
        $f_{\text{out},3}$ &0.15& 0.048 & 0.020\\
        $f_{\text{out},4}$ &0.15& 0.065 & 0.032\\
        $f_{\text{out},5}$ &0.15& 0.061 & 0.030\\
    \end{tabular}
\end{table}

\begin{figure*}
   \centering
   \includegraphics[width=1.0\textwidth]{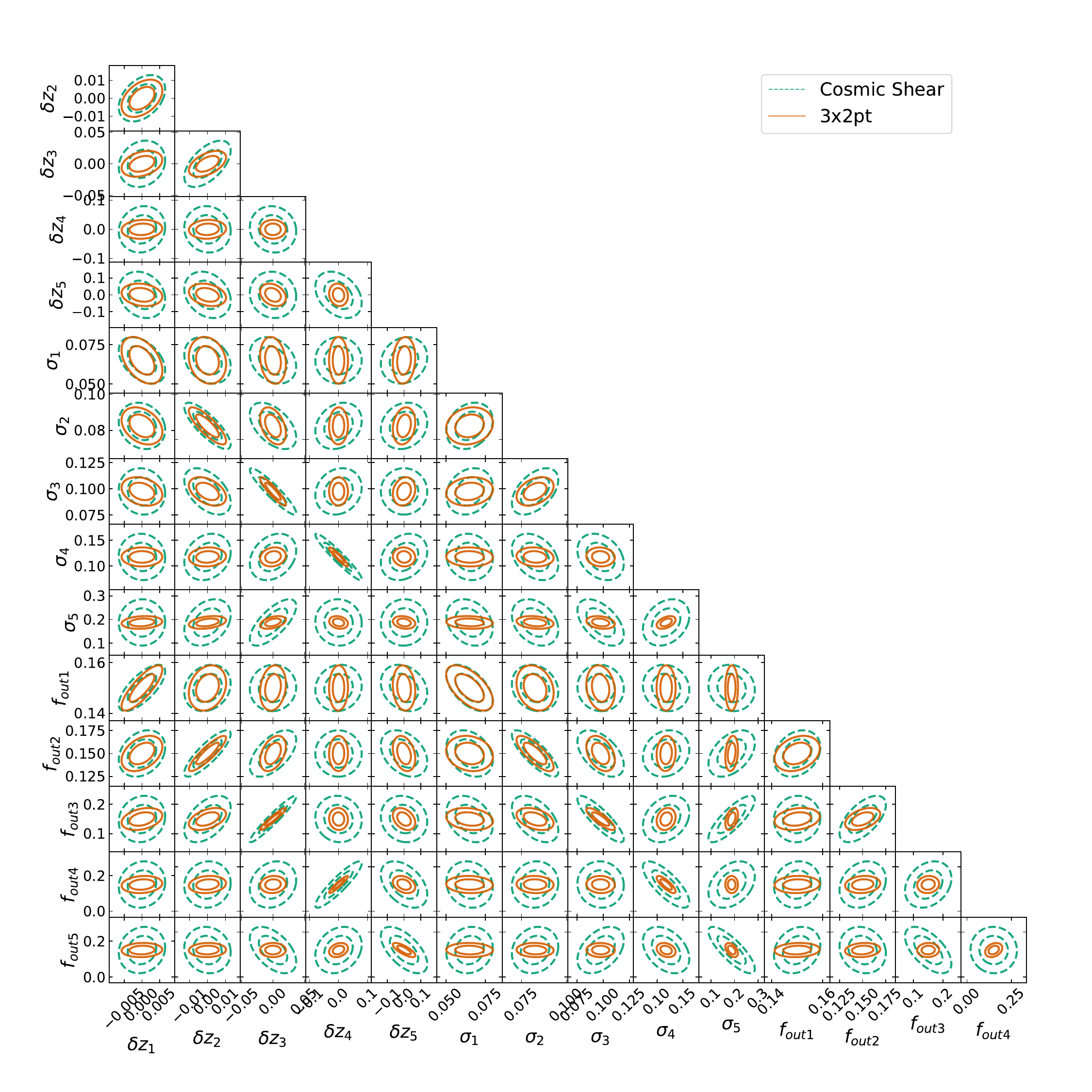}
   \caption{The 1--$\sigma$ and 2$\sigma$ contour on the bias, variance, and outlier parameters of the redshift distribution. We can see that a $3 \times 2$pt analysis is able to have better self--constraining power on the redshift parameters than the cosmic shear. We also notice that all redshift parameters are constrained to a much tighter range than their priors, suggesting that the choice of prior does not affect the results.}
   \label{fig:self_contraints}
\end{figure*}

In this section, we demonstrate the self--constraining power on the redshift parameters by the $3 \times 2$pt and cosmic shear analysis. In Fig.~\ref{fig:self_contraints}, we show the self--constraining power on the redshift parameters provided by the LSST Y10 $3 \times 2$pt and cosmic shear analysis. In Table~\ref{tab:self_constrain}, we list the $1\sigma$ of the posterior of the self--constraining power from LSST Y10 cosmic shear and $3 \times 2$pt analysis on the redshift parameters. The Most noticeable feature is that $3 \times 2$pt analysis is able to significantly better self--constraint its redshift parameters than the cosmic shear analysis. This is especially true for the variance parameters. The self--constraints on all redshift parameters are significantly tighter than their prior distribution, which suggest that the choice of prior width does not affect the overall conclusion of the paper. 

We also notice a significant degeneracy between redshift parameters of the same tomographic bins: for bin 1--4, a higher outlier rate is equivalent to a positive mean bias, and it is the opposite for bin 5. This correlation can be explained by the redshift distribution shown in Fig.~\ref{fig:source}.

\end{document}